\documentclass[twocolumns]{aa}
\usepackage{graphicx}
\begin{document}
\title{ISOCAM observations in the Lockman Hole - II}
   \subtitle{The 14.3 $\mu$m deep survey: data reduction, catalogue and source counts\thanks{Based on 
observations obtained with the {\sl Infrared Space Observatory}, an ESA science mission with instruments
and contributions funded by ESA Member States and the USA (NASA).}\thanks{Tables 2 and 3 are available 
in electronic form at the CDS via anonymous ftp to cdsarc.u-strasbg.fr (130.79.128.5)
or via http://cdsweb.u-strasbg.fr/cgi-bin/qcat?J/A+A/. Table 1 and Figures 1, 2, 9, 10, 11 
will only be published in the electronic version of the Journal.}}

   \author{G. Rodighiero\inst{1}
	  \and
 	  C. Lari\inst{2}
	   \and
          D. Fadda\inst{3}
          \and
	  A. Franceschini\inst{1}
	  \and
	  D. Elbaz\inst{4}
	  \and
          C. Cesarsky\inst{5}
          }

   \offprints{G. Rodighiero}

   \institute{
	Dipartimento di Astronomia, Universit\`a di Padova, Vicolo dell'Osservatorio 5,  I-35122 Padova, Italy
        \email{rodighiero@pd.astro.it}
         \and
	Istituto di Radioastronomia del CNR (IRA), via Gobetti 101, I-40129 Bologna, Italy
        \and
	Spitzer Science Center, California Institute of Technology, 
	Mail Code 220-6, Pasadena, CA 91126, USA
         \and
	CEA, DSM, DAPNIA, Service d'Astrophysique, F-91191 Gif-sur-Yvette Cedex, France 
         \and
	European Southern Observatory (ESO), Karl-Schwarzschild-Strasse, 2, 85748 Garching bei M\"unchen, Germany
         }
   \date{}

   \date{Received date; accepted date}

   \abstract{
We present a new analysis of the ISOCAM 14.3 $\mu$m deep survey in a 20$\times$20 
square arcmins area in the Lockman Hole. 
This survey is intermediate between the ultra-deep surveys 
and the shallow surveys in the ELAIS fields.
The data have been analyzed with the method presented by Lari et al. (2001).
We have produced a catalogue of 283 sources detected above the 5-$\sigma$ threshold, with fluxes in the interval 0.1-8 mJy. 
The catalogue is 90\% complete at 1 mJy. The positional accuracy, estimated from the 
cross-correlation of infrared and optical sources, is around 1.5 arcsec.
The search for the optical counterparts of the sources in the survey is performed 
on a medium-deep r' band optical image (5$\sigma$ depth of r'=25), making use of the 
radio detections when available. 
The photometry has been checked through simulations and by comparing the data with 
those presented in a shallower and more extended ISOCAM survey in the Lockman Hole, that
we have presented in a companion paper. 
Only 15\% of the 14.3 $\mu$m sources do not have an optical counterpart down to r'=25 mag.
We use the 6.7/14.3 $\mu$m colour as a star/galaxy separator, together with a
visual inspection of the optical image and an analysis of the observed Spectral Energy
Distribution of the ISOCAM sources. The stars in the sample turn out to be only 6\% of the sample.
We discuss the 14.3 $\mu$m counts of extragalactic sources, combining our catalogue with that 
obtained from the shallower ISOCAM survey.
The data in the two surveys are consistent, and our results fully support the claims in previous 
works 
for the existence of an 
evolving population of infrared galaxies, confirming the evident departure from non-evolutionary 
model predictions.

  \keywords{Infrared: galaxies -- Galaxies: photometry, statistics, evolution -- Cosmology: observations -- Methods: data analysis
               }

   }
  \authorrunning{G. Rodighiero et al.}
  \titlerunning{ISOCAM observations in the Lockman Hole II}
   \maketitle
%

\section{Introduction}

The recent space missions (COBE, IRAS and ISO) devoted to the study of the Universe at long wavelengths
have demonstrated that the infrared (IR) and sub-millimeter emission from galaxies is a
main component of the energy budget of the extra-galactic background (Hauser \& Dwek 2001).  
Actually, the cosmic infrared background (CIRB) includes an energy 
density comparable to that in the UV/optical background (Lagache et al., 1999; Bernstein et al. 2002) and 
is interpreted as the integrated emission by dust present in distant galaxies.
The principal processes generating the IR emission are star formation and AGN activities.
The relative contribution of the two is one of the key elements for the evolutionary
models proposed up to now (i.e. Rowan-Robinson et al. 2001, Franceschini et al. 2001,
Chary \& Elbaz 2001, Xu et al. 2003).
IR galaxy counts based on the IRAS data (Rowan-Robinson et al., 1984; Soifer et al., 1984) 
showed a marginally significant excess of faint sources with respect to 
``no-evolution'' models (Hacking et al., 1987; Franceschini et al. 1988; Lonsdale et al.,
1990;  Gregorich et al., 1995; Bertin et al, 1997), but did not provide
enough statistics and dynamic range in flux to discriminate between evolutionary scenarios.
More recently, the improved resolution and sensitivities of the Infrared Space Observatory
(ISO, Kessler et al. 1996) detectors have provided deeper data that have been used to constrain
the model predictions (such as source counts, redshift distributions and colours).
The studies conducted on the mid-IR population have shown that these sources show remarkable
properties compared with optically selected galaxy populations. The ISOCAM (Cesarsky et al. 1996)
14.3 $\mu$m source counts obtained by different surveys, over a wide flux range, indicate evidence
of strong evolution at flux densities fainter than $\sim$2 mJy (i.e. Elbaz et al. 1999;
Flores et al. 1999; Lari et al. 2001; Metcalfe et al. 2001).
Some evidence of evolution has also been detected by ISO at longer wavelengths
(e.g. at 175 $\mu$m --  Efstathiou et al. 2000, Dole et al. 2001, at 95 $\mu$m  -- Rodighiero et al. 2003,
Rodighiero \& Franceschini 2004).

In order to understand the nature of the mid-IR sources,
Franceschini et al. (2003) and Elbaz et al. (2002) have matched the statistical
and IR-spectral properties of the faint ISO sources detected at $\lambda_{eff}=14.3\mu$m
with the spectral intensity of the CIRB and argue that, due to their high luminosity
and moderate redshifts ($z\simeq$ 0.5 to 1.3), these mid-IR sources represent the main
contributors to the CIRB.
Some evidence (Elbaz et al. 2004) suggests that most of the stars in today's
galaxies were formed during an active infrared phase, enhanced by the local environment. 
The CIRB is interpreted as a ``fossil record'' of these luminous infrared events (Elbaz \&
Cesarsky 2003), and it reflects the influence of the star formation induced by the 
interaction among galaxies.
The same conclusion is reached by Franceschini et al. (2003), who 
compared the on-going star formation activity with the already formed stellar masses to estimate
the time scales $t_{sf}$ for the stellar build-up. From a morphological point of view, the faint ISOCAM 
galaxies appear to form a composite population, including moderately active but very massive 
spiral-like galaxies, and very luminous ongoing starburst irregular galaxies, in a 
continuous sequence. From the observed $t_{sf}$ and assuming typical 
starburst time scales, Franceschini et al. (2003) infer that only a fraction of the galactic
stars can be formed in each single starburst event, while several of such episodes during a 
protracted SF history are required for the whole galactic build-up.

The mentioned scenario is supported by the results of optical identifications 
of the faint sources detected in the deep ISOCAM surveys. Indeed, most of these
sources are identified with relatively bright optical counterparts with a median 
redshift of z$\sim 0.7$. They are mainly dusty starbursts with large star formation 
often triggered by interactions ($\sim$30\% of them show optical signatures of interactions,
Flores et al. 1999). 

To explore these subjects, ISO has deeply surveyed sky regions where the
HI absorption is very low. The Lockman Hole (Lockman et al. 1986), in particular,  
was selected for its high ecliptic latitude ($|\beta|>50$)  to keep the Zodiacal
dust emission at a minimum, and for the low cirrus emission.
This region presents the lowest HI column density in the sky, hence is
suited for the detection of faint IR sources.
The data that we have analysed in the present work are a combination
of a deep survey of 20$\times$20 square arcminutes within the Lockman Hole region, 
and a shallower ISOCAM survey of 40$\times$40 square arcminutes centred in the same
position and completely overlapping the deeper one.
The results on the complete Shallow region are presented in a separate
paper (Fadda et al. 2004, hereafter Paper I). Here we consider only the data in the smaller
and deeper combined area.
The deep and shallow surveys of the Lockman Hole were performed as
part of the ISOCAM guaranteed time extragalactic surveys
(IGTES). Together with deeper surveys from the IGTES, they were used
to build the number counts published in Elbaz et al. (1999). In this
paper, we present the new 14.3 $\mu$m catalogue, obtained by analyzing
the ISOCAM data with the technique described by Lari et al. (2001). 

With this deep data-set we will span the flux range ($\sim$0.5 - 1 mJy) where the slope 
of the ISOCAM source counts changes and strongly deviates from the Euclidean regime.
The properties of our sample will then provide strong constraints on the evolution 
models for the IR galaxy populations.

The Spitzer Space Observatory (Werner et al., 2004) has recently observed the same
region in the Lockman Hole in complementary spectral channels at 24, 70 and 160$\mu$m with MIPS
and in the near-IR with IRAC. This will provide additional extensive information on the Spectral
Energy Distributions (SEDs) of ISO sources, complementary to the ISOCAM 14.3 $\mu$m information,
given that the nearest Spitzer bands are centred at 8 and 24 $\mu$m. 

The paper is organized as follows. Section 2 gives a
summary of the ISOCAM observations.  Section 3 describes the method used 
and summarizes the steps of the reduction, the map projection and the source extraction.
The photometry is presented in Section 4, and the absolute calibration 
is presented in section 5.
In Section 6 the mid-IR colours are discussed and Section 7 describes the
optical identifications of the infrared sources, together with the position
accuracy. The full catalogue is finally presented in Section 8 and the
extra-galactic counts at the bright flux levels are discussed in Section
9. A summary of the work is then reported in Section 10.

Forthcoming papers will discuss the redshift distribution of the
sources using spectroscopic and photometric redshifts, the
cross-correlation with the radio data and the relationship between
star-formation and the mid-IR flux.

\section{Observations}
\label{obs}
\begin{figure}
\centering
\includegraphics[width=0.5\textwidth]{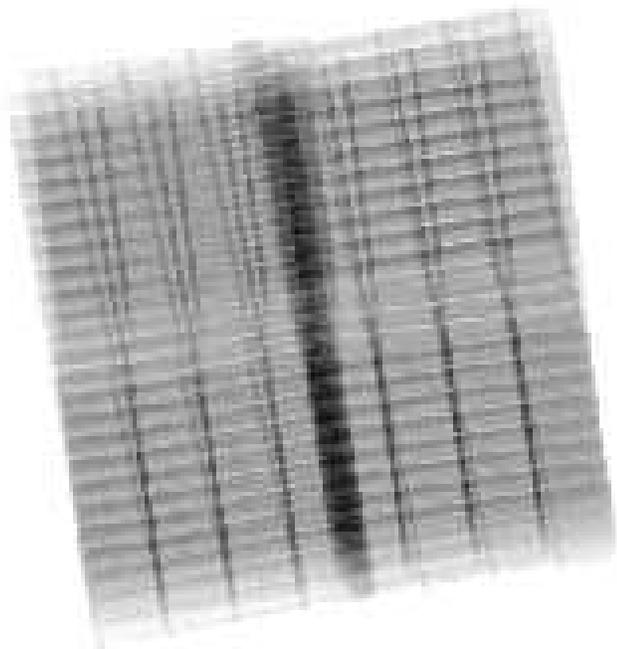}
\caption{Lockman Deep 14.3 $\mu$m sky coverage. Black regions correspond to $\sim$2000 seconds
of exposure time, grey regions to $\sim$700 seconds.}
\label{fig:sky_map}%
\end{figure}

The Lockman Hole, an area with the lowest HI column density and cirrus emission ($<$0.4MJy/sr, Lockman et al. 1986), has been 
observed by ISOCAM (on board the ISO satellite) at 6.7 and 14.3 $\mu$m over an area of 20$\times$20 
square arcminutes centred at 10:52:07 +57:21:02 (J2000), corresponding to the centre of the 
ROSAT HRI image. 
The field was deeply observed for a total of 45 ks at 14.3 $\mu$m (LW3 filter) and 70 ks at 6.75 $\mu$m
(LW2 filter). 
The LW3 observation was done in raster mode: the squared configuration is composed of four
sub-quadrants each repeated twice, for a total of 8 independent rasters.
The observation parameters are reported in Table \ref{tab:obsparam}.
In addition, a shallower survey at the same central position was done at 14.3$\mu$m on a region of 
40$\times$40 square arcminutes for a total exposure time of 55 ks (Paper I).
In this paper we present our new analysis and data reduction of the data in the combined area. 
The sky coverage is presented in Figure \ref{fig:sky_map}.
The same field has been observed with the ISOPHOT instrument, at 90 $\mu$m and 170 $\mu$m 
(Rodighiero et al. 2003, Kawara et al. 2004). These same sources will be observed  
by IRAC and MIPS in the Spitzer Guaranteed Time.

\begin{table}
 \caption[]{ISOCAM Deep Lockman Hole observation parameters.}
\label{tab:obsparam}
\begin{tabular}{l c}
   \hline
   \noalign{\smallskip}
   Parameter      &  LW3  \\
   \noalign{\smallskip}
   \hline
   \noalign{\smallskip}
   Band effective wavelength             & 14.3 $\mu$m \\
   Band width                            &  6   $\mu$m \\
   Detector gain                         &  2    \\
   Integration time                      &  5 s  \\
   Pixel field of view                   &  6''  \\
   Nr. of horizontal and vertical steps  &  11 $\times$ 4  \\
   Step sizes                            &  54'', 168'' \\
   Nr. of raster maps                    &  8    \\
   Total area covered                    &  0.14 deg$^{2}$ \\
   \noalign{\smallskip}
   \hline
\end{tabular}
\end{table}

In order to study the ISOCAM sources, a deep Sloan r' band image was
obtained with the Wide Field Camera (WFC) of the Isaac Newton Telescope
at La Palma, Spain, for a total of 4 hours of integration (Fadda et al., in preparation).

The images have been reduced using IRAF packages (see Paper I for details).
The photometric zero-point has been evaluated using the standard stars
in the night with the best transparency and every image has been scaled
to an image taken during this night.

The mean seeing is $\sim$1.3 arcsec, and the magnitude limit is around 25 mag
(computed within a circular aperture of 1.35xFWHM at a 5-$\sigma$ level).

We have run SExtractor (Bertin \& Arnouts 1999) on this image to get
the r' band magnitudes: we adopted a 3$\times$FWHM aperture magnitude and the 
$mag\_auto$ (i.e. the Kron fluxes) of SExtractor for extended sources.

\section{Data reduction}
\label{lari}
The ISOCAM data presented in this work have been reduced with the method discussed in Lari et al. (2001). 
This procedure has been recently successfully applied to other deep and shallow surveys 
(Gruppioni et al. 2002, Pozzi et al. 2003, Vaccari et al. 2004). The description of the data 
reduction has been described in the cited articles, and a general discussion of the strategy that
we used for the Lockman observations is also reported in Paper I.

\begin{figure*}
\centering
\includegraphics[width=1.0\textwidth]{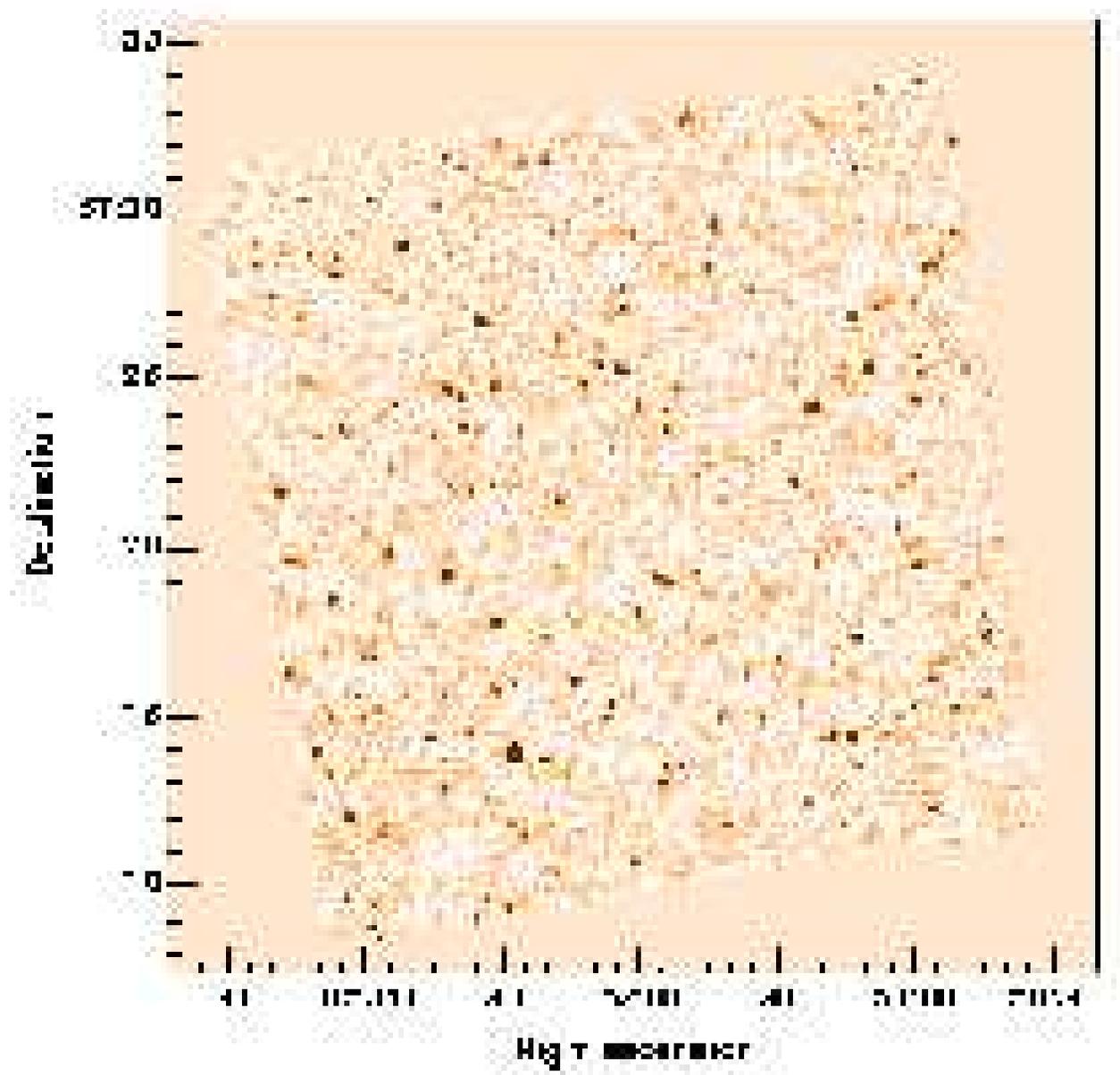}
\caption{Signal to noise map of the total field observed by ISOCAM in the
direction of the Lockman Hole. }
\label{fig:LH_map}%
\end{figure*}

\subsection{The reduction pipeline}
Summarizing, the reduction of ISOCAM data requires a careful treatment of various external and 
instrumental effects, in particular cosmic ray impacts ({\it glitches}) and detector hysteresis 
(i.e. the slow response of the detector to flux variations). 
To deal with all of them, the method discussed by Lari et al. (2001) was based on the assumption that 
the incoming flux of charged particles generates transient behaviours with two different time scales: 
a fast and a slow one. 

The method basically consists of looking at the time history of each detector pixel and 
identifying the stabilisation background level. Then it models the glitches, the background and the 
sources with all the transients over the whole pixel time history. 
 
The data are first corrected for short-time cosmic rays.  
The general background is then estimated as the stabilisation level along the whole 
time history of each pixel, and an initial guess of the fitting parameters is performed
(see Vaccari et al. 2004).
The signal as a function of time is finally processed, independently for  
each pixel. The fitting procedure models the transients along the time 
history, and the features on both short and long time scales produced by 
cosmic ray impacts.  Furthermore, the code recognizes sources (above a  
given threshold level) and recovers all the time histories $reconstructing$  
the local background as it would appear in the absence of glitches. 

The fitting algorithm starts with the brightest glitches  
identified in the pixel time history, assumes discontinuities at these positions, and tries  
to find a fit to the time history that satisfies the model 
assumed to describe the solid-state physics of the detector. 
By successive iterations, the parameters and the background for each  
pixel are adjusted to better fit the data, until the rms of the difference  
between model and real data is smaller than a given amount.  

After the first run of the automatic fitting procedure, the next step is  
the interactive mending of fitting failures. 

\subsection{Map projection}
Once a satisfying fit is obtained for all the pixels over the whole pixel 
history, the pipeline generates sky maps. An image for 
each raster position is created, by averaging the signals of all the  
readouts relative to that pointing for each pixel. The signal is then 
converted to flux units (mJy pixel$^{-1}$), glitches and bad data are masked 
and the images are then combined to create the final raster maps (one for  
each raster position). These images are projected and combined onto a sky map 
(raster image) using the projection algorithm available for 
ISOCAM data in the CIA package (Cam Interactive Analysis, Ott et al., 2001). 
When projecting the signal on the sky, we apply to the nominal astrometry
of each raster a median offset computed from the positions of those sources
in common with the Lockman Shallow catalogue (see the astrometric calibration
presented in Paper I).
 
The redundancy of Lockman ISOCAM observations allowed us to  
generate high resolution maps, rebinning the original data into a  
final map with pixel size of 2$\times$2 arcsec.
The detector signal is distributed in a uniform way between the
smaller pixels. This process allows a better determination of source positions. 
The signal to noise map of the 20'$\times$20' field observed by ISOCAM in the
direction of the Lockman Hole is shown in Figure \ref{fig:LH_map}.

\subsection{Source extraction}
\label{sec:extr}
As the adopted procedure (Lari et al. 2001) considers only the peak flux, we performed
the source extraction using 6$''$ pixels sampled at a distance of 2$''$.
Thus, the pixels of the image used for the extraction have a size 2$''\times$2$''$
(see also Paper I).
The source detection is performed on the signal-to-noise maps:
our procedure selects pixels above a low flux threshold (0.5 $\mu$Jy pixel$^{-1}$) using the 
IDL Astronomy Users Library task called $find$ (based on DAOPHOT's 
equivalent algorithm). 
Then we extract from the selected list only those objects with signal-to-noise ratio $>$ 3. 
 
In the final stage of our reduction we perform simulations in order to 
re-project the sources detected on the raster map onto the pixel time history. 
In this way we are able to check the different temporal positions supposed  
to contribute to the total flux of each source. This procedure provides two main
advantages: firstly, it allows us to recover the flux losses and obtain a cleaned measure
of the total flux; secondly, the visual inspection of the temporal pixel histories
improve the rejection of spurious detections (see Vaccari et al. 2004 for a more
comprehensive description of Lari's procedures).
We have simulated all sources detected with a signal-to-noise ratio $>$3.

To strengthen the reliability of the Lockman catalogue, we have combined our reduction procedure
with the performances of the PRETI method (Starck et al. 1999, see also Elbaz et al 2004 in prep.
for a comparison of PRETI and the technique by Lari et al., 2001). 
This tool is based on a wavelet analysis which allows the discrimination of mid-IR sources from cosmic ray 
impacts at the very limit of the instrument. It is called PRETI 
for Pattern REcognition Technique for ISOCAM data, because glitches with transient behaviours are isolated 
in the wavelet space, i.e. frequency space, where they present peculiar signatures in the form of patterns 
automatically identified and then reconstructed.
We have run PRETI on the ISOCAM 14.3 $\mu$m data used in this paper and extracted an independent list
of sources. We started our own reduction by checking these brightest PRETI detections
with simulations along the pixel time histories.
We then proceeded by checking all other fainter detections in our own source list.
This approach stabilizes the completeness at bright flux levels, and
it has been applied to ISOCAM ultra deep observations (Lari et al 2004, Elbaz et al 2004).

\section{Photometry}
The flux density $S_{\nu}$ at 14.3$\mu$m of a source is computed by applying a correction 
factor to the measured peak flux $f_s$ in order to have a measure 
of its ``total'' flux (Lari et al., 2001). $f_s$ is the value of the map 
in the position where the source is detected. It is computed from
a cubic interpolation of the data (in units of $\mu$Jy/pixel, being the pixel
of 6$\times$6 arcsec rebinned in 3$\times$3 subpixels).

The correction factor to obtain the total flux is computed by simulating a source (with a flux
similar to that measured) in the detected position (see the autosimulation
process in Lari et al 2001 and Vaccari et al. 2004). By applying this procedure
to all the sources in our list, we can deduce the total flux for each entry in our
catalogue.
Since most of the sources are distant faint sources and the pixel field of view
is quite large (six arcseconds), the method is applied to almost every source
with a few exceptions. For extended sources aperture photometry is used.

A comparison of the use of $f_s$ with respect to aperture photometry has been discussed
by Lari et al. (2001) and Vaccari et al. (2004).
The implications for the choice of using only the central brightest pixel instead of 
an aperture on the central position of the source, when computing the total fluxes,
will also be discussed in a forthcoming paper (Elbaz et al., 2004).

Extensive simulations in wider samples (e.g. ELAIS, Gruppioni et al. 2002)
have shown that various statistical effects (such as the fact that the actual position of
an infrared source is poorly known or the projection process itself) prevent
the accurate reconstruction of the total flux of a source. However, with simulations 
we can quantify this bias (called $q_{med}$) and statistically recover flux losses. 
In Paper I, we discuss the simulations performed in the Lockman Shallow ISOCAM survey:
the derived value of $q_{med}$ is 0.84. This means that the total flux computed after
the autosimulation has to be incremented by 16\%.

In our new analysis of the Lockman Deep observations, we choose to apply the same factor derived
for the Shallow case. We have separately quantified the effects of overlapping rasters
on the flux lost. 

The Shallow map presented in Paper I is composed of four rasters with only small
overlap. On the contrary, our Deep map is the combination of twelve independent rasters, 
divided in four quadrants, each made of three rasters completely overlapping (Section \ref{obs}).

\begin{figure}
\centering
\includegraphics[width=0.5\textwidth]{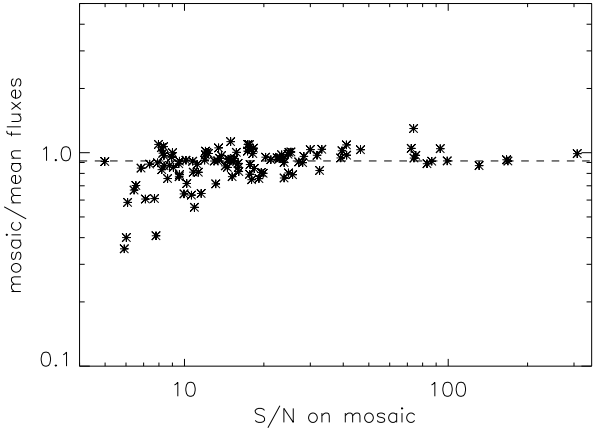}
\caption{115 sources with three independent detections 
have been used to compute the photometric bias caused
by the mosaicing of repeated rasters. The ratio between the 14.3 $\mu$m mean fluxes
and the corresponding mosaic fluxes are plotted as a function of their signal to noise ratio on the mosaic.
The horizontal line indicates the median ratio (bias\_mos=0.915) that we will
use to correct the total fluxes extracted from the mosaic. }
\label{fig:bias_mos1}%
\end{figure}

Performing the source extraction on the single rasters and on the combined map,
we can estimate the flux for each source in the single and in the mosaic maps.
In Figure \ref{fig:bias_mos1} we compare the ratio between the 14.3 $\mu$m   
mean fluxes computed from the single rasters and that from the final combined mosaic, 
as a function of their signal-to-noise ratio on the mosaic. 
We take into account only sources detected in at least three rasters.
This ratio is almost constant down to a
signal-to-noise $>$10.5, the scatter at fainter values being mainly due to the
noise that dominates the determination of fluxes in the single rasters.

As indicated by the plot, the fluxes extracted from the mosaic map tend to be
statistically lower than that measured on the single maps. This can be due to 
small fluctuations on the astrometric offsets applied to the rasters, and to the
projection itself: the mean effect is the broadening of the source in the 
mosaic, with a corresponding lower peak flux.

Considering only those sources with three independent detections, we can
constrain the mean correction factor for recovering the underestimated fluxes. 
This is done in Figure \ref{fig:bias_mos1}, where the horizontal line
shows the median value of the ratio, that is $bias\_mos$=0.915.  We adopt this
factor to further correct the total fluxes. 

In the following, ``measured fluxes'' will refer to the total fluxes
corrected for the flux bias ($q_{med}$) and the mosaicing bias ($bias\_mos$).

\section{Calibration of the LW3 deep catalogue}
\label{rel_aussel}

\begin{figure}
\centering
\includegraphics[width=0.5\textwidth]{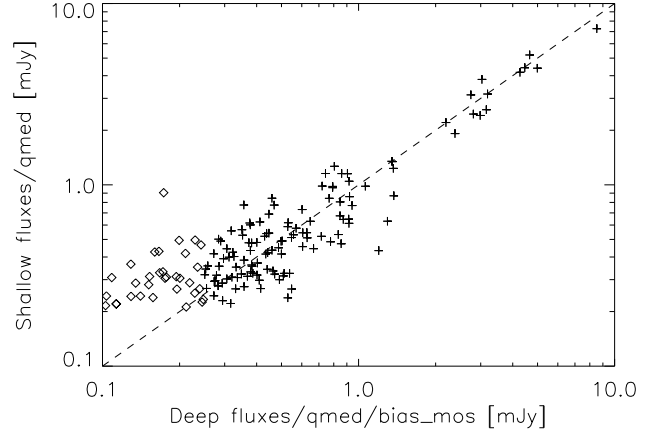}
\caption{Correlation of 14.3 $\mu$m total fluxes for the sources in common in the Shallow
and in the Deep catalogues. Fluxes in the Deep have been corrected for
the effect of the combination of repeated rasters ($bias\_mos$=0.915) and for the
same flux bias as in the Shallow catalogue ($q_{med}$=0.84).
For sources above 0.25 mJy (crosses) the correlation follows the unitary relation (dashed 
line).}
\label{fig:flux_sh1}%
\end{figure}

The corrections that we have applied to the 14.3 $\mu$m total fluxes ($q_{med}$ and $bias\_mos$) should
have gives the fluxes of the sources in the Deep catalogue on the same scale as those in the
Shallow catalogue. This is confirmed by directly comparing the total fluxes measured for the 
sources in common in the two samples. 
Figure \ref{fig:flux_sh1} shows that for sources brighter than 0.25 mJy (crosses) the correlation follows 
the unitary relation (dashed line). Below 0.25 mJy in the Shallow survey
only bright sources located over positive noise peaks are detected and there flux enhanced (see discussion
in Paper I).

In Paper I we have checked that the LW3 fluxes of the Lockman Shallow (corrected for the flux bias
$q_{med}$=0.84) are consistent with the predictions of stellar atmosphere models at 14.3 $\mu$m, when
considering the multi-wavelength photometry (optical, near- and mid-infrared) of the stars in the sample.
Here, we compared the 14.3 $\mu$m total fluxes with the predictions by Aussel \& Alexander (2001),
who performed a detailed study of mid-infrared emission from stars. They exploited large samples
from the IRAS Faint Source Catalog with counterparts in the 2MASS and Tycho-2 (Hog et al. 2000) catalogues.
One of their findings is that the $J-K$ colour of stars is extremely well correlated with the
$K-[12]$ colour, where $[12]$ is a magnitude scale constructed from the IRAS 12$\mu$m flux
(Omont et al. 1999).
This relation allows the accurate prediction of the 12 $\mu$m IRAS flux of a star, provided that its
$J-K$ is known and within a certain limit. Stellar atmosphere models (Lejeune, Cuisinier \& Buser 1998) 
indicate that the ratio of the 14.3 $\mu$m flux to the 12 $\mu$m flux of stars is constant, at least for 
spectral types hotter than K3. 
We have adopted this criterion to predict the 14.3 $\mu$m fluxes of the stars in the Lockman Shallow 
(see Paper I) and in the deeper area. The relations that we used in order to predict ISO 14.3 $\mu$m fluxes
are:
 
\begin{equation}
K-[LW3]=0.044+0.098(J-K),
\end{equation}
where $[LW3]$ is a magnitude scale defined as

\begin{equation}
[LW3]=3.202-2.5 $log$ (S_{LW3}[$mJy$]).
\end{equation}

\begin{figure}
\centering
\includegraphics[width=0.5\textwidth]{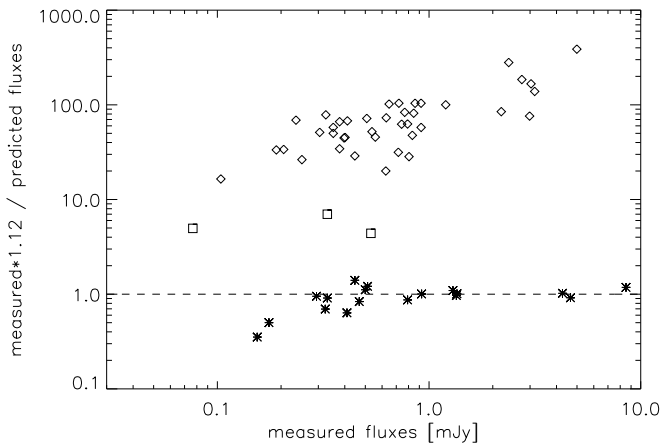}
\caption{The LW3 measured fluxes are compared with the corresponding ratio of
the measured over predicted fluxes. This makes sense for sources with 
reliable K and J magnitudes (see text for details). As expected, the ratio
is close to 1 for stars (asterisks), and greater for galaxies (diamonds). 
The intermediate objects (squares) should correspond
to elliptical galaxies.  }
\label{fig:flux_aussel}%
\end{figure}

We have obtained the $J$ and $K$ magnitudes for 61 sources in our Deep sample from the deep 
2MASS Survey of the Lockman Hole (Beichman et al. 2003).
Using the previous equations, we have computed the predicted 14.3 $\mu$m flux for these  
sources and compared it with the measured one. 
The result is presented in Figure \ref{fig:flux_aussel}: the 14.3 $\mu$m measured 
fluxes are compared with the corresponding ratio of the measured over the predicted fluxes. 
As expected, the ratio is close to 1 for stars (asterisks), and greater for galaxies 
(diamonds). The intermediate objects (squares) should correspond to elliptical galaxies.
We have verified that the 18 sources with ratios close to 1 are all stars on the optical image.  
The median of this ratio for stars is 1.12, as found in Paper I for the stars in the Shallow sample. 
Given the poor statistics (18 objects), this value is 
consistent within 1-$\sigma$ with that obtained on much wider samples for the ELAIS surveys: 
Gruppioni et al. (2002) and Vaccari et al. (2004) found a constant factor of 1.0974.
This $\sim 10\%$ should be attributable to discrepancies between the independently established
IRAS and ISO calibrations.

We decided to apply this factor to the fluxes when computing the galaxy counts (Section \ref{sec:counts}),
in order to compare different samples on the same flux scale.

\section{LW2/LW3 colour}
\label{colcol}

As mentioned in Section \ref{obs}, the Lockman Deep area has been observed both at 14.3 and
at 6.7 $\mu$m. We have applied the reduction procedure described in Section \ref{lari} only
to the longer wavelength ISOCAM data, however exploiting the PRETI pipeline we obtained 
a LW2-6.7 $\mu$m map and checked for the counterparts of the LW3-14.3 $\mu$m list (the same dataset has
also been used in Fadda et al. 2002).
Starting from the signal to noise map, we applied the same extraction tool as we did for the LW3
band (Section \ref{sec:extr}) and obtained a catalogue with the measure of the peak fluxes.

We computed the correction factor to derive the total fluxes using as a calibrator the
relations by Aussel \& Alexander (2002), as we did for the 14.3 $\mu$m (Section \ref{rel_aussel}).
In this case the 6.7$\mu$m predicted fluxes were obtained through the following expressions:

\begin{equation}
K-[LW2]=0.044+0.098(J-K),
\end{equation}
where $[LW2]$ is a magnitude scale defined as

\begin{equation}
[LW2]=4.860-2.5 $log$ (S_{LW2}[$mJy$]).
\end{equation}

\begin{figure}
\centering
\includegraphics[width=0.5\textwidth]{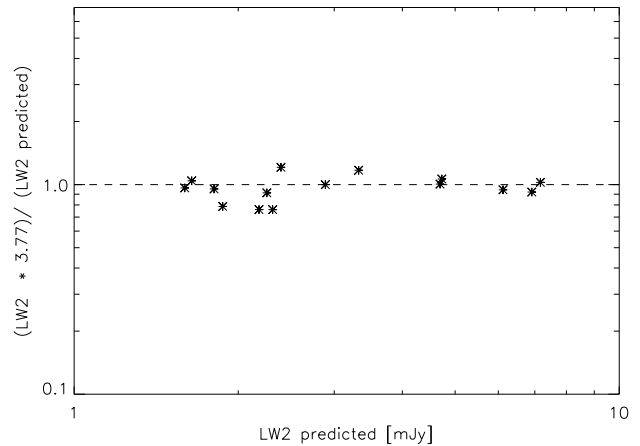}
\caption{We have calibrated the LW2-6.7 $\mu$m fluxes against model predictions 
(Aussel \& Alexander 2002).
The figure shows that for the stars with $J$ and $K$ counterparts, the 6.7 $\mu$m total fluxes 
are derived by multiplying for a factor 3.77 the corresponding peak fluxes.}
\label{fig:flux_lw23}%
\end{figure}

The 17 sources with $J$, $K$ and 6.7 $\mu$m detections and optical stellar shapes are used to calibrate the
6.7 $\mu$m total fluxes: we found the correction factor to be 3.77 (see figure \ref{fig:flux_lw23}). 
We did not perform other analyses to derive a more accurate absolute calibration. In fact
a relative scale is enough for our purpose, as we are mainly interested in using the 6.7/14.3 $\mu$m
colour as a star/galaxy separator.

\begin{figure}
\centering
\includegraphics[width=0.5\textwidth]{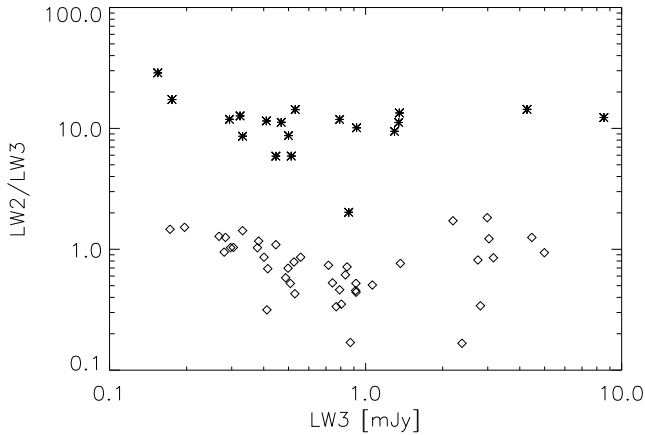}
\caption{For the 63 sources with both LW2 and LW3 detections, the 14.3 $\mu$m fluxes are plotted 
against the ratio of the 6.7 and 14.3 $\mu$m fluxes. As a first approximation,
this ratio can be used as a star/galaxy indicator. Sources with high LW2/LW3
should correspond to stars (asterisks in the figure). }
\label{fig:LW23}%
\end{figure}

By cross-correlating the LW2 and LW3 catalogues, we found 63 objects with both detections:
their colour-magnitude diagram is plotted in Figure \ref{fig:LW23}.

In the plot the regions occupied by stars (asterisks) and galaxies (open diamonds) are clearly
separated, as expected given the different spectra of the two classes.
The consistency of this approach is supported by the consideration that the 18 stars identified
with the LW2/LW3 ratio are exactly the same found from the predictions of the 14.3 $\mu$m fluxes
(see Section \ref{rel_aussel} and Figure \ref{fig:flux_aussel}).

The identifications of stars in the sample is important, given that these objects must be
neglected when computing galaxy counts (Section \ref{sec:counts}).

\section{Optical identifications}
\label{id_opt}

One of the best ways to look at the properties of faint infrared sources and study
their evolution is to check their optical morphologies.
Although only deep space observations allow a comprehensive view of the fainter population,
we have attempted to perform the optical identifications of the Lockman Deep sources.
Given the good quality of the optical R image (R$_{\lim}$=25 mag, Section \ref{obs})
we are able to make reliable associations with the infrared sources and the corresponding optical counterparts.
The mean separation between the optical and the infrared position is of the order of 1.5$''$, 
the maximum separation being $\sim 4''$ (see discussion below).

We assume the r' band catalogue as our reference optical catalogue, which is used to search for the optical 
counterparts of the ISO sources using the likelihood ratio technique described by Sutherland \& Saunders (1992). 
We adopt the procedure discussed in Pozzi et al. (2004): the likelihood ratio (LR) is the ratio between the 
probability that a given source at the observed position and with the measured magnitude is the true optical 
counterpart, and the probability that the same source is a chance background object. For each ISO source we adopted 
a mean positional error of 2 arcsec, and we assumed a value of 0.1 arcsec as the optical position uncertainty.
We choose a search radius of 10 arcsec from the infrared position to look for the possible optical counterparts.

For each optical candidate the reliability (REL) is computed by taking into account, if needed, the 
presence of other optical candidates for the corresponding ISO source (Sutherland \& Saunders 1992). Once the likelihood 
ratio has been calculated for all the optical candidates, one has to choose the best threshold value for LR (LR$_{th}$)
to discriminate between spurious and real identifications. As the LR threshold we adopted LR$_{th}$= 0.5.
With this value, all the optical counterparts of the ISOCAM sources with only one identification (the majority
 in our sample) and LR$ > LR_{th}$ have a reliability greater than 0.8 (we assumed a value of $Q$=0.9 for the 
probability that an optical counterpart of the ISOCAM source is brighter than the magnitude limit of the 
optical catalogue, see Ciliegi et al. 2003, for more details). 
Using a less conservative value of LR$_{th}$=0.2 we found the same number of ISO/optical associations.

Moreover, we have checked by visual inspection the optical associations of each infrared source. 
With stars and isolated objects the cross-correlation is unambiguous:
the shape and the peak of the infrared contours overlayed on the R image
have confirmed the associations.
However, in the case of extended ISO sources, where the confusion starts to play a role, 
the correlation between infrared and optical catalogues is much more uncertain: there 
are no unique associations. Several optical sources could lie inside the ISO detection 
and contribute to the infrared emission.
When more than one optical candidate with LR$_{th}>$0.5 is present for the same ISO source, 
we assume that the object with the highest likelihood ratio value is the real counterpart of the ISO source. 
However, when a radio emission is detected ($\sim$30 cases), the tight correlation 
with the infrared emission allows us to associate the ISO positions with the radio ones. 

\begin{figure}
\centering
\includegraphics[width=0.5\textwidth]{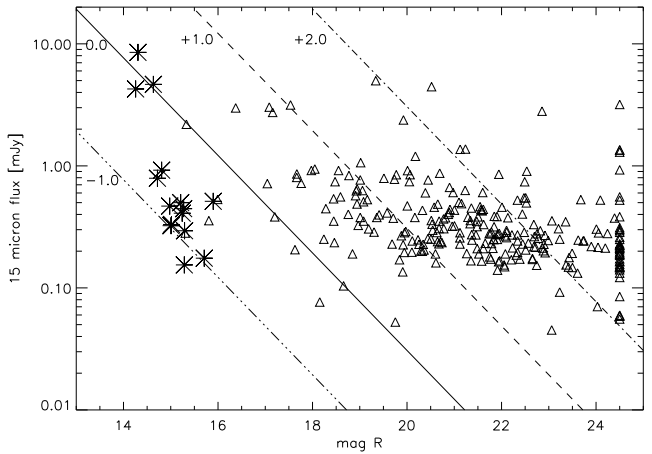}
\caption{LW3 fluxes as a function of the optical r' magnitudes 
for the Lockman Deep ISOCAM sources .
The lines indicate the relations where the infrared to optical ratio 
is constant ($S(15_{\mu m})/S_R$=-1, 0, +1, +2). 
The brightest objects are stars (marked as asterisks):
the two exceptions correspond to stars blended with galaxies.  
}
\label{fig:opt_col}%
\end{figure}

With this threshold value, 85\% of the ISO sources have a likely identification.
A percentage of 15\% of the 14.3 $\mu$m sources do not present any optical reliable identifications
down to the optical limits (r'$\sim$25 mag), and are marked as blank fields. 
As suggested by Gonzalez-Solares et al.(2004), blank fields are probably the extreme version 
of the objects with high infrared to optical ratios ($S(15_{\mu m})/S_R>\sim 10^3$). 
This is shown in Figure \ref{fig:opt_col}),
where we plot the LW3 14.3$\mu$m fluxes as a function of the optical r' magnitudes 
for the Lockman Deep ISOCAM sources .
The lines indicate the relations where the infrared to optical ratio 
is constant ($S(15_{\mu m})/S_R$=-1, 0, +1, +2). 
We do not find evidence for a connection between the r' magnitude and the 14.3 $\mu$m flux density.

\begin{figure*}
\centering
\includegraphics[width=4cm]{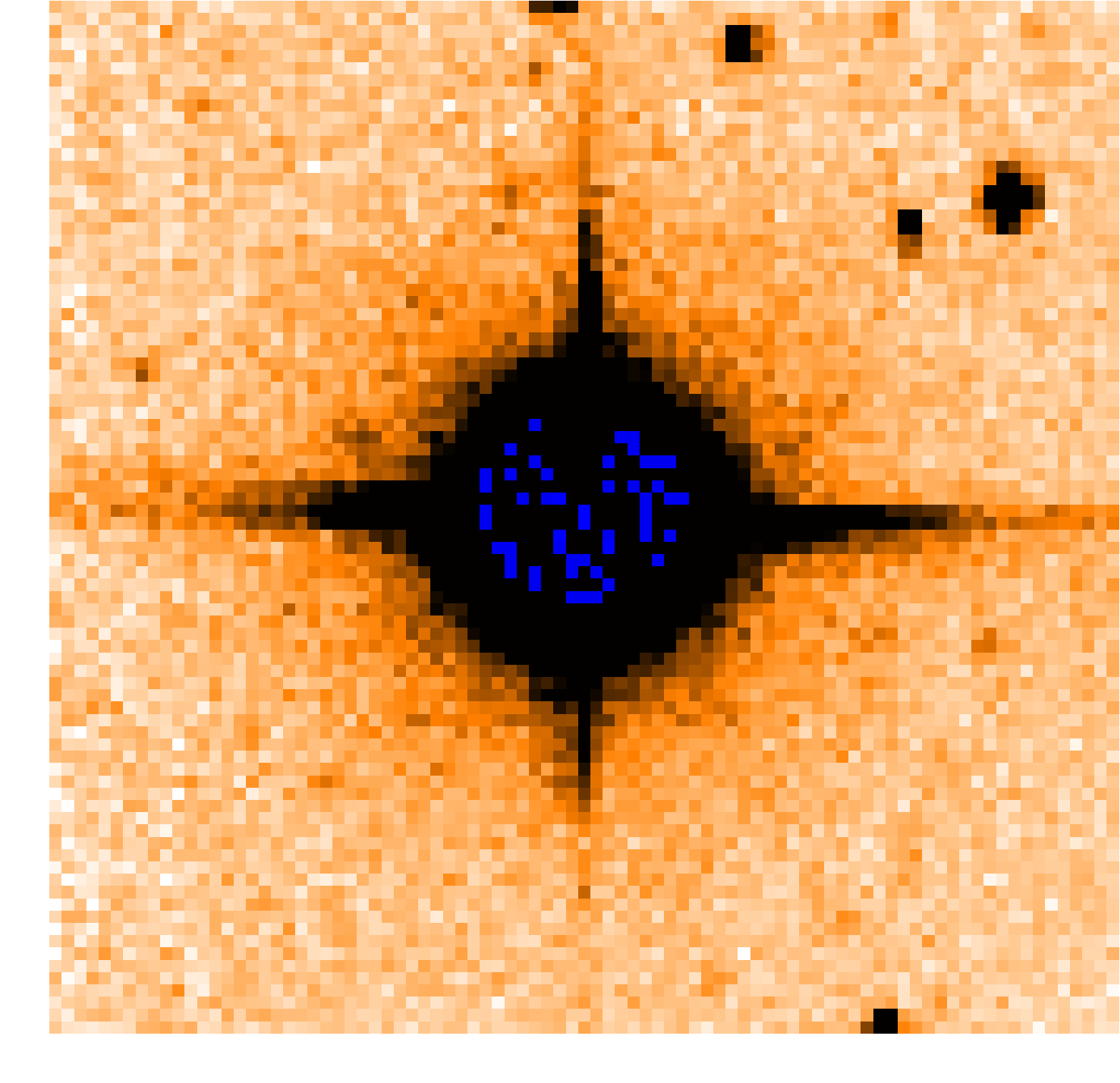}
\includegraphics[width=4cm]{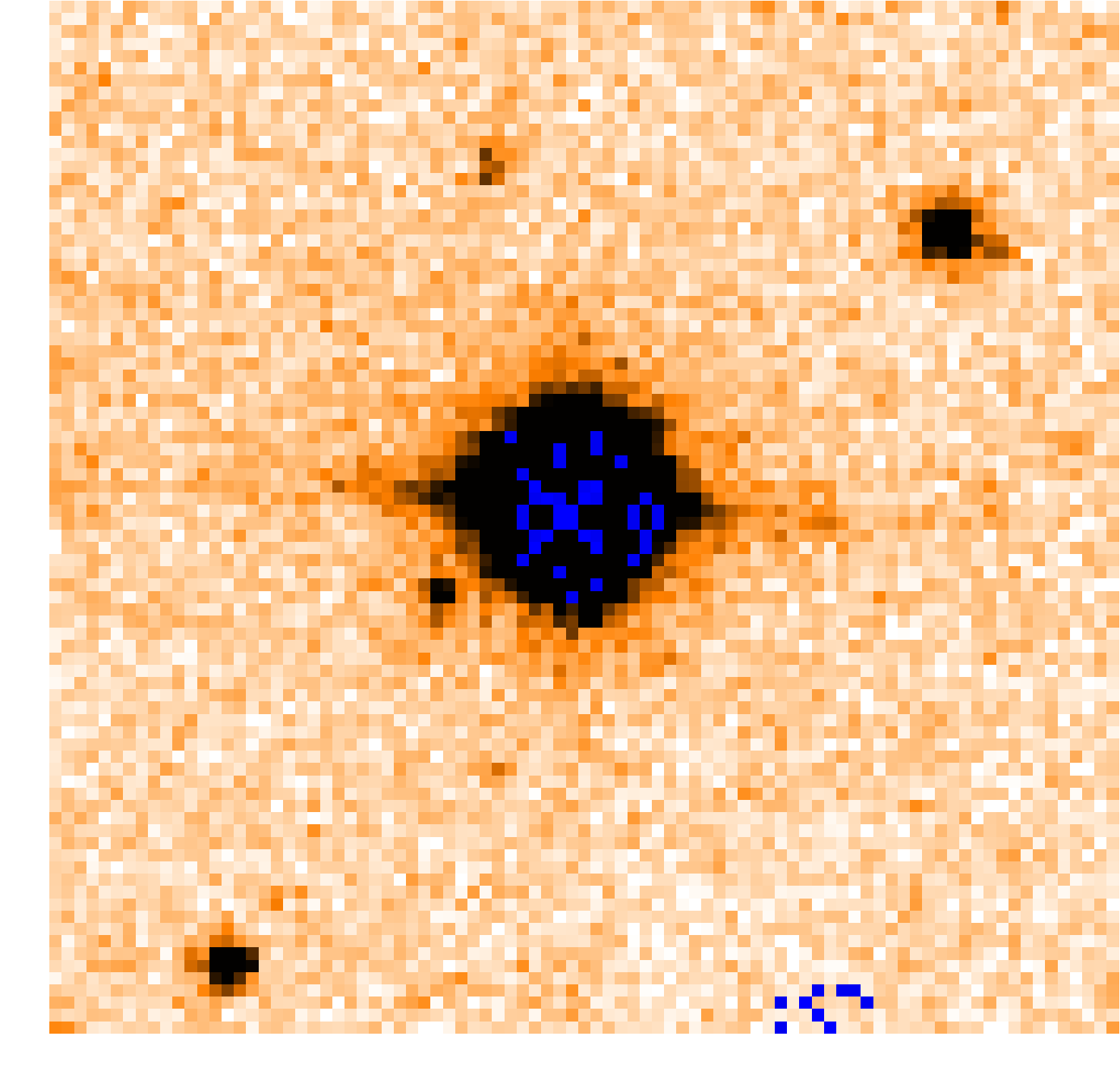}
\includegraphics[width=4cm]{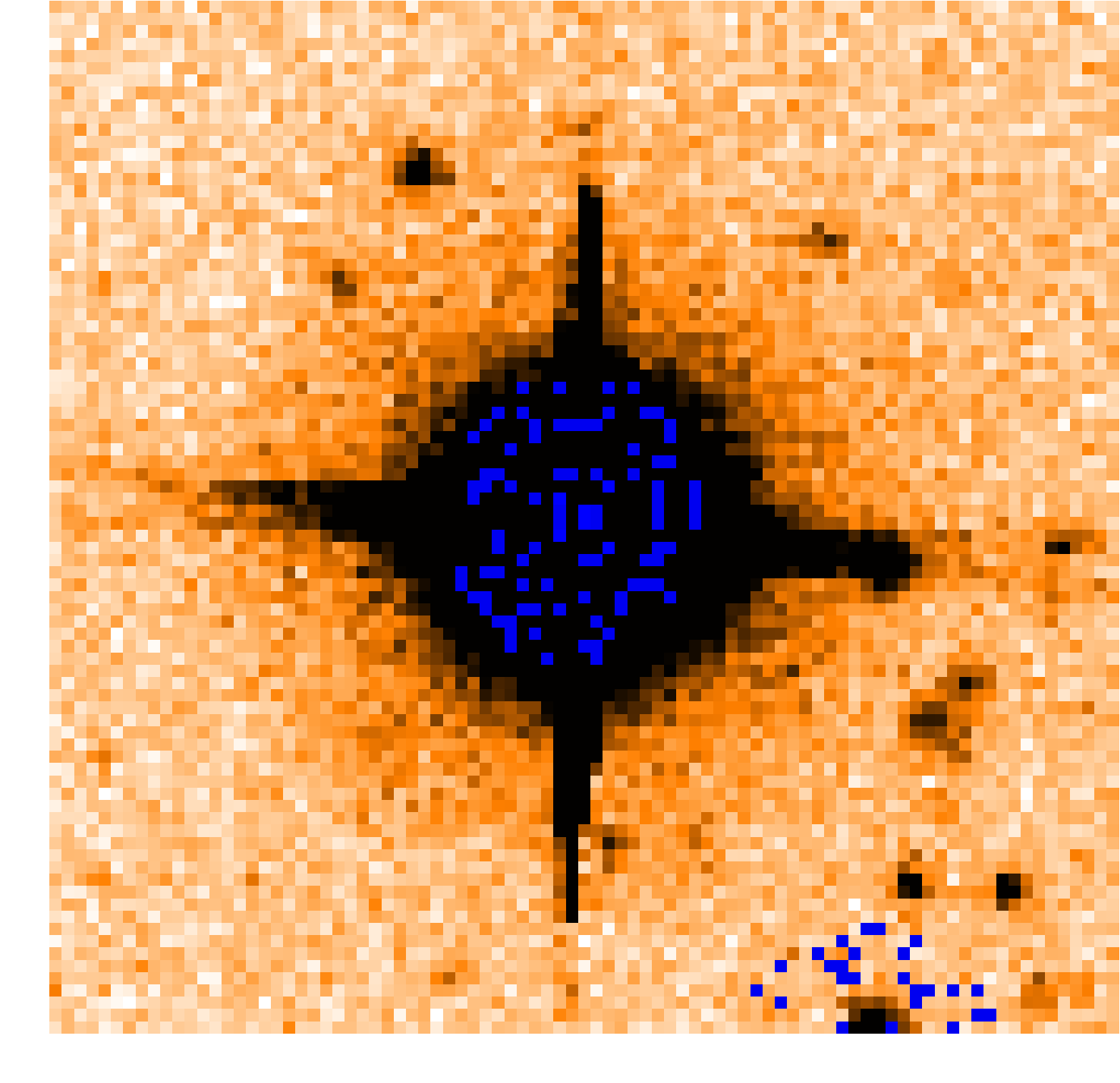}
\includegraphics[width=4cm]{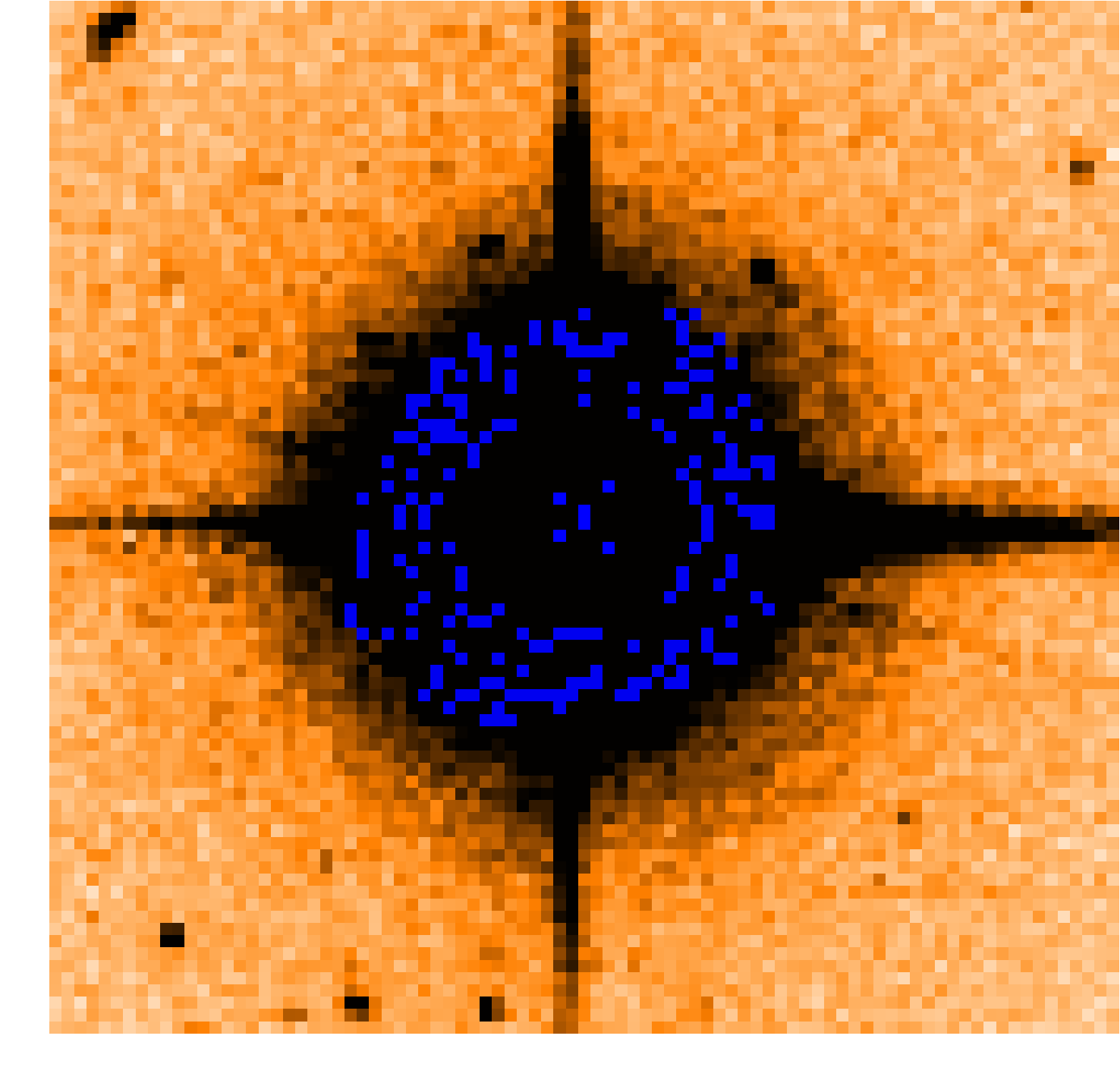}
\caption{Examples of few STARS in the 14.3 $\mu$m Lockman catalogue. 
We report the postage stamps of the R optical image overlayed with the 14.3 $\mu$m contours
(starting from 5-$\sigma$ at increasing signal-to-noise levels). Each map is $\sim 1\times 1$ 
square arcminutes. North is upward, East leftward.}
\label{fig:opt_1}%
\end{figure*}

\begin{figure*}
\centering
\includegraphics[width=4cm]{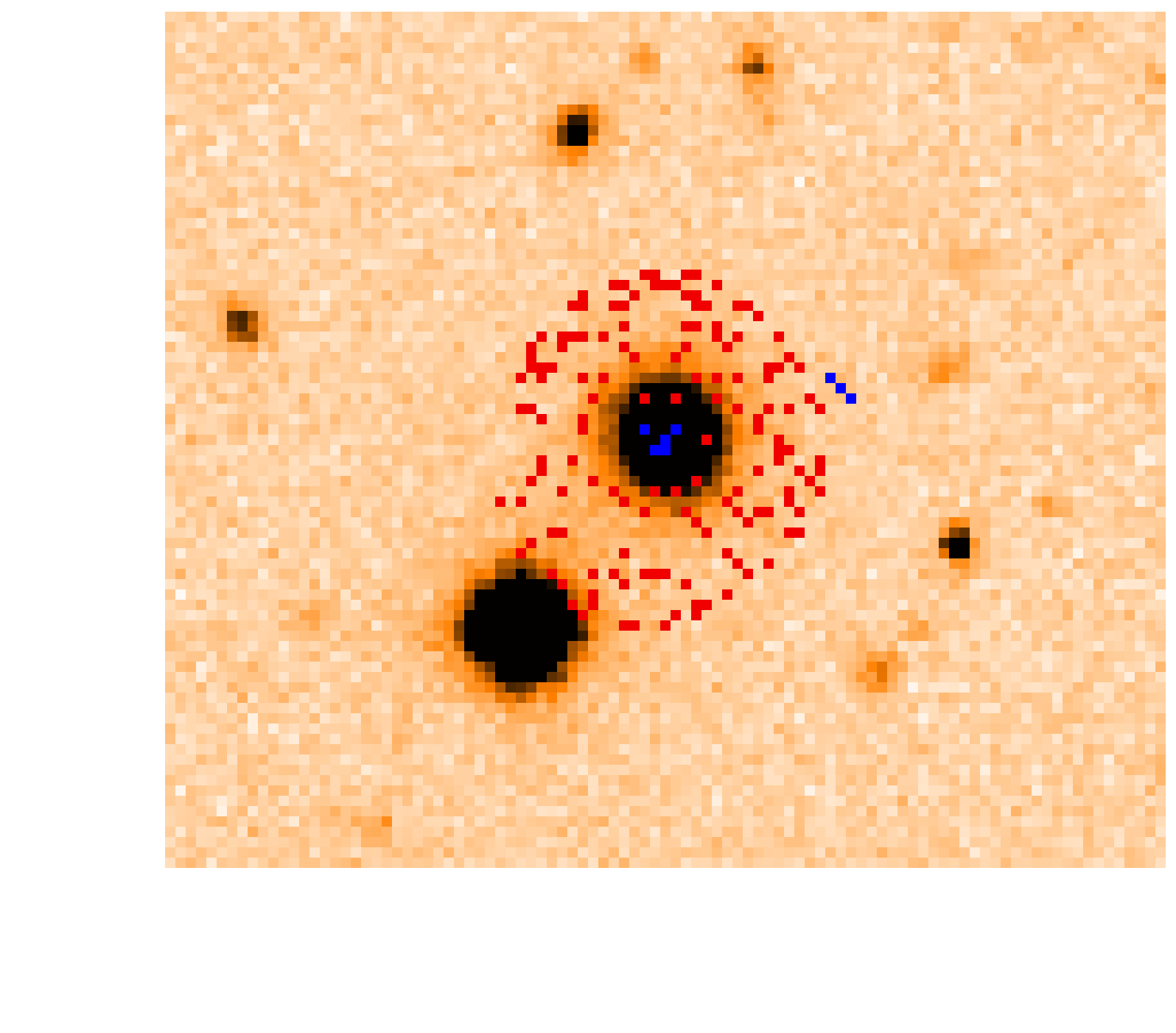} 
\includegraphics[width=4cm,height=3.4cm]{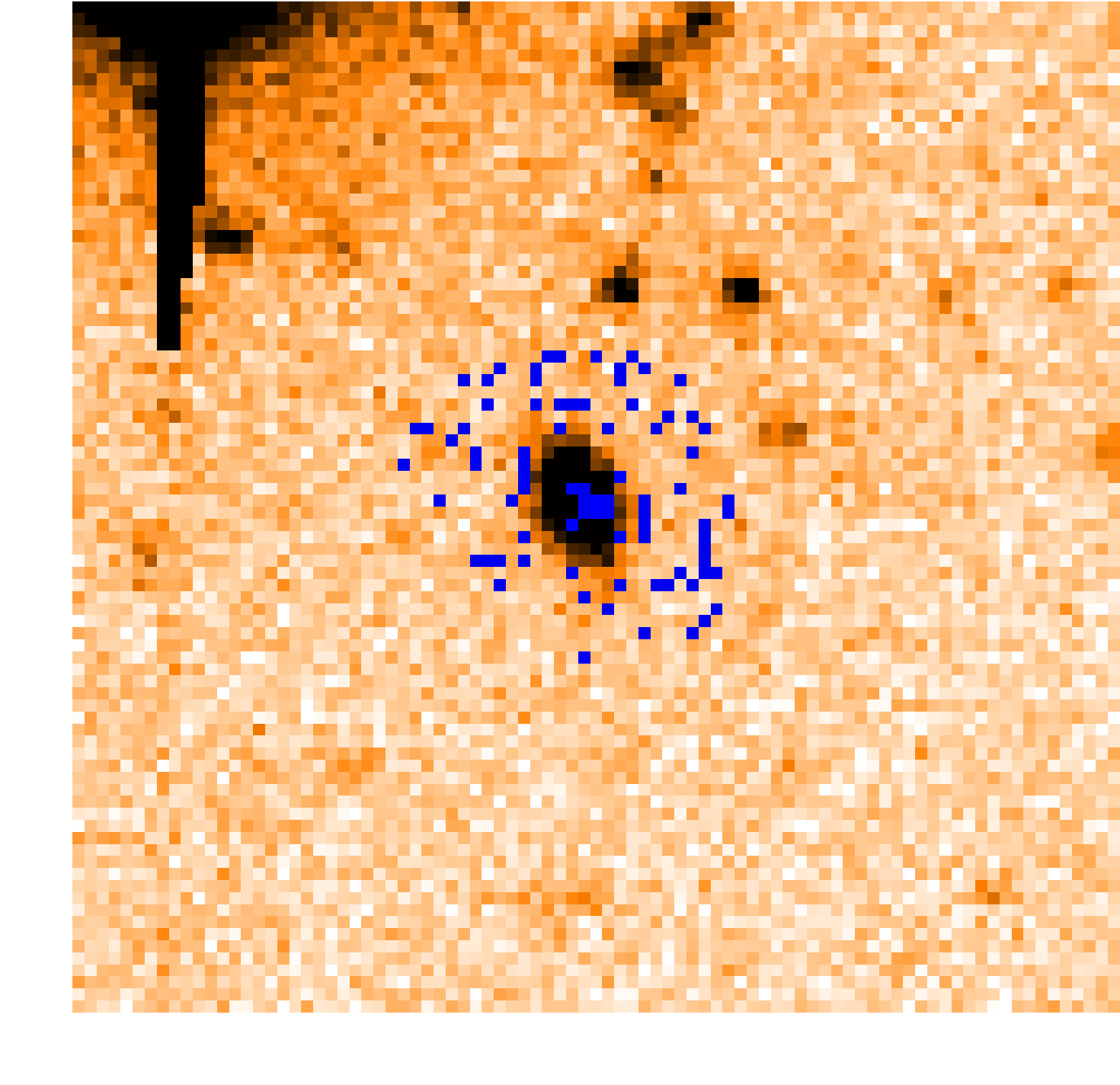} 
\includegraphics[width=4cm]{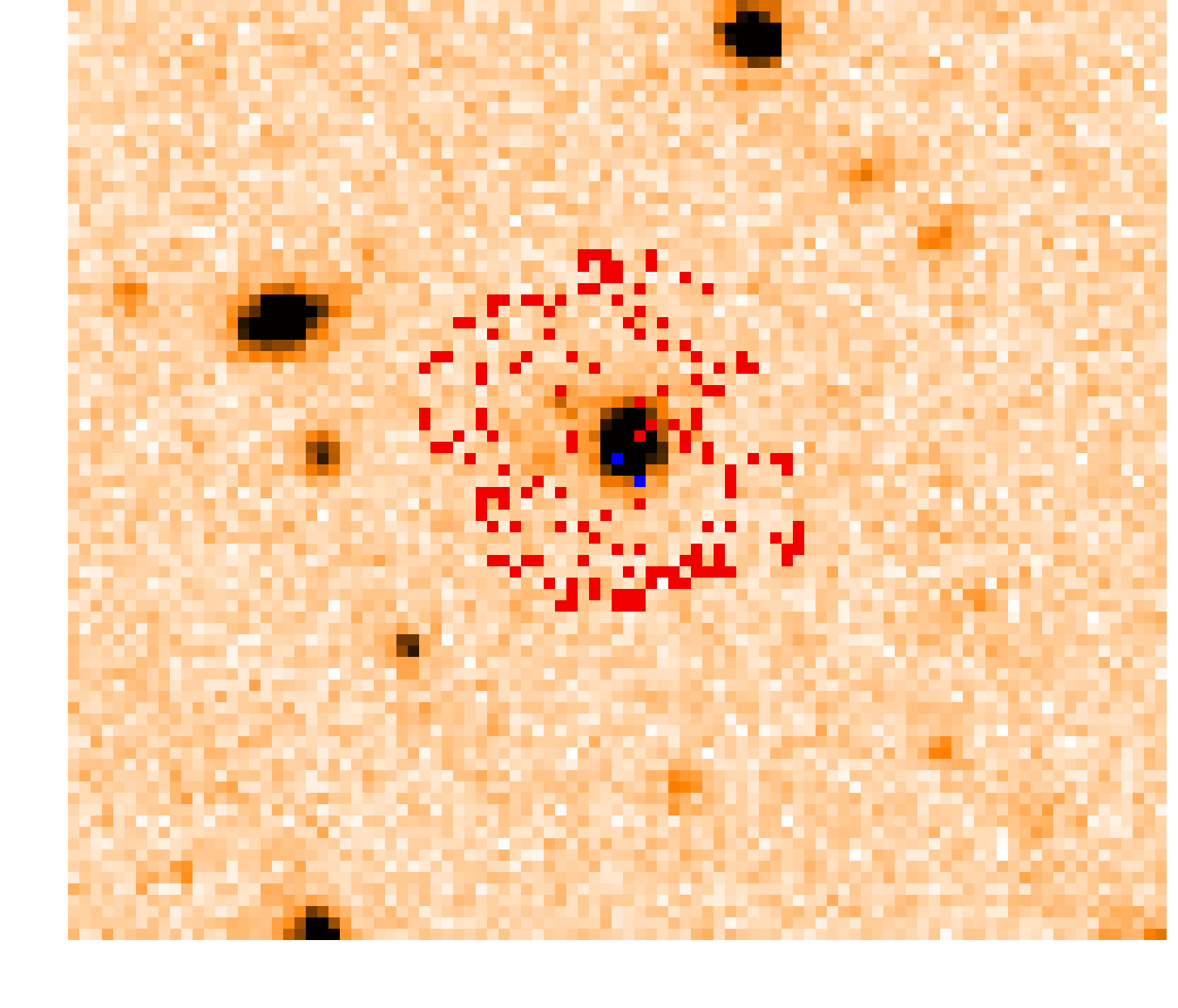} 
\includegraphics[width=4cm]{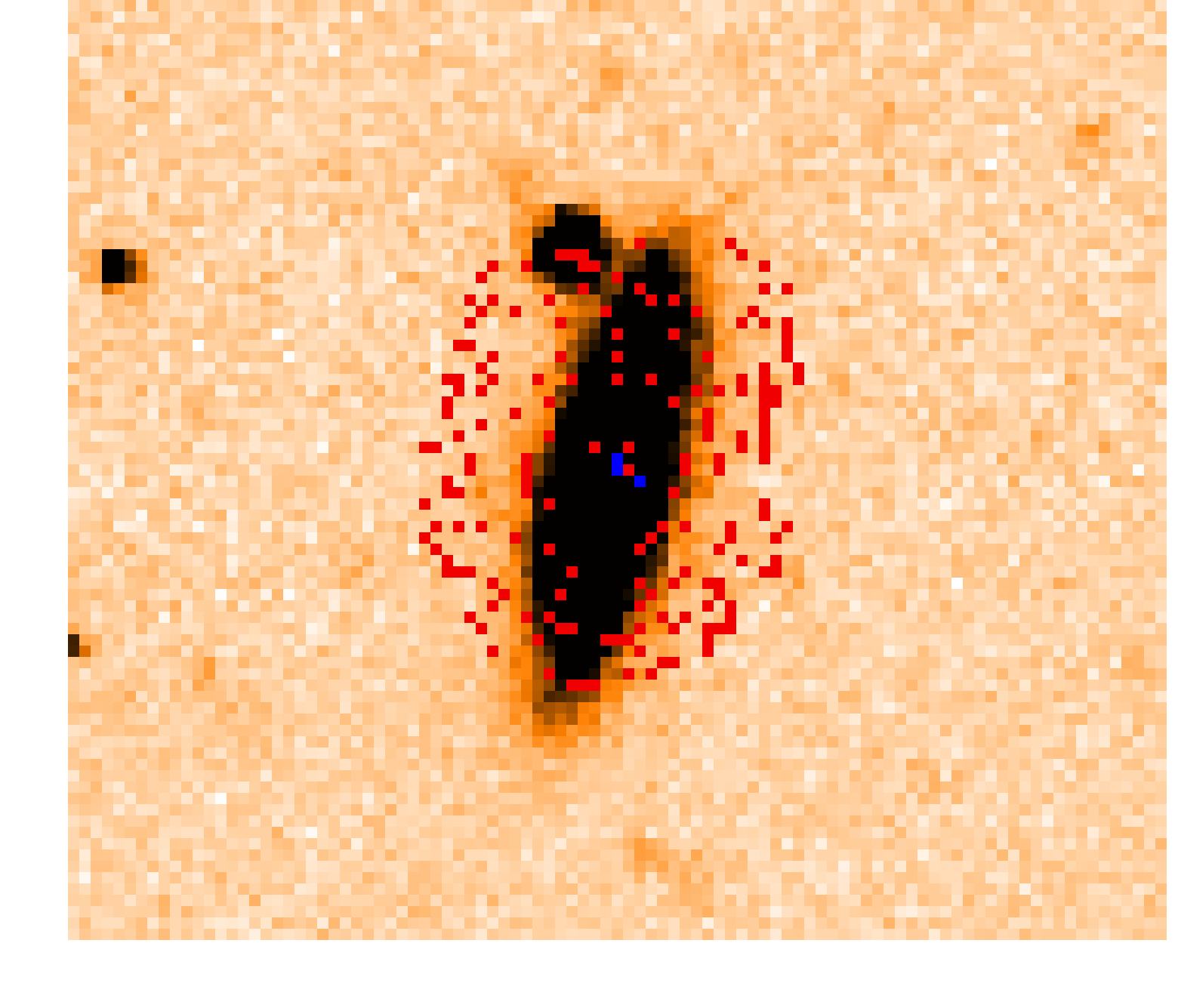} 
\includegraphics[width=4cm]{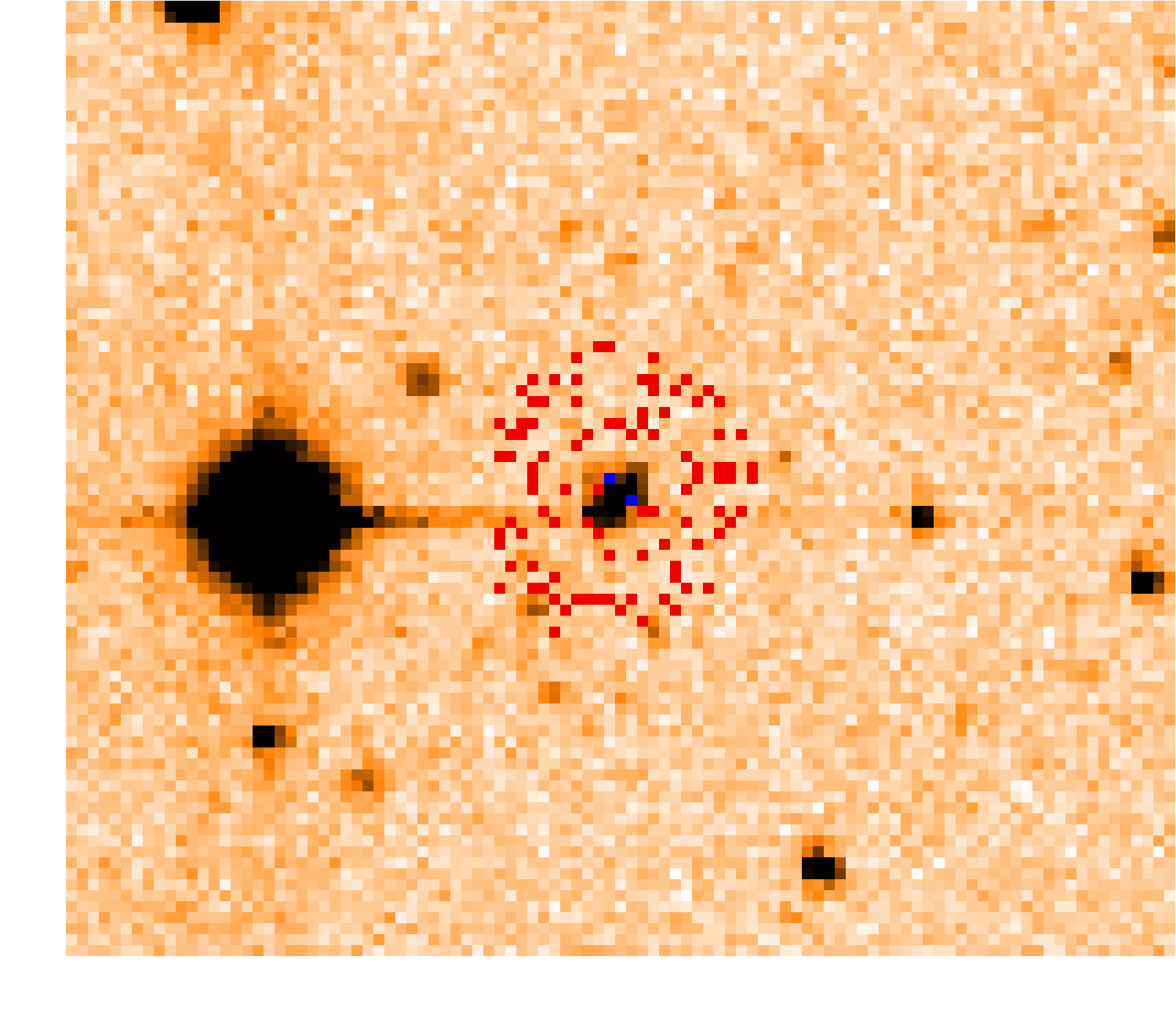} 
\includegraphics[width=4cm]{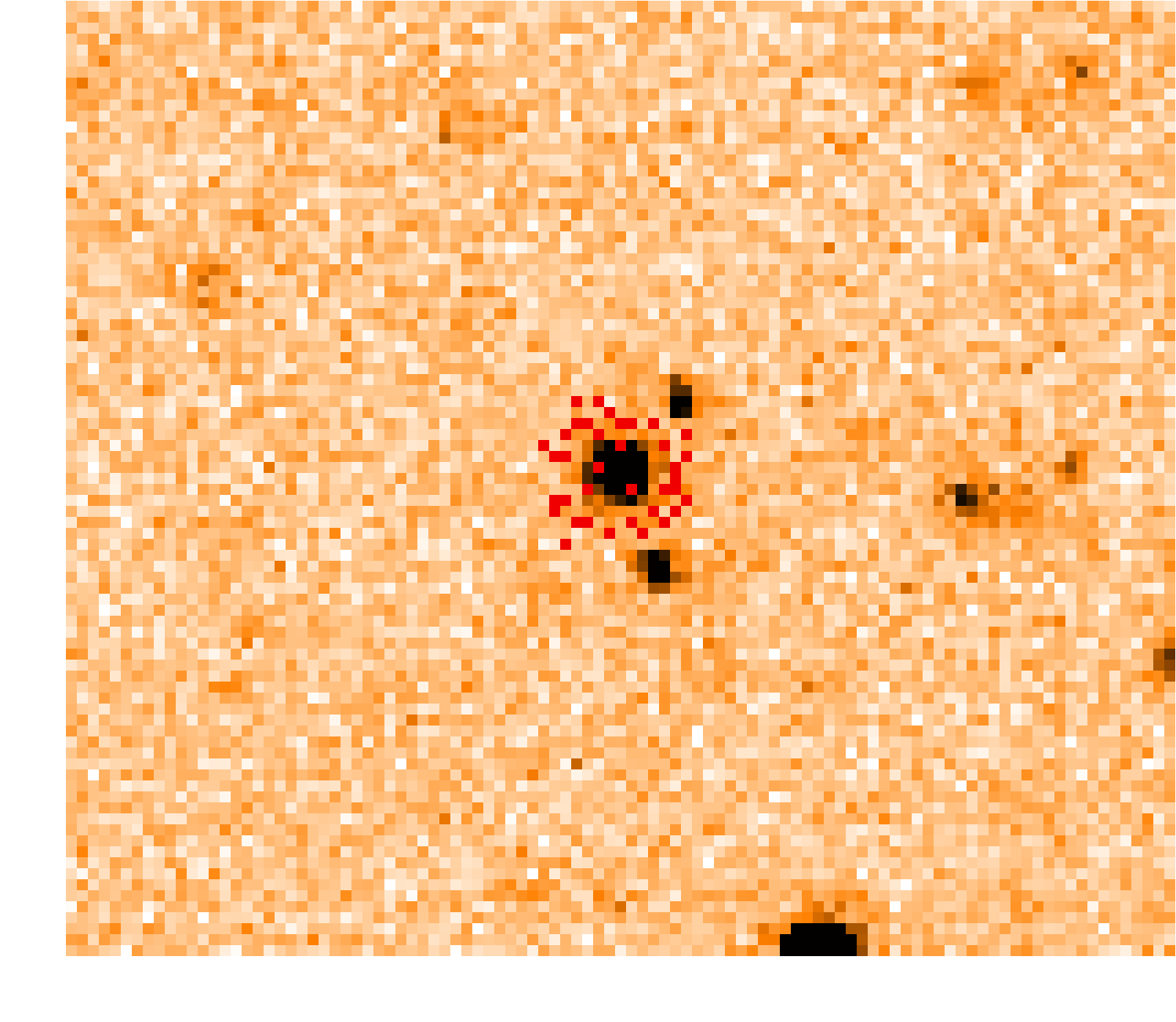} 
\includegraphics[width=4cm]{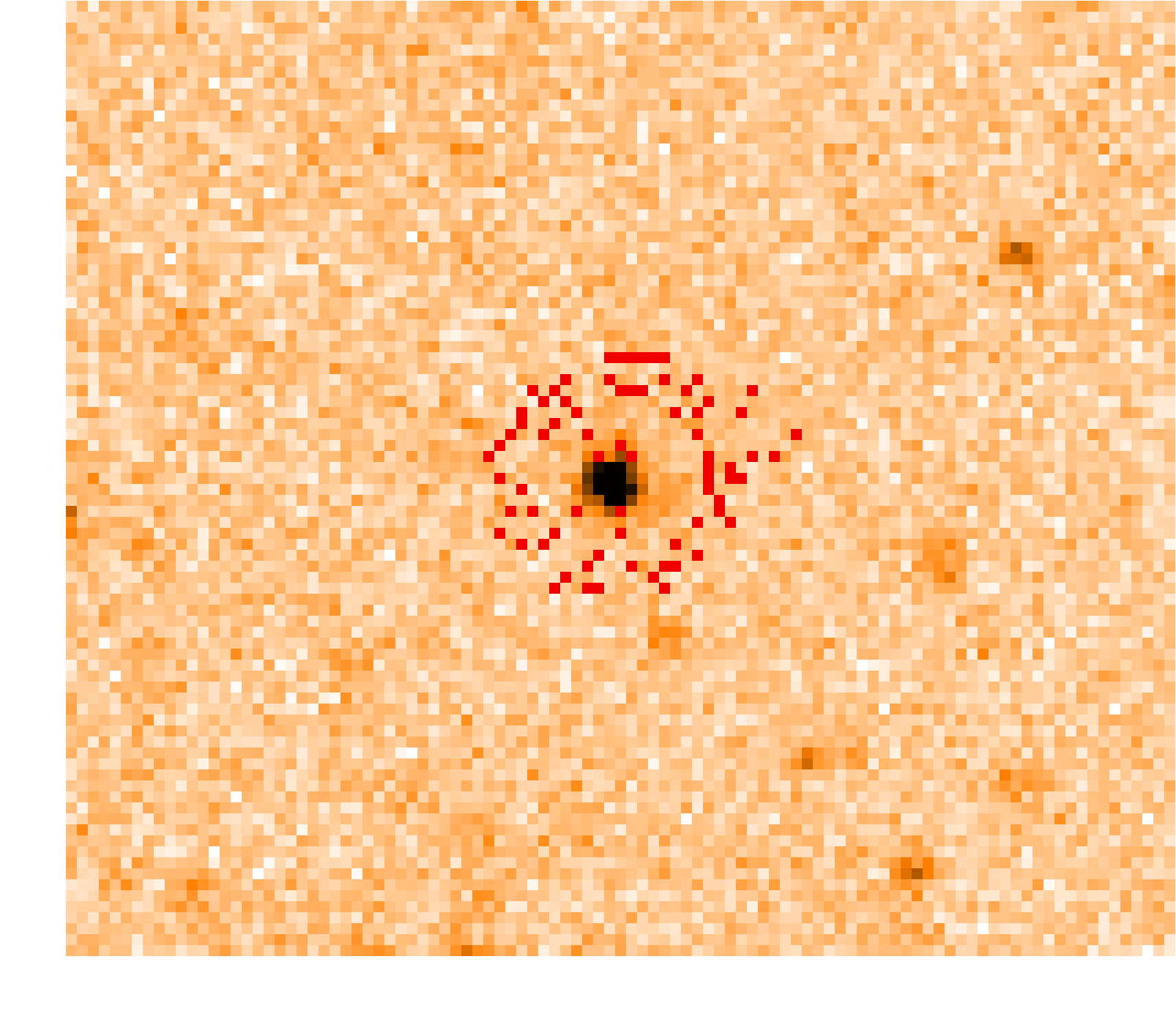} 
\includegraphics[width=4cm]{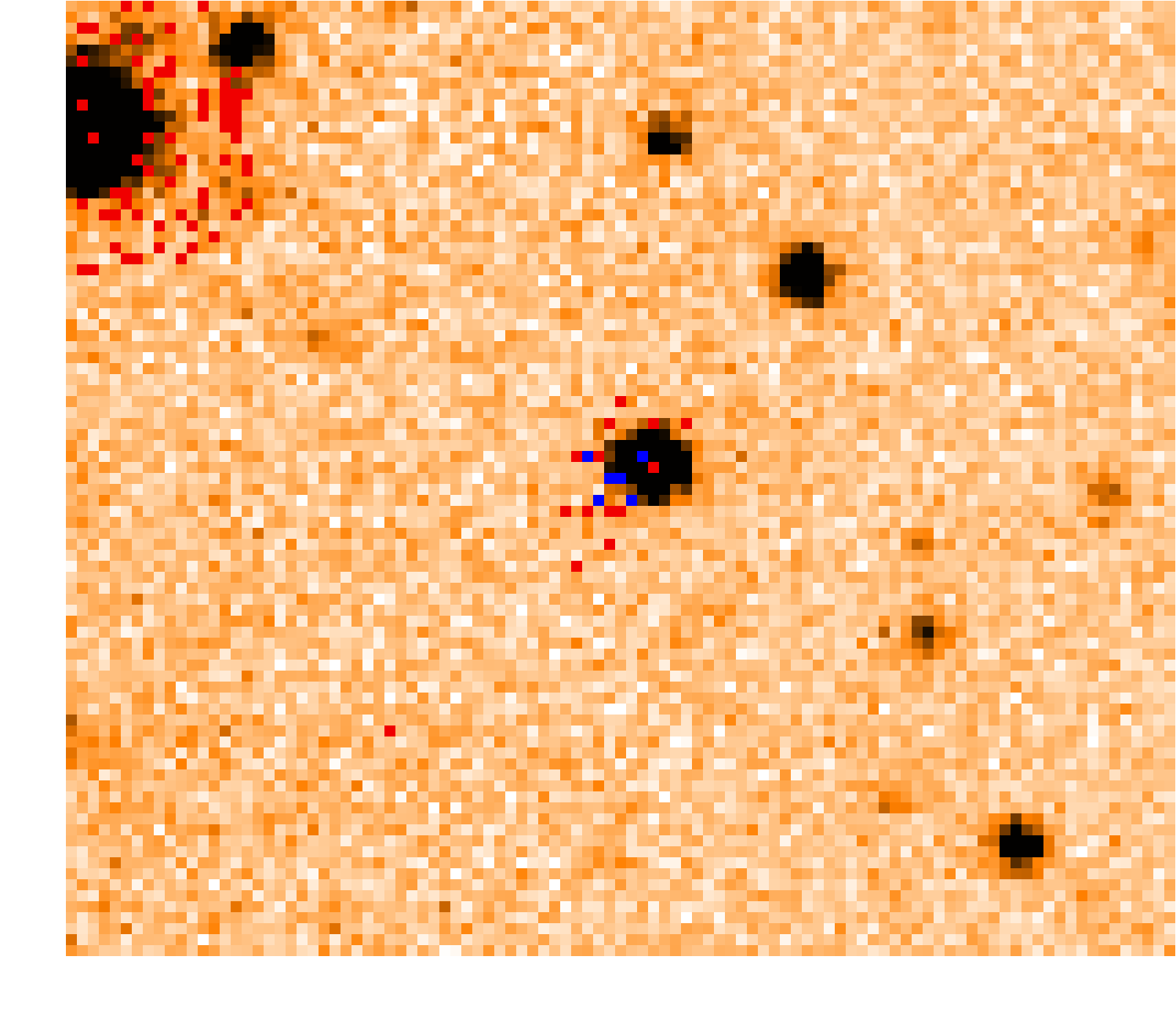} 
\caption{Examples of few ISOLATED OBJECTS in the 14.3 $\mu$m Lockman catalogue. For this class
of sources the optical identification is generally straightforward.
The reference for the maps legend is the same as in Figure \ref{fig:opt_1}.
}
\label{fig:opt_2}%
\end{figure*}

\begin{figure*}
\centering
\includegraphics[width=4cm]{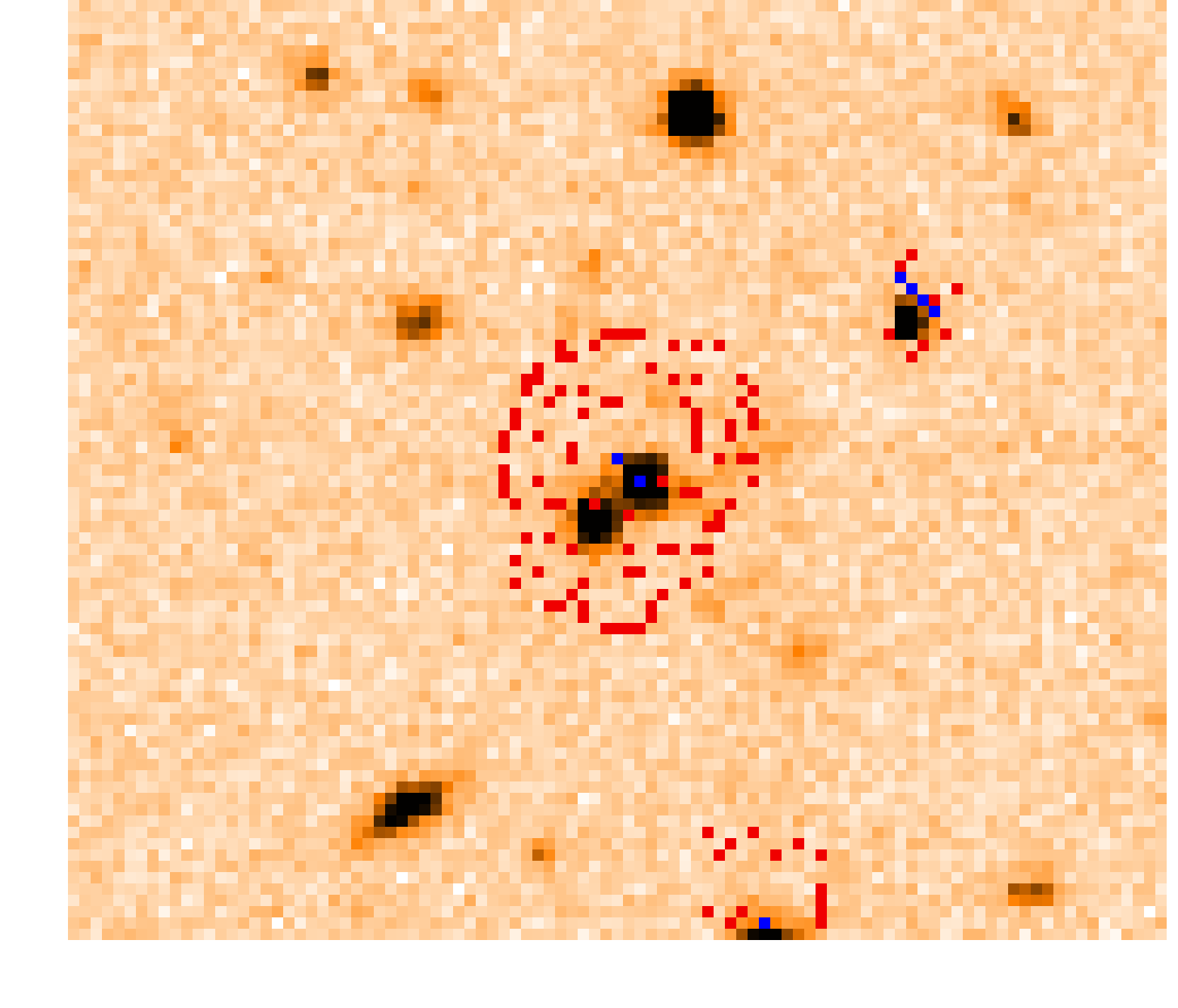}
\includegraphics[width=4cm]{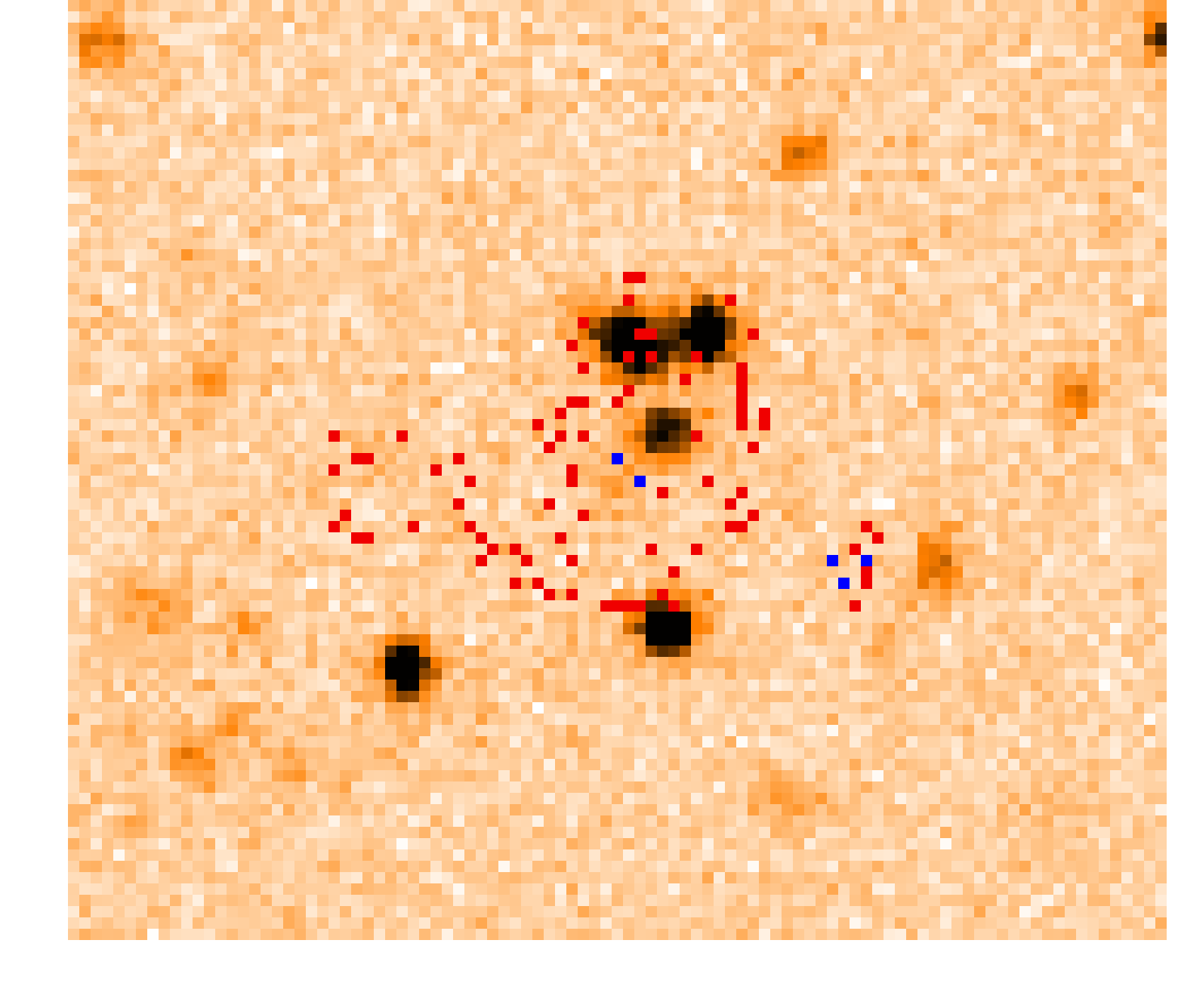}
\includegraphics[width=4cm]{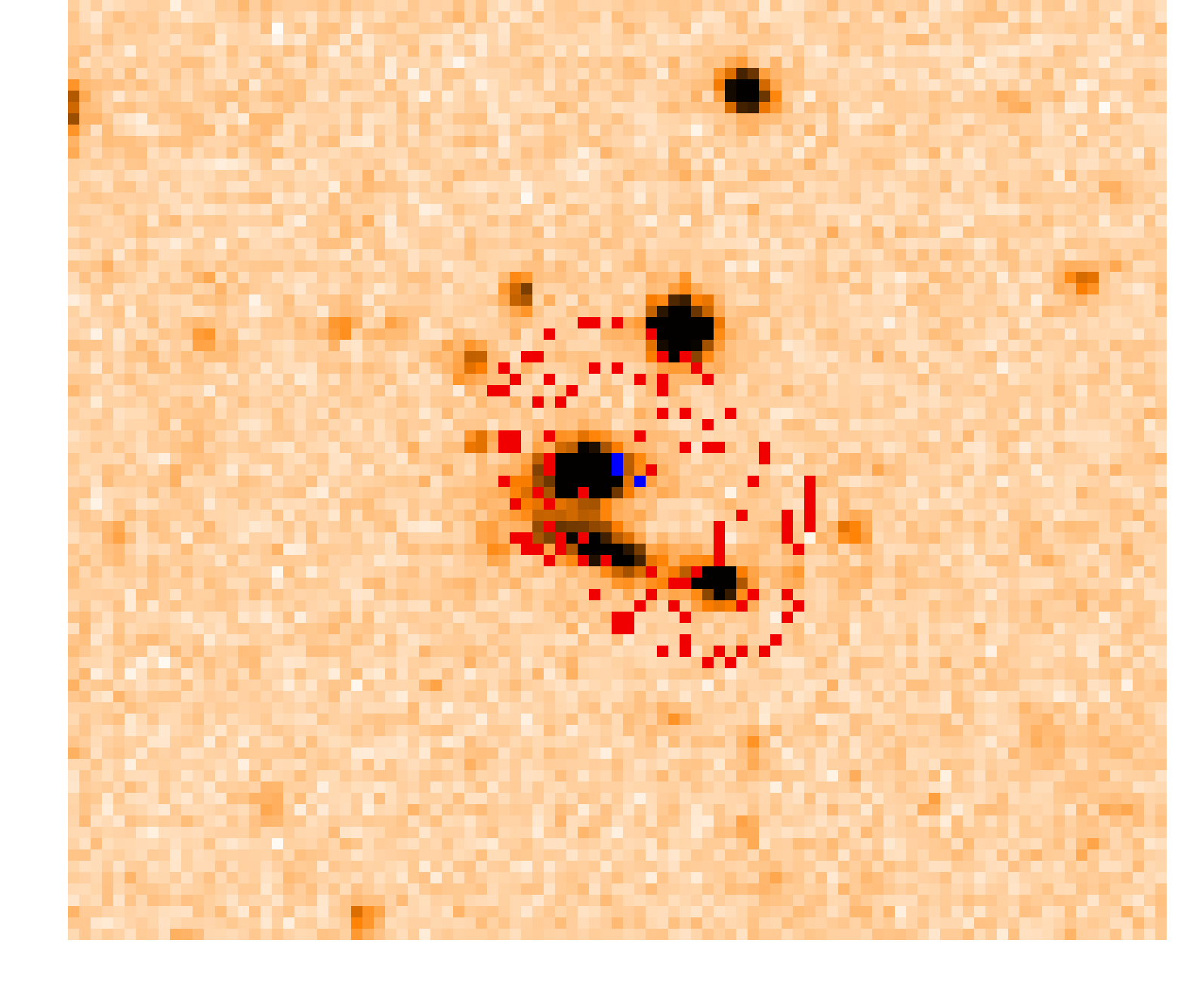}
\includegraphics[width=4cm]{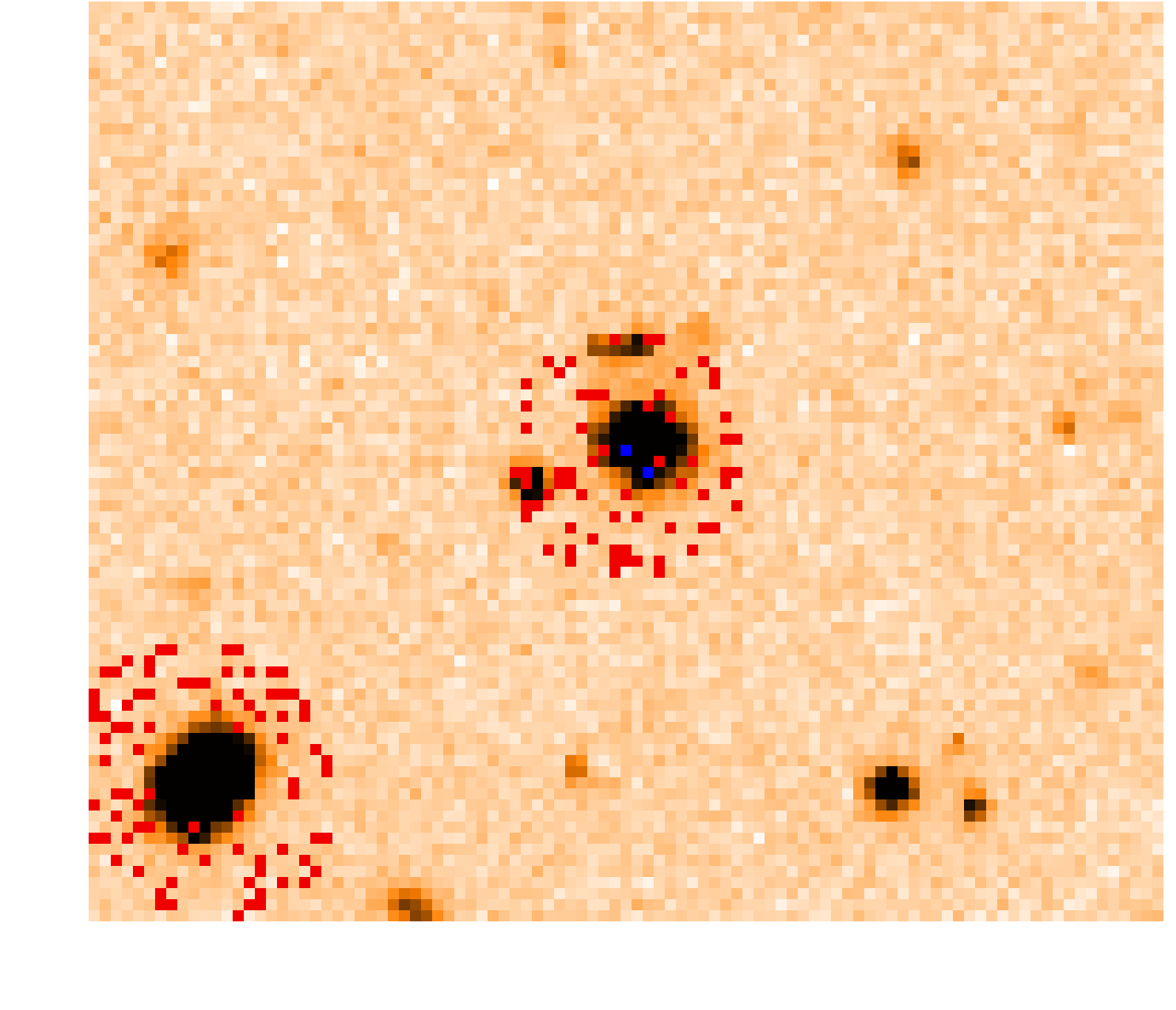}
\includegraphics[width=4cm]{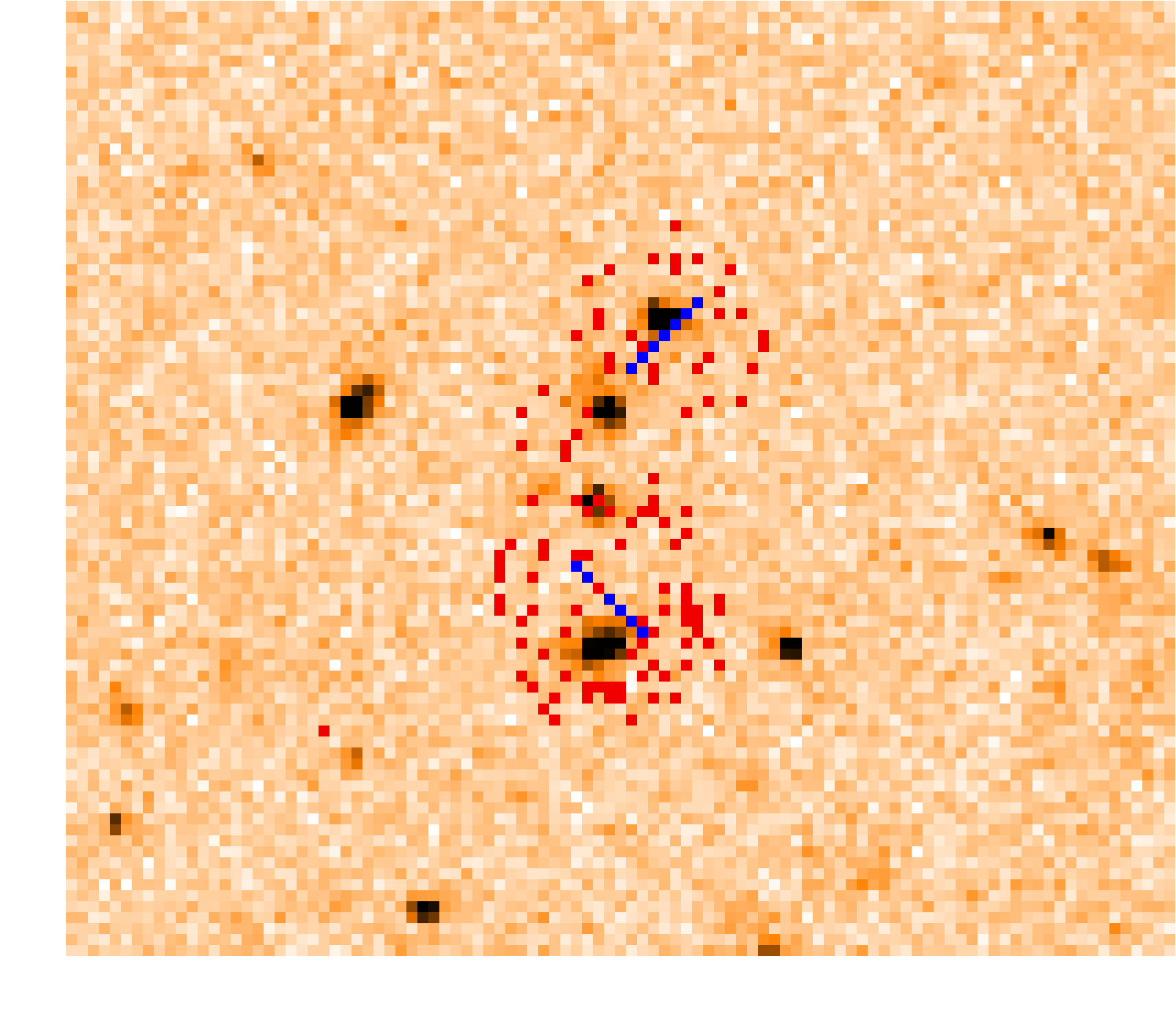}
\includegraphics[width=4cm]{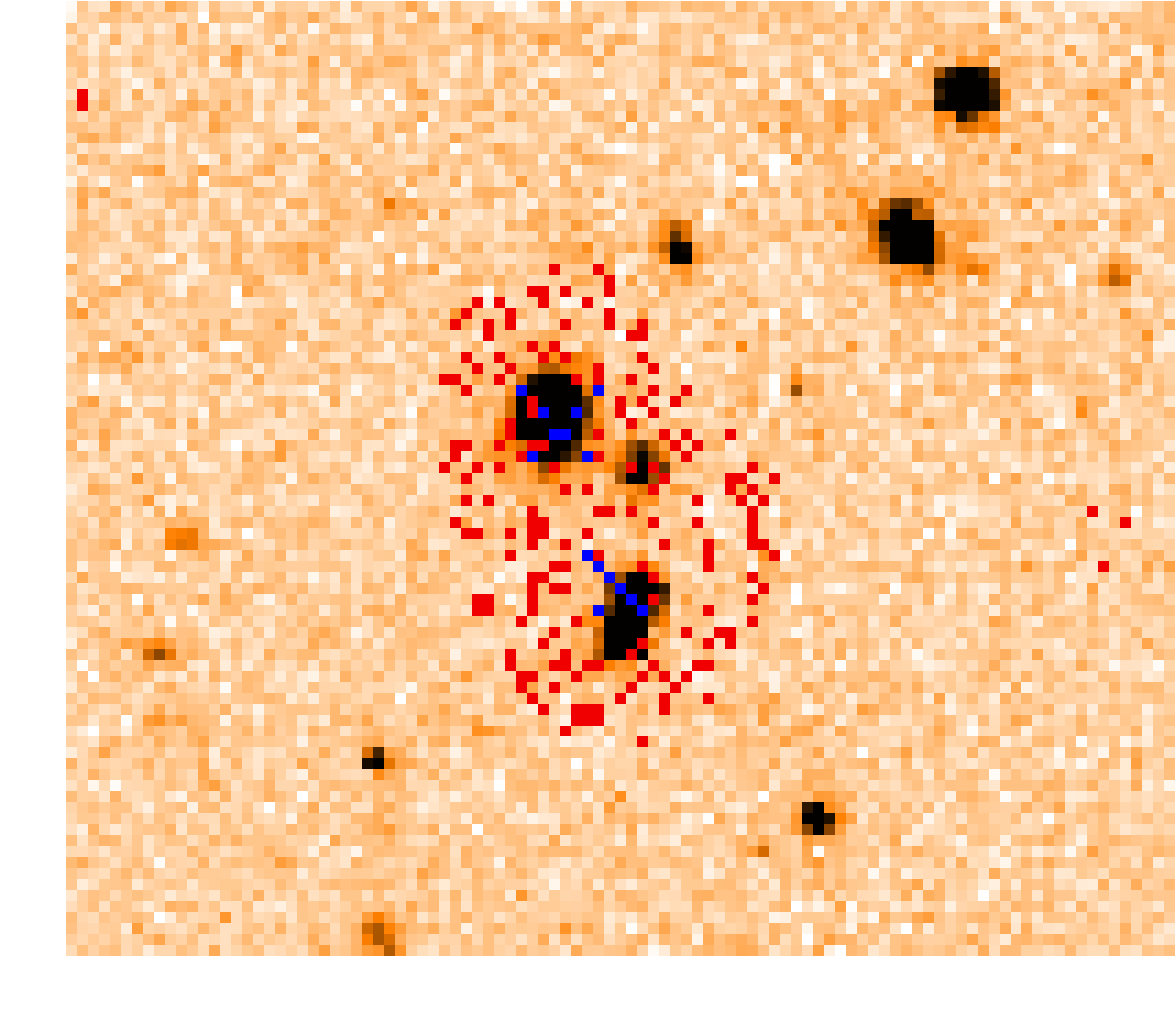}
\includegraphics[width=4cm]{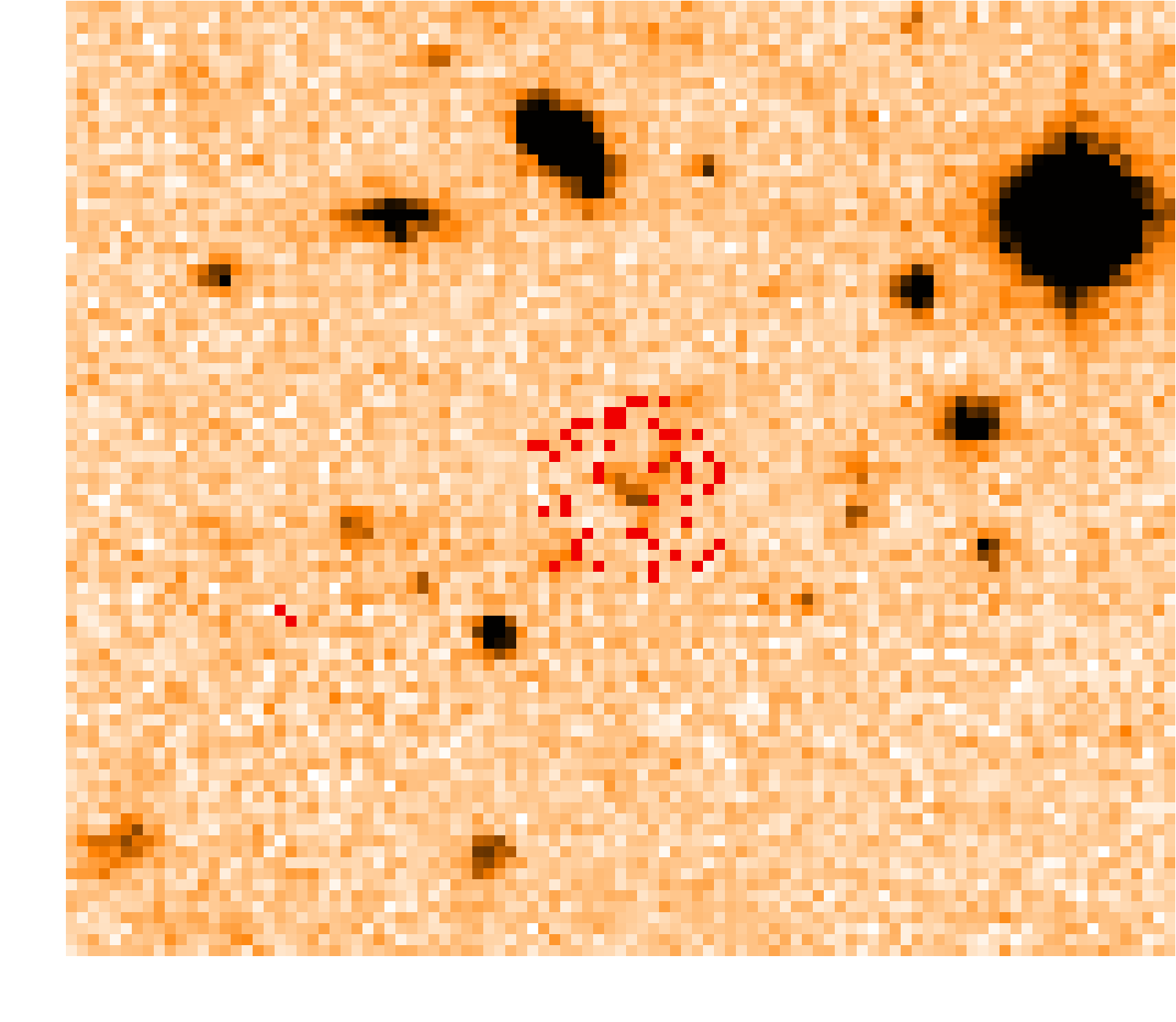}
\includegraphics[width=4cm]{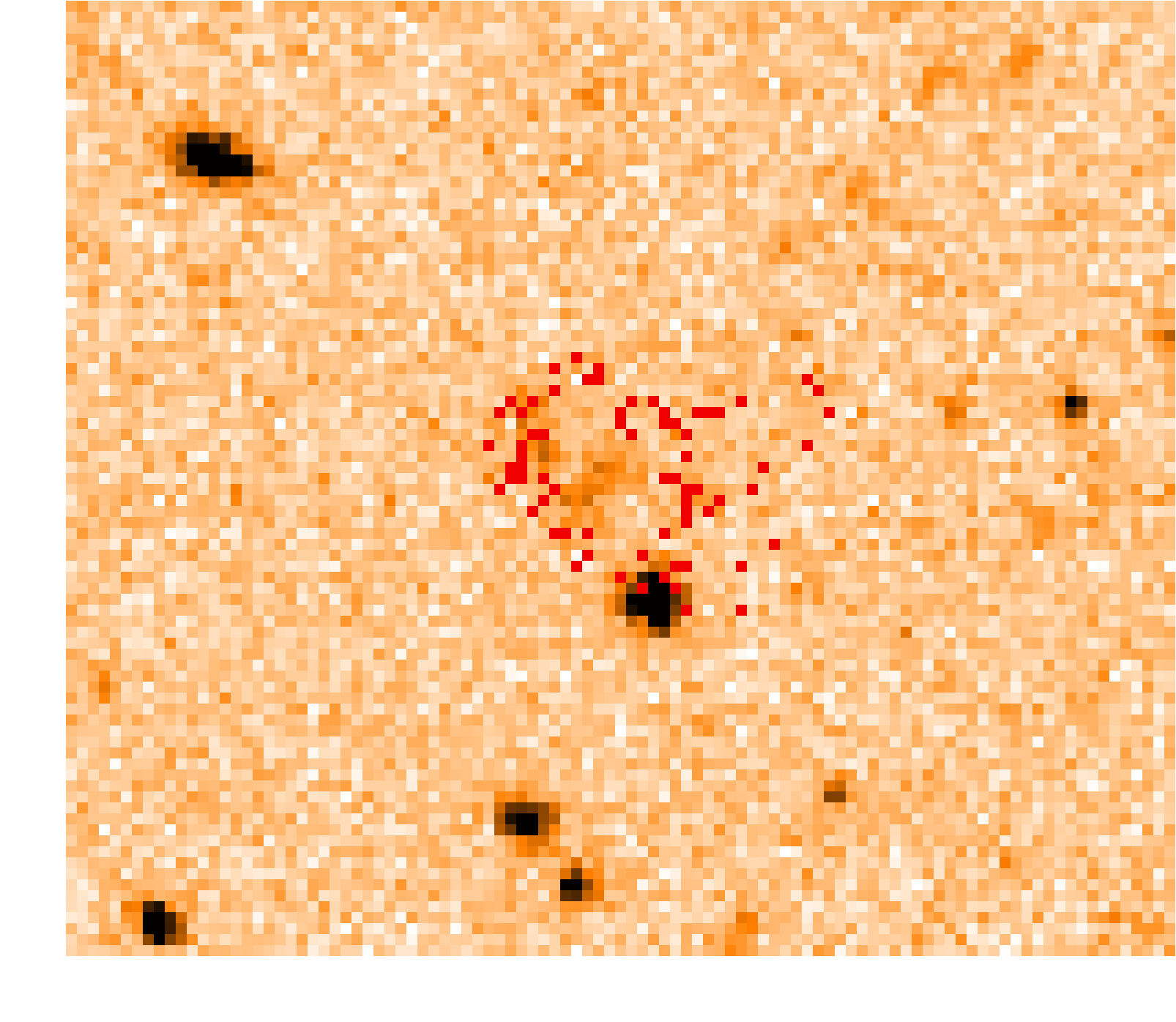}
\caption{Examples of a few BLENDED OBJECTS: as can be seen in the figure, they look like
mergers or small groups. For these sources in the catalogue list  we report the main optical 
associations within a search radius of 10 arcsec (from the major component to those with 
decreasing significance).
The reference for the maps legend is the same as in Figure \ref{fig:opt_1}.
}
\label{fig:opt_3}%
\end{figure*}

Apart from stars, we found that the infrared sources in our catalogue can be
separated into two main classes: isolated objects, and blended objects. 
The second class is mainly composed of interacting galaxies (mergers) and small groups.
In Figures \ref{fig:opt_1}-\ref{fig:opt_3} we present some examples  for each of the mentioned
morphological classes. 

\begin{figure}
\centering
\includegraphics[width=7cm]{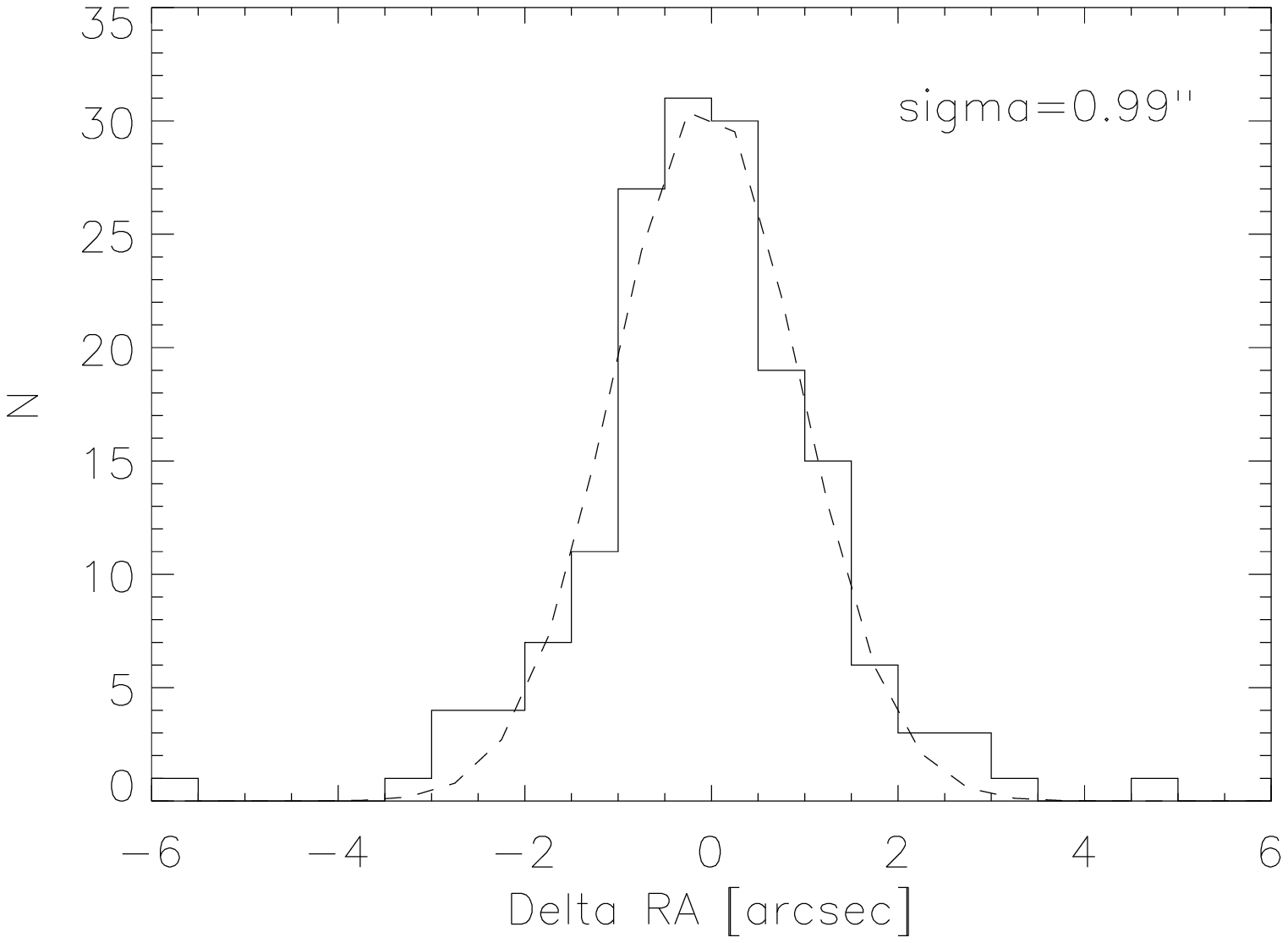}
\includegraphics[width=7cm]{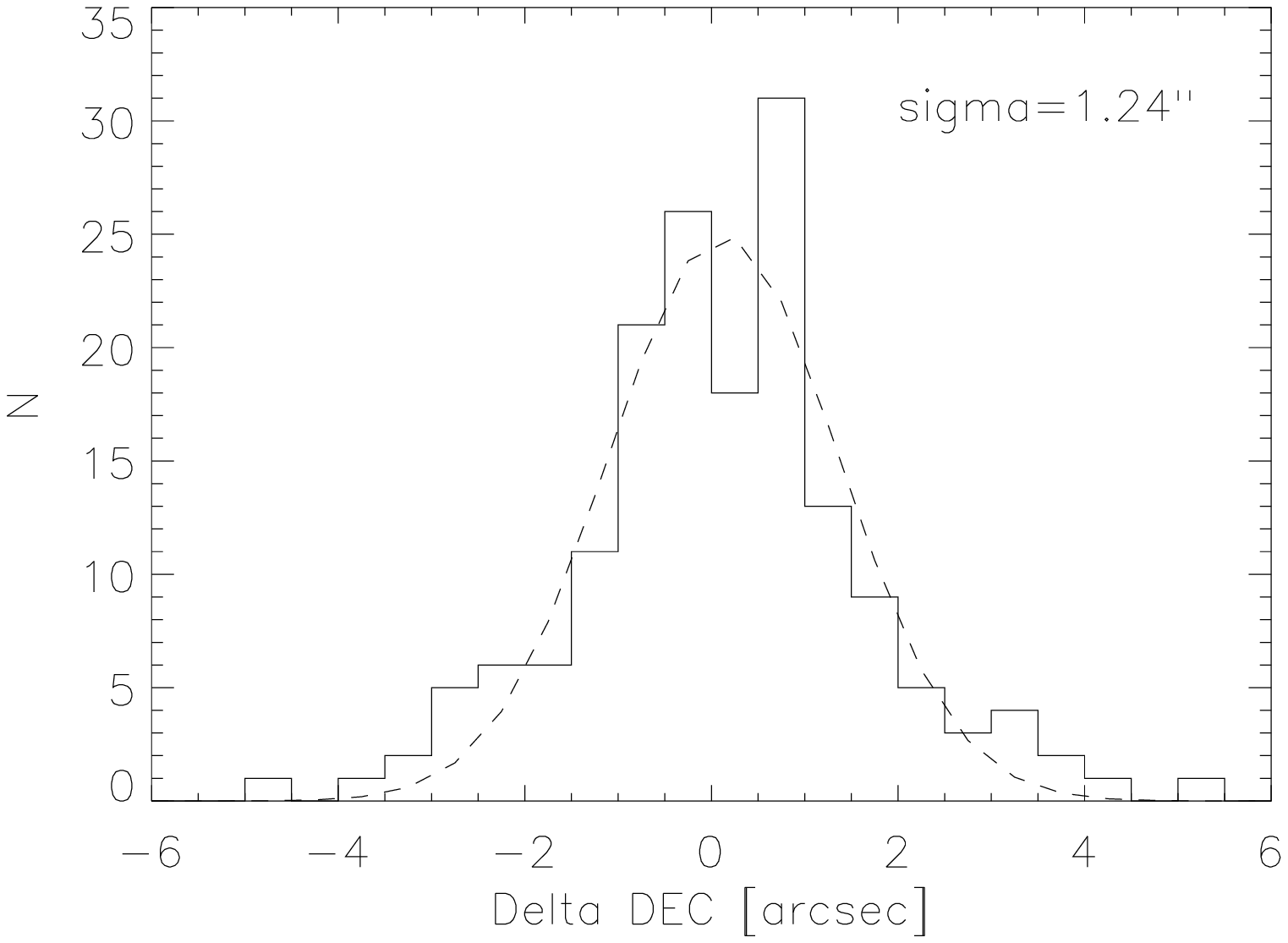}
\caption{Distribution of the difference in RA ($top$) and DEC ($bottom$) between the infrared coordinates
and the positions of the optical identifications. In these graphics only sources with highly reliable
optical counterparts (stars and isolated objects) have been considered. The distributions have been fitted 
with Gaussian functions, with dispersion of respectively 0.99'' and 1.24'' for RA and DEC. }
\label{fig:histo_pos}%
\end{figure}

We used stars and isolated objects with unambiguous optical identifications to measure
the astrometric uncertainties of the infrared sample. The results are plotted in Figure
\ref{fig:histo_pos}, where we have considered the distributions of the difference in right
ascension and declination between the infrared and the optical positions (determined as
discussed in Section \ref{id_opt}).
By fitting the histograms with a Gaussian function, the 3-$\sigma$ limits of the distributions are, 
respectively, $\sim$2.9 arcsec for the right ascension and $\sim$3.7 arcsec for the
declination. Of the 167 sources considered, $\sim 80\%$ lie within [-1.5, +1.5] arcsec 
both in RA and DEC.

\section{The catalogue}
The final catalogue obtained with our method contains 283 sources detected at 14.3 $\mu$m in 
a 20'$\times$20' region centred on the Lockman Hole. All sources have a signal-to-noise ratio
greater than 5. The entries are as detailed in the following:

\begin{itemize}
\item {\bf Number } : source identification number;
\item {\bf Name } : IAU source name, referring to the name of the satellite (ISO) and to the
identification of the survey (LHDS);
\item {\bf RA (J2000) } : Right Ascension at epoch J2000 in sexagesimal units;
\item {\bf DEC (J2000) } : Declination at epoch J2000 in sexagesimal units;
\item {\bf S/N } : signal to noise ratio;
\item {\bf Total Flux } : source total flux obtained from the peak flux or aperture photometry, 
expressed in mJy  (the corresponding errors have been computed as discussed by Gruppioni et al 2002
and Vaccari et al. 2004, mainly taking into account the contribution of two effects, namely
the autosimulation process and the noise present on the sky maps).
An asterisk before the flux value indicates that for this source aperture photometry has been
performed;
\item {\bf Radio Flux } : radio 1.4GHz flux density expressed in mJy (from De Ruiter et al. 1997
and Ciliegi et al. 2003);
\item {\bf S/G } : flag for star (1) or galaxy (0);
\item {\bf REL } : reliability of the optical counterpart, when available;
\item {\bf r' magnitude } : optical r' Sloan magnitudes in the Vega system.  
For objects with multiple optical counterparts we 
assume that the object with the highest likelihood ratio value is the real counterpart of the ISO source.
B.F. stands for Blank Field;
\item {\bf $\Delta(IR-opt)$} : distance in arcseconds between the infrared position and its
optical association.

\end{itemize} 

The complete catalogue is reported in Tables \ref{tabellona}-\ref{tabella_faint}.

\begin{figure}
\centering
\includegraphics[width=0.5\textwidth]{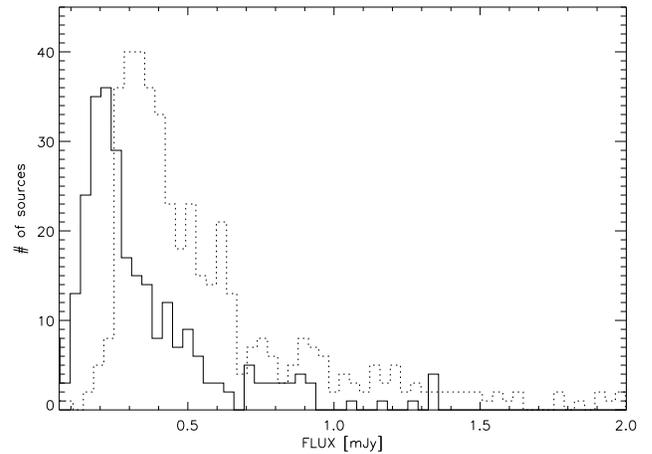}
\caption{Histogram Flux Distributions of the Lockman Deep sources
(solid line) compared to that in the Lockman Shallow (dotted line).}
\label{fig:histo_1}%
\end{figure}

The flux distribution of the catalogue sources is illustrated in Figure \ref{fig:histo_1}
as a solid line, and is compared to the distribution of the Lockman Shallow catalogue
(dotted line, see Paper I). 50\% of the sources have fluxes greater than 0.3 mJy, 
$\sim80\%$ greater then 0.19 mJy and $\sim90\%$ greater then 0.16 mJy.
As can be clearly seen from the two separately plotted histograms, sources
in non-repeated regions (i.e. the Shallow sample) increase in number down to about
0.3 mJy, at which flux their number per flux bin drop sharply. Conversely, the number
of sources per flux bin in repeated regions (i.e. the Deep sample) continues to increase down
to 0.15 mJy.
We note that the high reliability of the catalogue in the Shallow survey has been estimated
to be above $\sim$0.45 mJy (this level corresponds to the 20\% completeness limit, 
see discussion in Paper I). Given that the ratio of the
median exposure times ($t_{exp}$) of the Deep and of the Shallow maps is of the order of $\sim$4, and that
the sensitivity is proportional to $\sqrt{t_{exp}}$, we assume a cutoff of $\sim$0.22 mJy
as a strong confidence level for the catalogue presented in this paper. Below this flux,
a few spurious detections could contaminate the reliability of the source list.
For this reason we have split the catalogue into two different tables: Table \ref{tabellona}
lists the highly reliable detections. In Table \ref{tabella_faint} we report the faint
detections (flux$<$0.22 mJy).

Images in fits format and catalogues in ASCII format are made publicly available through
the world-wide-web\footnote{via anonymous ftp to cdsarc.u-strasbg.fr (130.79.128.5)
or via http://cdsweb.u-strasbg.fr/cgi-bin/qcat?J/A+A/, http://spider.ipac.caltech.edu/staff/fadda/lockman.html
and http://irsa.ipac.catech.edu/data/SPITZER/SWIRE} or directly on request from the authors.

\section{Extragalactic source counts}
\label{sec:counts}

\begin{figure}
\centering
\includegraphics[width=0.5\textwidth]{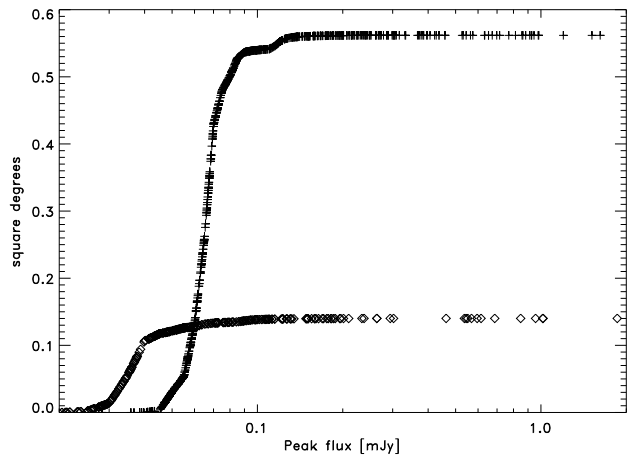}
\caption{Sky coverage as a function of the peak flux in the Lockman Deep 
and in the Lockman Shallow surveys.
Above $\sim$0.1 mJy the areal correction does not affect the two samples.
}
\label{area}%
\end{figure}

\begin{figure}
\centering
\includegraphics[width=.5\textwidth]{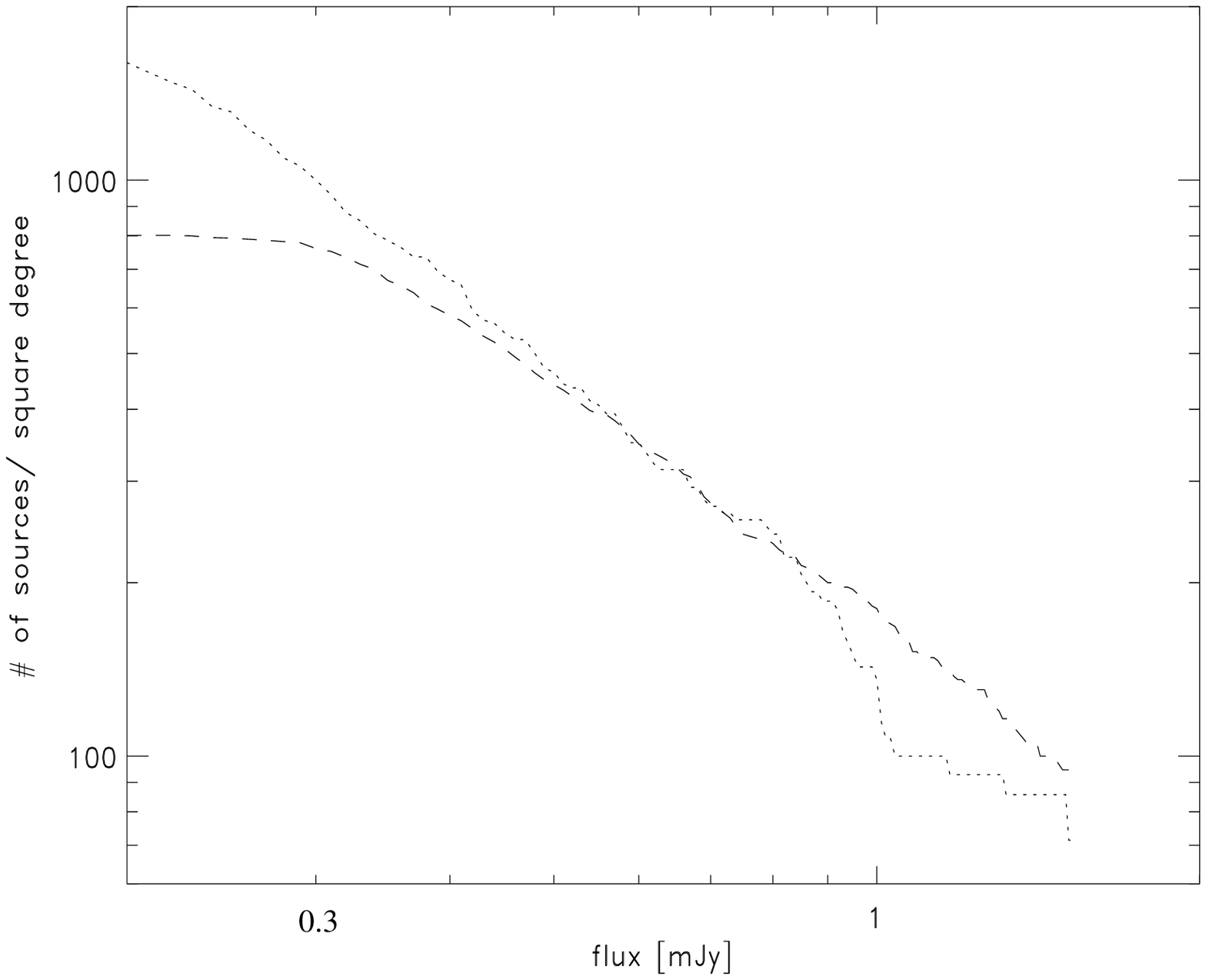}
\caption{Integral source counts for extragalactic ISOCAM sources detected at 14.3 $\mu$m 
in the Deep and Shallow survey, without areal and completeness correction. 
The dotted line refers to the Deep sample, the dashed line to the Shallow one.}  
\label{sinar}%
\end{figure}

In a forthcoming paper (Lari et al. 2004), we will discuss in detail the completeness 
and the source counts down to the fainter flux level reachable with the present 
and the Shallow samples ($\sim 0.2$ mJy). This is to be done by performing intensive 
simulations at different flux levels and spatial scales.
However, in this work we will attempt to characterize the statistical properties of our 
different 14.3 $\mu$m samples (Lockman Deep and Shallow) at the bright levels with the 
presently available data (catalogues and simulations in the shallower survey). 

We have computed the 14.3 $\mu$m source counts down to a flux level of 0.5 mJy. 
We excluded the stars from the computation.
The counts have been obtained by weighting each single source for the effective area 
corresponding to that source flux.
In order to compare the two surveys, we plot in Figure \ref{area} the respective sky coverages
as a function of the peak flux (these are obtained by computing the sky area in the noise map
with a signal lower than a given amount, spanning the range of the peak fluxes in the survey).
It is evident that, above 0.1 mJy in peak flux units, the area 
weight does not affect the counts in both surveys. This converts to about 0.5 mJy in total flux units,
meaning that above this flux level, in principle, the Deep and the Shallow catalogues should
have a similar incompleteness factor.
We can verify this assumption by directly comparing the cumulative source counts in the two
areas, without applying any correction. This is done in Figure \ref{sinar}: the integral
counts are in perfect agreement above 0.5 mJy. Below this level, the shallower sample 
(dashed line) is more affected by incompleteness compared to the deeper one (solid line).
Above $\sim0.9$ mJy the Lockman Deep statistics are insufficient (see Figure \ref{fig:histo_1})
and the computed counts are meaningless.
We then reasonably assume that, above 0.5 mJy, we can apply to both samples the incompleteness
derived through simulations in Paper I for the Lockman Shallow. 
At 1 mJy almost all the sources are detected, and thus the  samples are nearly complete ($\sim 90 \%$) at 
this level. At 0.7 mJy we expect our sample to be $> 70\%$ complete (see Figure 7 in Paper I). 
This completeness level is very close to that found for the ELAIS S2 field (Pozzi et al. 2003).

To compare our results with those reported in the literature (Gruppioni et al, 2002, Lari et al. 2004),
we multiplied the fluxes by the constant factor of 1.0974 (see discussion in Section \ref{rel_aussel})
in order to compare different samples on the same flux scale (i.e. the IRAS scale).
This factor is correlated to discrepancies between the independently established
IRAS and ISO calibrations.

The errors associated with the counts in each level
have been computed as $\sqrt{\sum_i{1/A^2_{eff}(S_i)}}$ (Gruppioni et al. 2002), where the
sum is for all the sources with flux density $S_i$ and $A_{eff}(S_i)$ is the effective area.
The contributions of each source to both the counts and the associated errors are
weighted for the area within which the source is detectable.
The errors represent in any case the Poissonian term of the uncertainties, and have to be
considered as lower limits to the total errors.

\begin{figure*}
\centering
\includegraphics[width=.9\textwidth]{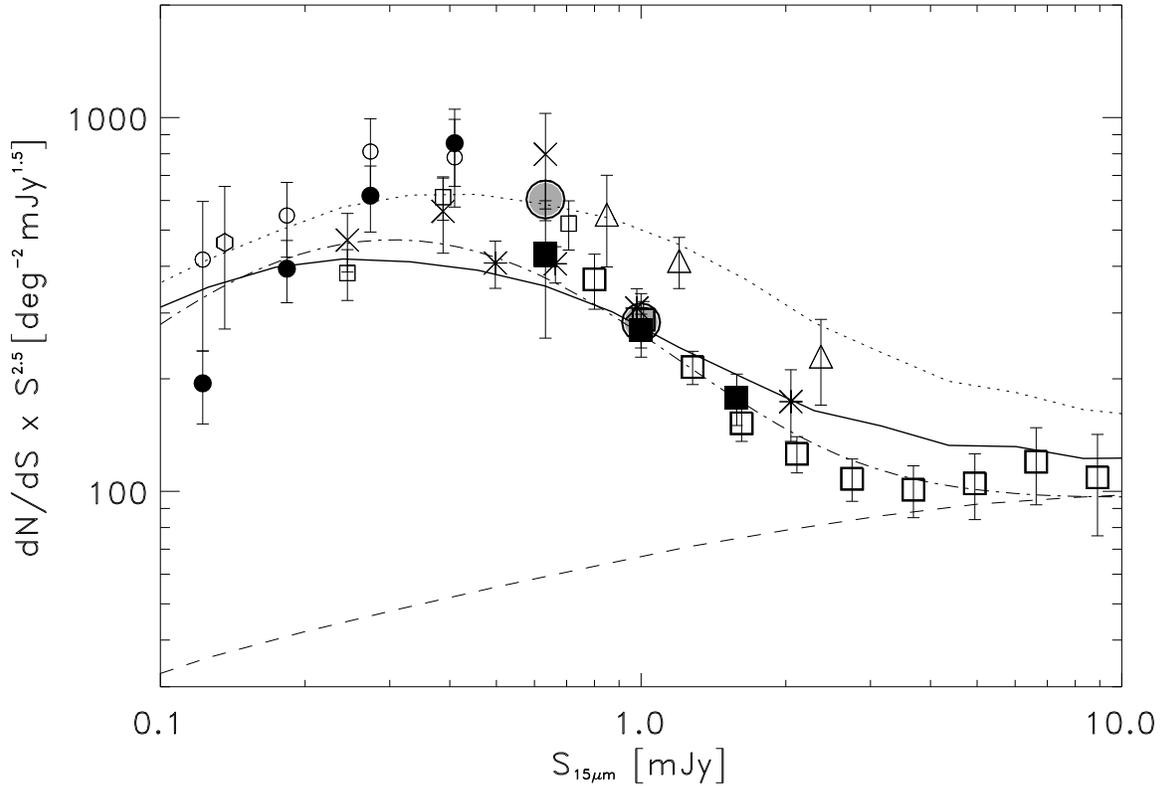}
\caption{14.3 $\mu$m differential counts normalized to the Euclidean
law ($N \propto S^{-2.5}$). Our estimates (Lockman Deep -- grey filled circles,
Lockman Shallow -- black filled squares) are compared with those
from other ISOCAM surveys (A2390 from Altieri, Metcalfe \& Kneib (1999) -- hexagon; 
ISO Hubble Deep Field North (HDF-N) from Aussel et al. (1999) -- filled circles; 
ISO Hubble Deep Field South (HDF-S) -- open circles, 
Marano Firback -- thin open squares, Marano Ultra-Deep -- crosses, Marano Deep -- asterisks,
and Lockman Deep -- triangles from Elbaz et al. (1999); 
ELAIS S1 from Gruppioni et al. 2002 -- thick empty squares).
The data are compared with model predictions: Franceschini et al. (2001, dotted line),
Franceschini et al. (2004 in prep., solid line), Pozzi et al. (2004, dot-dashed line).
The dashed line represents the expectations from a no-evolution model normalized to the 
IRAS 12 $\mu$m local luminosity function (Fang et al. 1998).
}  
\label{fig:cdiff}%
\end{figure*}

In Figure \ref{fig:cdiff} we report the differential 14.3 $\mu$m counts normalized to the Euclidean
law ($N \propto S^{-2.5}$). 
For the Lockman Deep survey, we have limited the computation to between 0.5 and 1 mJy. Below 0.5 mJy we need
to improve the simulations to get a more reliable completeness estimate,
above 1 mJy the number of bright sources drops (see Figure \ref{fig:histo_1}) and the
statistics are insufficient. 

The results presented in this work are in quite good agreement with those already published.
However, the counts by Elbaz et al. (1999) in the Lockman Deep region (open triangles
in Fig. \ref{fig:cdiff})  are systematically higher than ours.
On the one hand, in our analysis we have accurately identified stars by means of the optical r' band image,
and with an objective criterion based on the colour and on the SEDs of the LW3 sources
(Sections \ref{rel_aussel} and \ref{colcol}).
On the other hand, our procedure is different from that used by Elbaz et al. (1999).
In any case, it has been carefully checked by using the stellar fluxes, thus providing high
fidelity photometry.
The good agreement with the results of Gruppioni et al. (2002) in the ELAIS fields, 
could suggest that a re-analysis of the other deep- and ultra-deep fields might
slightly lower the corresponding ISOCAM counts over these flux ranges (this will be done in
a separate paper by Lari et al. 2004).

When the data are compared with predictions, they turn out to be consistent
with the expectations of a slightly modified version of the model by Franceschini et al. (2001).
This multi-wavelength evolution model was designed to reproduce in particular the observed statistics 
of the ISOCAM mid-IR selected sources, but it also accounts for data at other IR and sub-millimetric wavelengths. 
The model assumes the existence of three basic populations of cosmic sources 
characterized by different physical and evolutionary properties: 
a population of non-evolving quiescent spirals, a population of fast evolving sources
(including starburst galaxies and type-II AGNs) and a third component 
considered - but always statistically negligible - are type-I AGNs.
The fraction of the evolving starburst population in the local universe 
is assumed to be $\sim$10\%  of the total, consistent 
with the local observed fraction of interacting galaxies. 
Franceschini et al. (2001) have fitted the IRAS 12 $\mu$m local galaxy luminosity function (LLF)
with an analytic form and a given normalization, by summing the contribution of the three 
different basic populations (see discussion and details in their paper).
In the new configuration, the luminosity and density evolution rate are kept unchanged,
while the luminosity function of the evolving population has been slightly renormalized
downward to better fit the SWIRE/Spitzer 24 $\mu$m counts (Franceschini et al. 2004, in prep). 
The two solutions are plotted in Figure \ref{fig:cdiff}: the dotted line
traces the old predictions by Franceschini et al. (2001), the new ones are
marked as a solid line.
For comparison we report the recent evolution model by Pozzi et al. (2004, dot-dashed line),
which is built to reproduce the brighter ELAIS counts (Gruppioni et al. 2002).
The approach is similar to that presented by Franceschini et al. (2001),
however Pozzi et al. (2004) start from a direct determination of the 15 $\mu$m LLF from the ELAIS southern field.
Even if a better computation of the incompleteness of our catalogues is needed to constrain the faint end
of the distribution, our results fully support
the claims by previous works (Elbaz et al. 1999, Franceschini et al. 2001, Gruppioni et al. 2002) 
for the existence of an evolving population of infrared galaxies. Indeed, our data confirm 
the evident departure from non evolutionary model predictions: the dashed line in 
Figure \ref{fig:cdiff} represents the expectations from a no-evolution model normalized to the 
IRAS 12 $\mu$m local luminosity function (Fang et al. 1998).

\section{Summary and conclusions}

The reduction technique of Lari et al. (2001) has been applied to the ISOCAM 14.3 $\mu$m 
observation of a region of 0.11 square arcminutes in the direction of the Lockman Hole.
We produced a catalogue with 283 sources detected above the 5-$\sigma$ threshold with
fluxes in the interval 0.1-8 mJy. 
This survey is intermediate between the ultra-deep surveys (like the HDF-North) and 
the shallow surveys in the ELAIS fields (Rowan-Robinson et al. 2003). 
The catalogue is 90\% complete at 1 mJy. The positional accuracy, estimated from the 
cross-correlation of infrared and optical sources, is around 1.5 arcsec.

We have checked the calibration (essentially based on the cookbook factor, Blommaert et al. 2000)
comparing the 14.3 $\mu$m measured fluxes with model
predictions, and found a discrepancy of $\sim10\%$ between the ISO and the IRAS 
zero points. We have taken into account this factor when computing galaxy counts,
in order to compare different samples on the same flux scale (as done by
Gruppioni et al. 2002 in the shallower ELAIS fields).  

We found that the 6.7 $\mu$m$/$14.3 $\mu$m colour works well as a star$/$galaxy indicator: we have
combined this method with the visual inspection of an optical image to classify all the 
catalogue sources and reject stars when computing galaxy counts.

The optical counterparts of the sources in the survey have been found by looking at the
infrared contours overlayed on the optical image, making use of the radio detections when
available, and computing the maximum likelihood for every possible optical association. 
For infrared extended and/or blended objects, we reported the list of  
possible optical associations.
Only 15\% of the 14.3 $\mu$m sources turned out to be optically empty fields (no reliable
associations within a search radius of 10 arcseconds). 

We discussed the 14.3 $\mu$m counts of extra-galactic sources. We obtained the 
number densities from the sample presented in this paper (Lockman Deep) and
from the sample introduced in Paper I (Lockman Shallow). The two surveys are
fully consistent, and these  results
support previous findings for a strong evolution of the infrared population. However
a more detailed discussion will be presented in a forthcoming paper, given  
that the completeness of our sample drops at around $\sim$0.5 mJy and extensive
simulations are needed to recover the quantity of undetected sources as a function
of the flux level.

Forthcoming papers will also present the spectroscopic follow-up observations made with 
the CFHT and Keck telescopes, the multi-wavelength spectral energy distributions and a 
cross-correlation between the infrared and radio sources in the field.

Finally, the recently launched Spitzer Space Telescope (25th of August 2003, Fazio et al., 1999)
is not set to observe in the 14.3 $\mu$m channel in imaging mode (the closest ones are centred at 8 and 24 $\mu$m).
Only the IRS spectrograph can observe in this wavelength range, but it is not supposed to be
used for surveys.
This makes our set of data unique. However the planned observations of the Lockman Hole
with Spitzer in its Guaranteed Time and in the SWIRE Legacy program (Lonsdale et al., 2003)
will allow a more complete description of the SEDs of the objects detected by ISO,
and will also extend the redshift flux density coverage.

\begin{acknowledgements}
We thank F. Pozzi for providing her source counts model in electronic form and
C. Gruppioni for the reliabilities of the optical identifications.
We are in debt to the referee, Carol J. Lonsdale, for helpful comments that improved
the quality of the paper.
G.R. wants to thank J. Manners for a careful reading of the paper. 
This work was partly supported by the "POE" EC TMR Network Programme 
(HPRN-AT-2000-00138). 
 
\end{acknowledgements}

\begin{table*} [!ht]
\caption{LW3 source catalogue in the deep Lockman Hole: highly reliable detections.} 
\tiny
\begin{tabular}{clccccccccc} 
\hline 
\hline   
Nr  &    ID                    &      RA          &  DEC          & S/N   & LW3 Flux         &1.4Ghz &S/G& REL  & r'         &$\Delta(IR-opt)$ \\ 
~   &    ~                     &   (J2000)        & (J2000)       &~      & [mJy]            &[mJy]  &~  & ~    & [mag]     & ['']  \\   
\hline 
\hline   
1   &	ISO\_LHDS\_J105121+572415&  10:51:21.942	&  +57:24:15.48	&313    &	8.518$\pm$ 	0.93& --    & 1 &  0.99 &  10.75   &  0.84 \\
2   &	ISO\_LHDS\_J105303+571205&  10:53:03.768	&  +57:12:05.66	&168    &       4.991$\pm$ 	0.54&  0.22 & 0 &  0.99 &  19.34   &  0.78 \\
3   &	ISO\_LHDS\_J105227+571354&  10:52:27.580	&  +57:13:54.48	&169    &	*4.677$\pm$ 	0.49& --    &0  &  0.80 &  21.49   &  1.12 \\
4   &	ISO\_LHDS\_J105201+571044&  10:52:01.174	&  +57:10:44.29	&78	&	4.658$\pm$ 	0.51&  --   & 1 &  0.99 &  10.97   &  4.30 \\
5   &	ISO\_LHDS\_J105153+571900&  10:51:53.682	&  +57:19:00.07	&137    &	4.269$\pm$ 	0.46& --    & 1 &  0.99 &  10.66   &  2.57 \\
6   &	ISO\_LHDS\_J105228+570918&  10:52:28.591	&  +57:09:18.44	&62	&      *3.940$\pm$ 	0.35&  --   & 0 &  0.98 &  19.79   &  1.40 \\
7   &	ISO\_LHDS\_J105143+572937&  10:51:43.701	&  +57:29:37.57	&75	&	*3.520$\pm$ 	0.32&  0.21 & 0 &  0.99 &  16.39   &  0.78 \\
8   &	ISO\_LHDS\_J105242+571914&  10:52:42.468	&  +57:19:14.45	&95	&	3.155$\pm$ 	0.34&  0.22 & 0 &  0.99 &  17.55   &  0.50 \\
9   &	ISO\_LHDS\_J105318+572141&  10:53:18.929	&  +57:21:41.92	&86	&	*3.093$\pm$ 	0.30&  0.21 & 0 &  0.99 &  17.21   &  1.25 \\
10  &	ISO\_LHDS\_J105242+572444&  10:52:42.406	&  +57:24:44.72	&102   &	3.032$\pm$ 	0.33&  0.22 & 0 &  0.99 &  17.14   &  0.28 \\
11  &	ISO\_LHDS\_J105231+571751&  10:52:31.606	&  +57:17:51.90	&75    &	2.807$\pm$ 	0.31& --    & 0 &  0.96 &  22.99   &  0.61 \\
12  &	ISO\_LHDS\_J105113+571427&  10:51:13.442	&  +57:14:27.78	&83&	2.379$\pm$ 	0.26&  0.19 & 0 &  0.98 &  20.10   &  1.43 \\
13  &	ISO\_LHDS\_J105252+572901&  10:52:52.724	&  +57:29:01.01	&76&	2.193$\pm$ 	0.24&  0.23 &0  &  0.99 &  17.32   &  1.28 \\
14  &	ISO\_LHDS\_J105217+572127&  10:52:17.726	&  +57:21:27.63	&43&	1.373$\pm$ 	0.15&  0.26 &0  &  0.98 &  21.79   &  0.94 \\
15  &	ISO\_LHDS\_J105200+571805&  10:52:00.299	&  +57:18:05.87	&48&	1.367$\pm$ 	0.15& --    & 0 &  0.70 &  24.01   &  0.48 \\
16  &	ISO\_LHDS\_J105310+571357&  10:53:10.819	&  +57:13:57.10	&43&	1.354$\pm$ 	0.15& --    & 1 &  0.99 &  11.62   &  6.66 \\
17  &	ISO\_LHDS\_J105307+571826&  10:53:07.824	&  +57:18:26.78	&42&	1.346$\pm$ 	0.15& --    & 1 &  0.99 &  11.66   &  2.87 \\
18  &	ISO\_LHDS\_J105058+572512&  10:50:58.188	&  +57:25:12.67	&42&	1.296$\pm$ 	0.14& --    & 1 &  0.99 &  15.56   &  0.77 \\
19  &	ISO\_LHDS\_J105255+571952&  10:52:55.360	&  +57:19:52.10	&24&	*1.241$\pm$ 	0.12&  0.19 &0  &  0.54 &  24.30   &  2.43 \\
20  &	ISO\_LHDS\_J105258+570925&  10:52:58.243	&  +57:09:25.45	&26&	1.199$\pm$ 	0.13&  --   &0  &  0.99 &  17.93   &  1.13 \\
21  &	ISO\_LHDS\_J105256+570829&  10:52:56.814	&  +57:08:29.41	&13&      *0.957$\pm$      0.09&   0.21&0  &  0.98 &  16.88   &  3.61 \\
22  &	ISO\_LHDS\_J105237+571432&  10:52:37.666	&  +57:14:32.79	&30&	0.942$\pm$ 	0.10& --    &0  &  0.84 &  21.17   &  1.08 \\
23  &	ISO\_LHDS\_J105043+571730&  10:50:43.187	&  +57:17:30.98	&13&	*0.942$\pm$ 	0.06& --    & 0 &  0.98 &  18.99   &  3.70 \\
24  &	ISO\_LHDS\_J105258+570842&  10:52:58.849	&  +57:08:42.50	&14&	0.928$\pm$ 	0.11&  --   & 0 &  --   &  B.F.    &  --   \\
25  &	ISO\_LHDS\_J105200+571312&  10:52:00.010	&  +57:13:12.12	&24&	0.921$\pm$ 	0.10&  --   & 1 &  0.99 &  12.14   &  1.93 \\
26  &	ISO\_LHDS\_J105257+571515&  10:52:57.655	&  +57:15:15.80	&33&	0.917$\pm$ 	0.10&  0.27 &0  &  0.99 &  18.03   &  0.14 \\
27  &	ISO\_LHDS\_J105218+572617&  10:52:18.026	&  +57:26:17.21	&30&	0.916$\pm$ 	0.10& --    &0  &  0.99 &  18.62   &  0.94 \\
28  &	ISO\_LHDS\_J105135+572738&  10:51:35.482	&  +57:27:38.80	&26&	0.914$\pm$ 	0.10&  0.27 &0  &  0.92 &  22.73   &  0.56 \\
29  &	ISO\_LHDS\_J105227+571414&  10:52:27.440	&  +57:14:14.69	&24&	*0.913$\pm$ 	0.09& --    & 0 &  0.99 &  19.85   &  0.63 \\
30  &	ISO\_LHDS\_J105213+571605&  10:52:13.475	&  +57:16:05.03	&35&	0.905$\pm$ 	0.10&  0.22 &0  &  0.96 &  22.52   &  0.69 \\
31  &	ISO\_LHDS\_J105126+572200&  10:51:26.558	&  +57:22:00.71	&23&	*0.880$\pm$ 	0.08& --    &0  &  0.92 &  19.47   &  1.67 \\
32  &	ISO\_LHDS\_J105135+572959&  10:51:35.984	&  +57:29:59.86	&31&	0.871$\pm$ 	0.09& --    &0  &  0.91 &  21.82   &  1.47 \\
33  &	ISO\_LHDS\_J105239+572432&  10:52:39.569	&  +57:24:32.09	&25&	0.858$\pm$ 	0.09&  0.27 & 0 &  0.99 &  17.73   &  0.54 \\
34  &	ISO\_LHDS\_J105306+572807&  10:53:06.764	&  +57:28:07.49	&23&	*0.856$\pm$ 	0.06& --    & 0 &  0.97 &  22.02   &  1.08 \\
35  &	ISO\_LHDS\_J105113+571723&  10:51:13.297	&  +57:17:23.08	&27&	0.854$\pm$ 	0.09& --    &0  &    -- &  B.F.    &   --  \\   
36  &	ISO\_LHDS\_J105234+572643&  10:52:34.978	&  +57:26:43.88	&25&	0.847$\pm$ 	0.09& --    & 0 &  0.59 &  21.17   &  1.30 \\
37  &	ISO\_LHDS\_J105255+572223&  10:52:55.105	&  +57:22:23.69	&27&	0.833$\pm$ 	0.09& --    & 0 &  0.99 &  18.83   &  0.84 \\
38  &	ISO\_LHDS\_J105056+571632&  10:50:56.774	&  +57:16:32.39	&29&	*0.805$\pm$ 	0.08&  0.24 &0  &  0.71 &  20.49   &  1.72 \\
39  &	ISO\_LHDS\_J105225+571130&  10:52:25.642	&  +57:11:30.38	&19&	0.804$\pm$ 	0.09&  0.26 & 0 &  0.99 &  17.73   &  1.30 \\
40  &	ISO\_LHDS\_J105308+571322&  10:53:08.251	&  +57:13:22.44	&21&	*0.795$\pm$ 	0.09&  --   & 0 &  0.99 &  19.57   &  0.87 \\
41  &	ISO\_LHDS\_J105142+571503&  10:51:42.063	&  +57:15:03.95	&23&	*0.794$\pm$ 	0.08&  0.26 & 0 &  0.90 &  18.58   &  2.18 \\
42  &	ISO\_LHDS\_J105250+572608&  10:52:50.254	&  +57:26:08.44	&21&	0.791$\pm$ 	0.09& --    &1  &  0.99 &  11.83   &  5.24 \\
43  &	ISO\_LHDS\_J105315+571939&  10:53:15.736	&  +57:19:39.18	&19&	*0.768$\pm$ 	0.06& --    &0  &  0.90 &  20.73   &  2.88 \\
44  &	ISO\_LHDS\_J105307+571500&  10:53:07.870	&  +57:15:00.84	&18&	0.767$\pm$ 	0.09&  0.28 &0  &  0.90 &  21.68   &  1.01 \\
45  &	ISO\_LHDS\_J105235+572652&  10:52:35.376	&  +57:26:52.60	&26&	0.767$\pm$ 	0.08&  0.23 &0  &  0.98 &  19.16   &  0.74 \\
46  &	ISO\_LHDS\_J105059+571932&  10:50:59.454	&  +57:19:32.29	&23&	0.743$\pm$ 	0.08& --    &0  &  0.82 &  20.56   &  3.72 \\
47  &	ISO\_LHDS\_J105259+573017&  10:52:59.750	&  +57:30:17.30	&18&	0.741$\pm$ 	0.08& --    & 0 &  0.90 &  22.36   &  2.81 \\
48  &	ISO\_LHDS\_J105142+572124&  10:51:42.923	&  +57:21:24.50	&17&	*0.732$\pm$ 	0.06& --    & 0 &  0.99 &  18.65   &  2.10 \\
49  &	ISO\_LHDS\_J105242+573138&  10:52:42.959	&  +57:31:38.58	&12&	0.716$\pm$ 	0.09& --    & 0 &  0.99 &  17.07   &  0.96 \\
50  &	ISO\_LHDS\_J105215+571319&  10:52:15.328	&  +57:13:19.02	&18&	0.666$\pm$ 	0.07&  --   &0  &  0.92 &  21.38   &  0.96 \\
51  &	ISO\_LHDS\_J105215+572634&  10:52:15.140	&  +57:26:34.09	&23&	0.647$\pm$ 	0.07& --    & 0 &  0.98 &  19.31   &  1.41 \\
52  &	ISO\_LHDS\_J105109+572525&  10:51:09.478	&  +57:25:25.14	&22&	0.630$\pm$ 	0.07& --    & 0 &  0.99 &  21.09   &  1.03 \\
53  &	ISO\_LHDS\_J105053+572425&  10:50:53.410	&  +57:24:25.78	&15&	0.627$\pm$ 	0.07&  0.23 & 0 &  0.99 &  19.05   &  0.26 \\
54  &	ISO\_LHDS\_J105110+572143&  10:51:10.919	&  +57:21:43.18	&16&	*0.616$\pm$ 	0.06& --    &1  &  0.99 &  19.35   &  2.23 \\
55  &	ISO\_LHDS\_J105305+572331&  10:53:05.289	&  +57:23:31.37	&23&	0.604$\pm$ 	0.06& --    & 0 &  0.97 &  21.67   &  0.97 \\
56  &	ISO\_LHDS\_J105058+573354&  10:50:58.401	&  +57:33:54.59	&15&	0.604$\pm$ 	0.07& --    & 0 &  0.89 &  23.17   &  2.04 \\
57  &	ISO\_LHDS\_J105233+570933&  10:52:33.252	&  +57:09:33.92	&9      &      *0.580$\pm$ 	0.07&  --   &0  &  0.99 &  19.25   &  2.07 \\
58  &	ISO\_LHDS\_J105314+573021&  10:53:14.378	&  +57:30:21.49	&13&	0.557$\pm$ 	0.06&  0.22 & 0 &  0.84 &  23.71   &  2.10 \\
59  &	ISO\_LHDS\_J105121+571939&  10:51:21.361	&  +57:19:39.07	&20&	0.547$\pm$ 	0.06& --    &0  &  0.71 &  23.62   &  1.79 \\
60  &	ISO\_LHDS\_J105113+572655&  10:51:13.383	&  +57:26:55.25	&16&	0.547$\pm$ 	0.06&  0.25 &0  &  0.84 &  23.11   &  1.35 \\
61  &	ISO\_LHDS\_J105206+571524&  10:52:06.316	&  +57:15:24.51	&18&	0.537$\pm$ 	0.06& --    & 0 &  0.99 &  18.47   &  0.78 \\
62  &	ISO\_LHDS\_J105215+572942&  10:52:15.711	&  +57:29:42.65	&19&	0.529$\pm$ 	0.06& --    & 0 &  0.96 &  22.63   &  0.53 \\
63  &	ISO\_LHDS\_J105203+572707&  10:52:03.629	&  +57:27:07.90	&17&	0.525$\pm$ 	0.06& --    & 0 &  0.78 &  24.22   &  0.89 \\
64  &	ISO\_LHDS\_J105321+572726&  10:53:21.295	&  +57:27:26.24	&15&	0.523$\pm$ 	0.06& --    & 0 &  0.92 &  22.38   &  3.15 \\
65  &	ISO\_LHDS\_J105300+571341&  10:53:00.581	&  +57:13:41.45	&17&	0.512$\pm$ 	0.06&  --   & 1 &  0.99 &  18.63   &  2.93 \\
66  &	ISO\_LHDS\_J105205+572522&  10:52:05.040	&  +57:25:22.54	&18&	0.508$\pm$ 	0.06& --    & 0 &  0.99 &  19.13   &  1.44 \\
67  &	ISO\_LHDS\_J105218+571024&  10:52:18.301	&  +57:10:24.17	&11&      *0.505$\pm$ 	0.04&  --   & 0 &  0.98 &  21.48   &  2.16 \\
68  &	ISO\_LHDS\_J105245+571424&  10:52:45.868	&  +57:14:24.61	&13&	0.502$\pm$ 	0.06& --    & 0 &  0.99 &  21.15   &  0.44 \\
69  &	ISO\_LHDS\_J105207+571500&  10:52:07.250	&  +57:15:00.18	&16&	0.499$\pm$ 	0.05& --    &1  &  0.99 &  16.29   &  1.36 \\
70  &	ISO\_LHDS\_J105148+573248&  10:51:48.860	&  +57:32:48.16	&11&	0.497$\pm$ 	0.06&  0.19 & 0 &  0.95 &  22.26   &  1.28 \\
71  &	ISO\_LHDS\_J105225+571339&  10:52:25.738	&  +57:13:39.61	&16&	0.490$\pm$ 	0.05&  --   & 0 &  0.99 &  21.26   &  1.56 \\
72  &	ISO\_LHDS\_J105301+571456&  10:53:01.153	&  +57:14:56.64	&14&	0.486$\pm$ 	0.05& --    & 0 &  0.94 &  20.93   &  1.26 \\
73  &	ISO\_LHDS\_J105117+571426&  10:51:17.842	&  +57:14:26.13	&16&	*0.485$\pm$ 	0.05& --    & 0 &  0.91 &  22.05   &  1.28 \\
74  &	ISO\_LHDS\_J105237+572148&  10:52:37.549	&  +57:21:48.46	&17&	*0.472$\pm$ 	0.05& --    & 0 &  0.99 &  21.01   &  1.45 \\
75  &	ISO\_LHDS\_J105051+572922&  10:50:51.113	&  +57:29:22.96	&13&	0.470$\pm$ 	0.05& --    & 0 &  0.98 &  20.04   &  1.31 \\
76  &	ISO\_LHDS\_J105251+572736&  10:52:51.846	&  +57:27:36.93	&15&	0.468$\pm$ 	0.05& --    & 1 &  0.99 &  12.26   &  1.70 \\
77  &	ISO\_LHDS\_J105047+571856&  10:50:47.763	&  +57:18:56.10	&12&	0.463$\pm$ 	0.05& --    &0  &  0.88 &  22.74   &  1.62 \\
78  &	ISO\_LHDS\_J105125+572543&  10:51:25.227	&  +57:25:43.70	&15&	0.458$\pm$ 	0.05& --    &0  &  0.94 &  21.92   &  1.97 \\
79  &	ISO\_LHDS\_J105058+572544&  10:50:58.844	&  +57:25:44.55	&14&	0.458$\pm$ 	0.05& --    &0  &  0.98 &  19.58   &  0.87 \\
80  &	ISO\_LHDS\_J105155+571301&  10:51:55.113	&  +57:13:01.90	&19&	0.451$\pm$ 	0.05&  --   & 0 &  0.97 &  21.29   &  1.19 \\
81  &	ISO\_LHDS\_J105233+571233&  10:52:33.447	&  +57:12:33.89	&13&	0.446$\pm$ 	0.05&  --   & 1 &  0.99 &  16.74   &  1.09 \\
82  &	ISO\_LHDS\_J105228+571146&  10:52:28.161	&  +57:11:46.84	&12&	0.445$\pm$ 	0.05&  --   & 0 &  0.99 &  18.37   &  1.72 \\
83  &	ISO\_LHDS\_J105253+572419&  10:52:53.804	&  +57:24:19.28	&15&	0.437$\pm$ 	0.05& --    & 0 &  0.98 &  21.65   &  1.64 \\
84  &	ISO\_LHDS\_J105059+572427&  10:50:59.000	&  +57:24:27.00	&11&	*0.435$\pm$ 	0.05& --    & 1 &  0.99 &  15.93   &  2.11 \\
85  &	ISO\_LHDS\_J105151+572448&  10:51:51.930	&  +57:24:48.59	&14&	0.433$\pm$ 	0.05& --    & 0 &  0.99 &  21.07   &  1.53 \\
86  &	ISO\_LHDS\_J105112+571334&  10:51:12.590	&  +57:13:34.31	&17&	0.432$\pm$ 	0.05&  --   & 0 &  0.99 &  20.40   &  1.18 \\
87  &	ISO\_LHDS\_J105055+571219&  10:50:55.894	&  +57:12:19.64	&9	&	0.430$\pm$ 	0.06&  --   & 0 &  --   &  B.F.    &  --   \\
88  &	ISO\_LHDS\_J105200+572409&  10:52:00.170	&  +57:24:09.97	&16&	0.429$\pm$ 	0.05& --    & 0 &  0.93 &  21.59   &  2.64 \\
89  &	ISO\_LHDS\_J105212+572453&  10:52:12.577	&  +57:24:53.24	&16&	0.414$\pm$ 	0.05&  0.23 &0  &  0.99 &  20.99   &  0.61 \\
90  &	ISO\_LHDS\_J105103+572039&  10:51:03.580	&  +57:20:39.57	&20&	0.411$\pm$ 	0.04& --    & 0 &  0.99 &  19.63   &  2.53 \\
91  &	ISO\_LHDS\_J105240+571601&  10:52:40.248	&  +57:16:01.36	&9	&	0.403$\pm$ 	0.05& --    &0  &  0.96 &  22.42   &  1.77 \\
92  &	ISO\_LHDS\_J105151+571332&  10:51:51.303	&  +57:13:32.56	&14&	0.401$\pm$ 	0.04&  --   & 0 &  0.99 &  18.83   &  0.25 \\
93  &	ISO\_LHDS\_J105154+572333&  10:51:54.055	&  +57:23:33.42	&13&	0.401$\pm$ 	0.05& --    & 0 &  0.99 &  20.93   &  0.52 \\
94  &	ISO\_LHDS\_J105051+573208&  10:50:51.067	&  +57:32:08.58	&10&	0.396$\pm$ 	0.05& --    & 0 &  0.99 &  19.41   &  2.88 \\
95  &	ISO\_LHDS\_J105304+570934&  10:53:04.084	&  +57:09:34.41	&8 &	0.386$\pm$ 	0.05&  --   &0  &  0.98 &  21.84   &  2.00 \\
96  &	ISO\_LHDS\_J105244+573014&  10:52:44.880	&  +57:30:14.57	&12&	0.384$\pm$ 	0.04& --    &0  &  0.79 &  22.60   &  0.96 \\
97  &	ISO\_LHDS\_J105220+571347&  10:52:20.868	&  +57:13:47.88	&12&	0.383$\pm$ 	0.04&  --   & 0 &  0.52 &  24.89   &  2.06 \\
98  &	ISO\_LHDS\_J105156+571910&  10:51:56.008	&  +57:19:10.61	&13&	0.382$\pm$ 	0.04& --    & 0 &  0.99 &  19.31   &  0.38 \\
99  &	ISO\_LHDS\_J105238+573123&  10:52:38.902	&  +57:31:23.67	&8 &	0.379$\pm$ 	0.05& --    & 0 &  0.98 &  21.25   &  2.36 \\
100 &	ISO\_LHDS\_J105049+572527&  10:50:49.312	&  +57:25:27.69	&8 &	0.378$\pm$ 	0.05& --    & 0 &  0.99 &  18.91   &  0.71 \\
\hline 
\hline 
\end{tabular} 
\normalsize
\label{tabellona}
\end{table*}

\begin{table*}
\tiny
\begin{tabular}{clccccccccc} 
\hline 
\hline   
Nr  &    ID                    &      RA          &  DEC          & S/N   & LW3 Flux         &1.4Ghz &S/G& REL  & r'         &$\Delta(IR-opt)$ \\ 
~   &    ~                     &   (J2000)        & (J2000)       &~      & [mJy]            &[mJy]  &~  & ~    & [mag]     & ['']  \\   
\hline 
\hline   
101 &	ISO\_LHDS\_J105129+571606&  10:51:29.824	&  +57:16:06.38	&10&	0.378$\pm$ 	0.05& --    & 0 &  0.68 &  22.88   &  0.61 \\
102 &	ISO\_LHDS\_J105258+572749&  10:52:58.276	&  +57:27:49.79	&9 &	0.378$\pm$ 	0.05& --    & 0 &  0.99 &  19.39   &  1.06 \\
103 &	ISO\_LHDS\_J105154+572408&  10:51:54.323	&  +57:24:08.67	&11&	0.375$\pm$ 	0.04& --    & 0 &  0.92 &  22.09   &  2.72 \\
104 &	ISO\_LHDS\_J105104+573311&  10:51:04.303	&  +57:33:11.26	&5 &	0.375$\pm$ 	0.06& --    &0  &  0.77 &  21.31   &  1.85 \\
105 &	ISO\_LHDS\_J105235+570903&  10:52:35.647	&  +57:09:03.47	&6 &	0.373$\pm$ 	0.06&  --   & 0 &  --   &  B.F.    &  --   \\
106 &	ISO\_LHDS\_J105106+573336&  10:51:06.829	&  +57:33:36.75	&10&	0.373$\pm$ 	0.05& --    & 0 &   --  &  B.F.    &   --  \\
107 &	ISO\_LHDS\_J105243+571253&  10:52:43.040	&  +57:12:53.66	&8 &	*0.367$\pm$ 	0.03&  --   &0  &  0.97 &  21.47   &  1.63 \\
108 &	ISO\_LHDS\_J105232+572450&  10:52:32.019	&  +57:24:50.20	&14&	0.367$\pm$ 	0.04& --    &0  &  0.98 &  21.74   &  1.37 \\
109 &	ISO\_LHDS\_J105255+571149&  10:52:55.723	&  +57:11:49.60	&10&	0.357$\pm$ 	0.04&  --   & 0 &  0.96 &  21.40   &  0.75 \\
110 &	ISO\_LHDS\_J105324+572823&  10:53:24.802	&  +57:28:23.09	&12&	0.357$\pm$ 	0.04& --    & 0 &  0.97 &  20.61   &  0.37 \\
111 &	ISO\_LHDS\_J105149+573303&  10:51:49.991	&  +57:33:03.80	&7 &	0.355$\pm$ 	0.05& --    & 1 &  0.58 &  19.63   &  4.59 \\
112 &	ISO\_LHDS\_J105100+571519&  10:51:00.123	&  +57:15:19.37	&10&	0.354$\pm$ 	0.04& --    & 0 &  --   &  B.F.    &   --  \\ 
113 &	ISO\_LHDS\_J105203+572518&  10:52:03.339	&  +57:25:18.80	&14&	0.352$\pm$ 	0.04& --    & 0 &  0.71 &  18.59   &  3.25 \\
114 &	ISO\_LHDS\_J105128+572854&  10:51:28.991	&  +57:28:54.17	&11&	0.352$\pm$ 	0.04& --    & 0 &  0.99 &  20.74   &  0.51 \\
115 &	ISO\_LHDS\_J105313+572845&  10:53:13.044	&  +57:28:45.12	&14&	0.352$\pm$ 	0.04& --    & 0 &  0.99 &  19.11   &  0.39 \\
116 &	ISO\_LHDS\_J105112+571214&  10:51:12.518	&  +57:12:14.97	&10&	0.351$\pm$ 	0.04&  --   & 0 &  0.68 &  21.86   &  2.56 \\
117 &	ISO\_LHDS\_J105136+572515&  10:51:36.070	&  +57:25:15.37	&10&	0.348$\pm$ 	0.04& --    & 0 &  0.93 &  23.50   &  0.49 \\
118 &	ISO\_LHDS\_J105154+571330&  10:51:54.132	&  +57:13:30.25	&10&	*0.345$\pm$ 	0.03&  --   &0  &  --   &  B.F.    &   -- \\
119 &	ISO\_LHDS\_J105315+571826&  10:53:15.048	&  +57:18:26.31	&11&	0.336$\pm$ 	0.04& --    & 0 &  0.99 &  21.08   &  1.87 \\
120 &	ISO\_LHDS\_J105104+572737&  10:51:04.259	&  +57:27:37.69	&10&	0.332$\pm$ 	0.04& --    & 0 &  0.99 &  20.49   &  1.77 \\
121 &	ISO\_LHDS\_J105309+571700&  10:53:09.110	&  +57:17:00.65	&8 &	0.330$\pm$ 	0.04& --    & 0 &  0.95 &  15.63   &  3.40 \\
122 &	ISO\_LHDS\_J105057+572603&  10:50:57.371	&  +57:26:03.37	&8 &	0.330$\pm$ 	0.04& --    &1  &  0.99 &  15.89   &  1.66 \\
123 &	ISO\_LHDS\_J105301+571539&  10:53:01.868	&  +57:15:39.80	&12&	0.330$\pm$ 	0.04& --    &0  &  0.90 &  21.95   &  2.58 \\
124 &	ISO\_LHDS\_J105111+573311&  10:51:11.679	&  +57:33:11.18	&7 &	0.325$\pm$ 	0.05& --    & 0 &  0.98 &  18.97   &  3.96 \\
125 &	ISO\_LHDS\_J105201+572920&  10:52:01.924	&  +57:29:20.31	&9 &	0.323$\pm$ 	0.04& --    & 1 &  0.99 &  16.17   &  2.07 \\
126 &	ISO\_LHDS\_J105232+572158&  10:52:32.387	&  +57:21:58.22	&8 &	0.322$\pm$ 	0.04& --    &0  &  0.98 &  21.06   &  1.22 \\
127 &	ISO\_LHDS\_J105141+571150&  10:51:41.256	&  +57:11:50.81	&8 &	0.320$\pm$ 	0.04&  --   & 0 &  0.99 &  21.03   &  1.21 \\
128 &	ISO\_LHDS\_J105112+573118&  10:51:12.755	&  +57:31:18.77	&11&	0.318$\pm$ 	0.04& --    & 0 &  0.98 &  21.52   &  1.23 \\
129 &	ISO\_LHDS\_J105220+573132&  10:52:20.959	&  +57:31:32.42	&7 &	0.317$\pm$ 	0.04& --    &0  &  0.63 &  22.78   &  1.13 \\
130 &	ISO\_LHDS\_J105049+572901&  10:50:49.965	&  +57:29:01.07	&7 &	0.311$\pm$ 	0.04& --    & 0 &  0.00 &  25.06   &  7.62 \\
131 &	ISO\_LHDS\_J105051+571515&  10:50:51.243	&  +57:15:15.64	&10&	0.306$\pm$ 	0.04& --    & 0 &  0.99 &  20.72   &  1.87 \\
132 &	ISO\_LHDS\_J105056+573218&  10:50:56.706	&  +57:32:18.16	&6 &	0.305$\pm$ 	0.04& --    & 0 &  0.89 &  23.09   &  2.09 \\
133 &	ISO\_LHDS\_J105123+571228&  10:51:23.486	&  +57:12:28.15	&8 &	0.305$\pm$ 	0.04&  --   & 0 &  --   &  B.F.    &  --   \\
134 &	ISO\_LHDS\_J105300+571651&  10:53:00.781	&  +57:16:51.31	&9 &	0.304$\pm$ 	0.04& --    & 0 &  0.99 &  19.24   &  0.67 \\
135 &	ISO\_LHDS\_J105054+572843&  10:50:54.099	&  +57:28:43.72	&9 &	0.303$\pm$ 	0.04& --    & 0 &  0.77 &  24.17   &  1.22 \\
136 &	ISO\_LHDS\_J105119+571954&  10:51:19.219	&  +57:19:54.88	&10&	0.302$\pm$ 	0.03& --    & 0 &  --   &  B.F.    &   --  \\ 
137 &	ISO\_LHDS\_J105218+571020&  10:52:18.649	&  +57:10:20.62	&8 &      *0.297$\pm$ 	0.02&  --   & 0 &  --   &  B.F.    &   --  \\
138 &	ISO\_LHDS\_J105253+571602&  10:52:53.039	&  +57:16:02.33	&11&	0.297$\pm$ 	0.03& --    &0  &  0.99 &  20.62   &  0.96 \\
139 &	ISO\_LHDS\_J105204+573030&  10:52:04.502	&  +57:30:30.18	&10&	0.296$\pm$ 	0.03& --    & 0 &  0.87 &  22.17   &  2.60 \\
140 &	ISO\_LHDS\_J105120+573238&  10:51:20.285	&  +57:32:38.41	&8 &	0.295$\pm$ 	0.04& --    & 0 &  0.26 &  25.25   &  1.83 \\
141 &	ISO\_LHDS\_J105231+571549&  10:52:31.435	&  +57:15:49.93	&9 &	0.294$\pm$ 	0.03& --    &1  &  0.99 &  18.63   &  3.92 \\
142 &	ISO\_LHDS\_J105143+571836&  10:51:43.377	&  +57:18:36.16	&10&	0.294$\pm$ 	0.03& --    & 0 &  0.96 &  22.84   &  0.93 \\
143 &	ISO\_LHDS\_J105235+572330&  10:52:35.173	&  +57:23:30.25	&8 &	0.289$\pm$ 	0.04& --    & 0 &  0.99 &  19.95   &  0.81 \\
144 &	ISO\_LHDS\_J105059+572715&  10:50:59.864	&  +57:27:15.85	&8 &	0.287$\pm$ 	0.04& --    &0  &  0.87 &  22.92   &  1.32 \\
145 &	ISO\_LHDS\_J105148+571631&  10:51:48.243	&  +57:16:31.69	&8&	0.286$\pm$ 	0.04& --    &0  &  0.90 &  19.76   &  2.70 \\
146 &	ISO\_LHDS\_J105042+571705&  10:50:42.627	&  +57:17:05.05	&8&	0.285$\pm$ 	0.04& --    & 0 &  --   &  B.F.    &   --  \\ 
147 &	ISO\_LHDS\_J105139+573108&  10:51:39.363	&  +57:31:08.45	&9&	0.285$\pm$ 	0.03& --    & 0 &  0.95 &  22.57   &  1.45 \\
148 &	ISO\_LHDS\_J105314+571933&  10:53:14.232	&  +57:19:33.71	&6&	0.285$\pm$ 	0.04& --    & 0 &  0.72 &  25.00   &  1.63 \\
149 &	ISO\_LHDS\_J105310+572324&  10:53:10.719	&  +57:23:24.65	&9&	0.284$\pm$ 	0.03& --    & 0 &  0.96 &  20.64   &  3.73 \\
150 &	ISO\_LHDS\_J105231+573204&  10:52:31.176	&  +57:32:04.74	&5&	0.283$\pm$ 	0.05& --    & 0 &  0.99 &  22.75   &  0.64 \\
151 &	ISO\_LHDS\_J105247+572117&  10:52:47.938	&  +57:21:17.67	&6&	0.283$\pm$ 	0.04& --    & 0 &  0.99 &  20.01   &  1.23 \\
152 &	ISO\_LHDS\_J105044+571539&  10:50:44.609	&  +57:15:39.66	&7&	0.282$\pm$ 	0.04& --    &0  &  --   &  B.F.    &   --  \\ 
153 &	ISO\_LHDS\_J105242+572846&  10:52:42.144	&  +57:28:46.06	&7&	0.281$\pm$ 	0.04& --    & 0 &  0.67 &  15.63   &  5.44 \\
154 &	ISO\_LHDS\_J105219+571528&  10:52:19.309	&  +57:15:28.09	&7&	0.280$\pm$ 	0.04& --    &  0&  0.94 &  22.00   &  0.93 \\
155 &	ISO\_LHDS\_J105311+571238&  10:53:11.205	&  +57:12:38.29	&8&	0.279$\pm$ 	0.04&  --   & 0 &  0.63 &  23.11   &  3.04 \\
156 &	ISO\_LHDS\_J105159+572542&  10:51:59.209	&  +57:25:42.45	&8&	0.278$\pm$ 	0.04& --    &0  &  0.93 &  21.82   &  3.68 \\
157 &	ISO\_LHDS\_J105219+572922&  10:52:19.548	&  +57:29:22.13	&10	&	0.278$\pm$ 	0.03& --    & 0 &  0.95 &  21.70   &  2.47 \\
158 &	ISO\_LHDS\_J105115+571150&  10:51:15.499	&  +57:11:50.91	&6&	0.278$\pm$ 	0.04&  --   & 0 &  0.78 &  22.79   &  3.21 \\
159 &	ISO\_LHDS\_J105151+572806&  10:51:51.333	&  +57:28:06.46	&8&	0.273$\pm$ 	0.03& --    &0  &  0.65 &  22.66   &  2.09 \\
160 &	ISO\_LHDS\_J105100+571955&  10:51:00.971	&  +57:19:55.27	&8&	0.273$\pm$ 	0.03& --    & 0 &  0.61 &  21.89   &  2.13 \\
161 &	ISO\_LHDS\_J105106+571429&  10:51:06.149	&  +57:14:29.43	&8&	0.271$\pm$ 	0.03& --    &0  &  0.91 &  22.48   &  1.39 \\
162 &	ISO\_LHDS\_J105225+572246&  10:52:25.062	&  +57:22:46.04	&9&	0.271$\pm$ 	0.03&  0.29 & 0 &  0.99 &  21.50   &  1.53 \\
163 &	ISO\_LHDS\_J105057+572821&  10:50:57.516	&  +57:28:21.41	&8&	0.271$\pm$ 	0.03& --    & 0 &  0.77 &  22.36   &  4.49 \\
164 &	ISO\_LHDS\_J105232+572837&  10:52:32.520	&  +57:28:37.47	&7&	0.269$\pm$ 	0.03& --    &0  &  0.62 &  23.46   &  3.61 \\
165 &	ISO\_LHDS\_J105057+571333&  10:50:57.441	&  +57:13:33.28	&7&	*0.268$\pm$ 	0.03&  --   & 0 &  0.97 &  19.78   &  3.79 \\
166 &	ISO\_LHDS\_J105048+572520&  10:50:48.092	&  +57:25:20.68	&8&	0.267$\pm$ 	0.04& --    & 0 &  0.99 &  19.77   &  1.15 \\
167 &	ISO\_LHDS\_J105108+572711&  10:51:08.091	&  +57:27:11.26	&10	&	0.266$\pm$ 	0.03& --    &0  &  0.97 &  22.26   &  0.27 \\
168 &	ISO\_LHDS\_J105257+571154&  10:52:57.416	&  +57:11:54.47	&7&	0.263$\pm$ 	0.03&  --   & 0 &  0.96 &  22.93   &  0.81 \\
169 &	ISO\_LHDS\_J105237+571342&  10:52:37.978	&  +57:13:42.31	&8&	0.260$\pm$ 	0.03&  --   & 0 &  0.11 &  24.87   &  4.62 \\
170 &	ISO\_LHDS\_J105144+571720&  10:51:44.428	&  +57:17:20.51	&8&	0.258$\pm$ 	0.03&  0.27 & 0 &  0.94 &  22.07   &  1.05 \\
171 &	ISO\_LHDS\_J105236+573104&  10:52:36.111	&  +57:31:04.19	&5&	0.257$\pm$ 	0.04& --    & 0 &  0.98 &  21.04   &  1.42 \\
172 &	ISO\_LHDS\_J105226+573130&  10:52:26.002	&  +57:31:30.96	&6&	0.255$\pm$ 	0.04& --    & 0 &  0.70 &  22.66   &  3.87 \\
173 &	ISO\_LHDS\_J105219+571854&  10:52:19.678	&  +57:18:54.14	&8&	0.253$\pm$ 	0.03& --    & 0 &  0.97 &  20.44   &  1.90 \\
174 &	ISO\_LHDS\_J105141+571953&  10:51:41.196	&  +57:19:53.12	&9&	0.250$\pm$ 	0.03&  0.22 &0  &  0.85 &  22.01   &  2.25 \\
175 &	ISO\_LHDS\_J105218+572150&  10:52:18.898	&  +57:21:50.09	&7&	0.250$\pm$ 	0.03& --    & 0 &  0.95 &  18.31   &  2.73 \\
176 &	ISO\_LHDS\_J105124+571523&  10:51:24.341	&  +57:15:23.87	&8&	0.250$\pm$ 	0.03& --    & 0 &  0.97 &  20.60   &  3.42 \\
177 &	ISO\_LHDS\_J105108+573343&  10:51:08.231	&  +57:33:43.40	&6&	0.250$\pm$ 	0.04& --    & 0 &  0.89 &  23.17   &  2.04 \\
178 &	ISO\_LHDS\_J105213+571138&  10:52:13.742	&  +57:11:38.72	&7&	0.249$\pm$ 	0.03&  --   &0  &  0.51 &  22.37   &  3.20 \\
179 &	ISO\_LHDS\_J105130+572304&  10:51:30.287	&  +57:23:04.71	&6&	0.249$\pm$ 	0.04& --    & 0 &  --   &  B.F.    &  --   \\
180 &	ISO\_LHDS\_J105125+572901&  10:51:25.158	&  +57:29:01.69	&5&	0.248$\pm$ 	0.04& --    & 1 &  0.98 &  21.76   &  0.14 \\
181 &	ISO\_LHDS\_J105118+571500&  10:51:18.171	&  +57:15:00.87	&7&	0.247$\pm$ 	0.03& --    &0  &  --   &  B.F.    &  --   \\ 
182 &	ISO\_LHDS\_J105119+571424&  10:51:19.649	&  +57:14:24.12	&7&	0.246$\pm$ 	0.03& --    & 0 &  0.73 &  23.76   &  1.43 \\
183 &	ISO\_LHDS\_J105317+571620&  10:53:17.192	&  +57:16:20.06	&8&	0.246$\pm$ 	0.03& --    &0  &  --   &  B.F.    &   --  \\ 
184 &	ISO\_LHDS\_J105227+573125&  10:52:27.095	&  +57:31:25.52	&7&	0.244$\pm$ 	0.03& --    & 0 &  --   &  B.F.    &   --  \\
185 &	ISO\_LHDS\_J105129+571644&  10:51:29.792	&  +57:16:44.01	&7&	0.243$\pm$ 	0.03& --    & 0 &  --   &  B.F.    &  --   \\
186 &	ISO\_LHDS\_J105227+571557&  10:52:27.426	&  +57:15:57.50	&7&	0.242$\pm$ 	0.03& --    &0  &  0.56 &  23.91   &  1.74 \\
187 &	ISO\_LHDS\_J105251+571540&  10:52:51.528	&  +57:15:40.16	&7&	0.242$\pm$ 	0.03& --    &0  &  0.53 &  23.36   &  2.90 \\
188 &	ISO\_LHDS\_J105319+571850&  10:53:19.229	&  +57:18:50.11	&6&	0.242$\pm$ 	0.03&  0.20 & 0 &  0.99 &  19.05   &  2.58 \\
189 &	ISO\_LHDS\_J105228+571152&  10:52:28.369	&  +57:11:52.85	&12	&      *0.242$\pm$ 	0.01&  --   & 0 &  0.93 &  18.37   &  4.66 \\
190 &	ISO\_LHDS\_J105055+571608&  10:50:55.811	&  +57:16:08.53	&8&	0.239$\pm$ 	0.03& --    &0  &  0.98 &  20.65   &  1.72 \\
191 &	ISO\_LHDS\_J105209+573022&  10:52:09.222	&  +57:30:22.13	&9&	0.239$\pm$ 	0.03& --    &0  &  0.90 &  21.00   &  2.73 \\
192 &	ISO\_LHDS\_J105307+572841&  10:53:07.022	&  +57:28:41.38	&7&	0.235$\pm$ 	0.03& --    & 0 &  0.99 &  18.50   &  2.10 \\
193 &	ISO\_LHDS\_J105110+571603&  10:51:10.634	&  +57:16:03.92	&8&	0.235$\pm$ 	0.03&  0.30 &0  &  0.99 &  20.78   &  2.01 \\
194 &	ISO\_LHDS\_J105226+572330&  10:52:26.515	&  +57:23:30.35	&7&	0.233$\pm$ 	0.03& --    &0  &  0.62 &  23.10   &  2.38 \\
195 &	ISO\_LHDS\_J105159+572424&  10:51:59.789	&  +57:24:24.09	&9&	0.232$\pm$ 	0.03& --    & 0 &  0.94 &  21.69   &  1.08 \\
196 &	ISO\_LHDS\_J105055+572823&  10:50:55.235	&  +57:28:23.14	&8&	0.231$\pm$ 	0.03& --    &0  &  0.93 &  22.11   &  2.66 \\
197 &	ISO\_LHDS\_J105229+573024&  10:52:29.266	&  +57:30:24.09	&6&	0.229$\pm$ 	0.03& --    & 0 &  0.99 &  20.66   &  0.93 \\
198 &	ISO\_LHDS\_J105313+572457&  10:53:13.369	&  +57:24:57.57	&7&	0.229$\pm$ 	0.03& --    &0  &  0.79 &  23.79   &  0.44 \\
199 &	ISO\_LHDS\_J105122+571842&  10:51:22.215	&  +57:18:42.31	&9&	0.228$\pm$ 	0.03& --    & 0 &  0.57 &  23.40   &  2.68 \\
200 &	ISO\_LHDS\_J105116+572012&  10:51:16.186	&  +57:20:12.94	&7&	0.227$\pm$ 	0.03& --    &0  &  0.85 &  20.32   &  3.19 \\
201 &	ISO\_LHDS\_J105044+571729&  10:50:44.748	&  +57:17:29.43	&7&	0.227$\pm$ 	0.03& --    & 0 &  0.79 &  22.73   &  3.85 \\
202 &	ISO\_LHDS\_J105243+571545&  10:52:43.090	&  +57:15:45.12	&6&	0.227$\pm$ 	0.03& --    &0  &  0.99 &  20.37   &  0.73 \\
203 &	ISO\_LHDS\_J105246+571325&  10:52:46.650	&  +57:13:25.67	&8&	0.225$\pm$ 	0.03&  --   & 0 &  --   &  B.F.    &   --  \\
204 &	ISO\_LHDS\_J105220+571722&  10:52:20.501	&  +57:17:22.74	&6&	0.225$\pm$ 	0.03& --    & 0 &  0.97 &  20.01   &  2.99 \\
205 &	ISO\_LHDS\_J105315+572449&  10:53:15.416	&  +57:24:49.63	&7&	0.223$\pm$ 	0.03& --    & 0 &  --   &  B.F.    &   --  \\
206 &	ISO\_LHDS\_J105219+571154&  10:52:19.979	&  +57:11:54.68	&6&	0.221$\pm$ 	0.03&  --   & 0 &  --   &  B.F.    &   --  \\
\hline 
\hline 
\end{tabular} 
\normalsize
\end{table*}

\begin{table*}
\caption{LW3 source catalogue in the deep Lockman Hole: faint detections.} 
\tiny
\begin{tabular}{clccccccccc} 
\hline 
\hline   
Nr  &    ID                    &      RA          &  DEC          & S/N   & LW3 Flux         &1.4Ghz &S/G& REL  & r'         &$\Delta(IR-opt)$ \\ 
~   &    ~                     &   (J2000)        & (J2000)       &~      & [mJy]            &[mJy]  &~  & ~    & [mag]     & ['']  \\   
\hline 
\hline   
207 &	ISO\_LHDS\_J105147+572214&  10:51:47.452	&  +57:22:14.47	&6&	0.218$\pm$ 	0.03& --    & 0 &  0.85 &  23.67   &  1.70 \\
208 &	ISO\_LHDS\_J105145+572821&  10:51:45.117	&  +57:28:21.20	&7&	0.217$\pm$ 	0.03& --    & 0 &  0.56 &  23.09   &  3.60 \\
209 &	ISO\_LHDS\_J105242+571346&  10:52:42.653	&  +57:13:46.36	&6&	0.215$\pm$ 	0.03&  --   & 0 &  0.76 &  24.18   &  0.63 \\
210 &	ISO\_LHDS\_J105208+572556&  10:52:08.249	&  +57:25:56.04	&9&	0.215$\pm$ 	0.03& --    &0  &  0.96 &  22.81   &  0.56 \\
211 &	ISO\_LHDS\_J105258+571308&  10:52:58.413	&  +57:13:08.81	&7&	0.214$\pm$ 	0.03&  --   & 0 &  --   &  B.F.    &   --  \\
212 &	ISO\_LHDS\_J105258+571712&  10:52:58.481	&  +57:17:12.47	&7&	0.214$\pm$ 	0.03& --    & 0 &  0.50 &  23.35   &  4.40 \\
213 &	ISO\_LHDS\_J105238+572335&  10:52:38.760	&  +57:23:35.88	&8&	0.213$\pm$ 	0.03& --    &0  &  0.61 &  23.48   &  1.38 \\
214 &	ISO\_LHDS\_J105241+571311&  10:52:41.380	&  +57:13:11.67	&5&	0.212$\pm$ 	0.03&  --   & 0 &  0.52 &  24.41   &  2.88 \\
215 &	ISO\_LHDS\_J105133+571500&  10:51:33.352	&  +57:15:00.27	&7&	0.212$\pm$ 	0.03& --    &0  &  0.96 &  22.91   &  0.52 \\
216 &	ISO\_LHDS\_J105209+572527&  10:52:09.157	&  +57:25:27.52	&9&	0.211$\pm$ 	0.02& --    & 0 &  --   &  B.F.    &   --  \\
217 &	ISO\_LHDS\_J105313+571238&  10:53:13.021	&  +57:12:38.82	&6&	0.211$\pm$ 	0.03&  --   & 0 &  0.98 &  21.82   &  1.27 \\
218 &	ISO\_LHDS\_J105136+571615&  10:51:36.660	&  +57:16:15.82	&8&	0.210$\pm$ 	0.03& --    &0  &  0.98 &  21.64   &  0.95 \\
219 &	ISO\_LHDS\_J105248+571205&  10:52:48.276	&  +57:12:05.15	&6&	0.210$\pm$ 	0.03&  --   & 0 &  0.98 &  21.51   &  0.72 \\
220 &	ISO\_LHDS\_J105258+571647&  10:52:58.635	&  +57:16:47.57	&6&	0.205$\pm$ 	0.03& --    & 0 &  0.27 &  22.36   &  6.16 \\
221 &	ISO\_LHDS\_J105325+572909&  10:53:25.217	&  +57:29:09.08	&7&	0.205$\pm$ 	0.03&  0.20 & 0 &  0.99 &  17.67   &  1.92 \\
222 &	ISO\_LHDS\_J105148+571917&  10:51:48.081	&  +57:19:17.37	&6&	0.204$\pm$ 	0.03& --    & 0 &  0.96 &  20.76   &  2.35 \\
223 &	ISO\_LHDS\_J105300+572028&  10:53:00.179	&  +57:20:28.66	&5&	0.204$\pm$ 	0.03& --    & 0 &  0.89 &  22.70   &  3.04 \\
224 &	ISO\_LHDS\_J105134+572919&  10:51:34.445	&  +57:29:19.66	&6&	0.201$\pm$ 	0.03& --    &0  &  0.71 &  23.66   &  1.62 \\
225 &	ISO\_LHDS\_J105058+571827&  10:50:58.041	&  +57:18:27.51	&6&	0.200$\pm$ 	0.03& --    & 0 &  0.99 &  20.35   &  1.13 \\
226 &	ISO\_LHDS\_J105100+572430&  10:51:00.426	&  +57:24:30.51	&5&	0.199$\pm$ 	0.03& --    & 0 &  0.99 &  19.60   &  2.19 \\
227 &	ISO\_LHDS\_J105159+571953&  10:51:59.513	&  +57:19:53.77	&7&	0.197$\pm$ 	0.03& --    & 0 &  --   &  B.F.    &   --  \\ 
228 &	ISO\_LHDS\_J105059+572209&  10:50:59.928	&  +57:22:09.13	&6&	0.196$\pm$ 	0.03& --    & 0 &  0.92 &  20.17   &  4.25 \\
229 &	ISO\_LHDS\_J105128+572735&  10:51:28.453	&  +57:27:35.04	&8&	0.195$\pm$ 	0.02& --    &0  &  0.74 &  22.05   &  3.92 \\
230 &	ISO\_LHDS\_J105058+572018&  10:50:58.185	&  +57:20:18.88	&6&	0.195$\pm$ 	0.03& --    & 0 &  0.86 &  22.37   &  1.17 \\
231 &	ISO\_LHDS\_J105141+571603&  10:51:41.289	&  +57:16:03.62	&8&	0.194$\pm$ 	0.02& --    & 0 &  0.74 &  23.00   &  4.16 \\
232 &	ISO\_LHDS\_J105108+571403&  10:51:08.218	&  +57:14:03.73	&6&	0.193$\pm$ 	0.03& --    & 0 &  --   &  B.F.    &   --  \\ 
233 &	ISO\_LHDS\_J105314+572507&  10:53:14.805	&  +57:25:07.55	&6&	0.192$\pm$ 	0.03& --    & 0 &  0.89 &  23.07   &  1.94 \\
234 &	ISO\_LHDS\_J105312+571441&  10:53:12.472	&  +57:14:41.83	&5&	0.191$\pm$ 	0.03& --    & 0 &  0.90 &  21.13   &  3.70 \\
235 &	ISO\_LHDS\_J105112+572629&  10:51:12.453	&  +57:26:29.73	&7&	0.190$\pm$ 	0.02& --    &0  &  0.76 &  22.06   &  4.03 \\
236 &	ISO\_LHDS\_J105202+571536&  10:52:02.000	&  +57:15:36.90	&6&	0.190$\pm$ 	0.03& --    &0  &  0.53 &  22.82   &  4.20 \\
237 &	ISO\_LHDS\_J105245+572323&  10:52:45.596	&  +57:23:23.12	&5&	0.190$\pm$ 	0.03& --    & 0 &  0.99 &  18.84   &  0.76 \\
238 &	ISO\_LHDS\_J105145+572229&  10:51:45.038	&  +57:22:29.88	&5&	0.190$\pm$ 	0.03& --    & 0 &  0.98 &  21.47   &  2.07 \\
239 &	ISO\_LHDS\_J105144+572414&  10:51:44.867	&  +57:24:14.00	&5&	0.189$\pm$ 	0.03& --    & 0 &  --   &  B.F.    &   --  \\ 
240 &	ISO\_LHDS\_J105240+572143&  10:52:40.781	&  +57:21:43.64	&6&	0.188$\pm$ 	0.03& --    & 0 &  --   &  B.F.    &   --  \\ 
241 &	ISO\_LHDS\_J105215+571349&  10:52:15.232	&  +57:13:49.97	&6&	*0.186$\pm$ 	0.02&  --   & 1 &  0.99 &  16.43   &  3.92 \\
242 &	ISO\_LHDS\_J105103+571501&  10:51:03.043	&  +57:15:01.62	&6&	0.185$\pm$ 	0.03& --    &0  &  0.97 &  22.34   &  1.31 \\
243 &	ISO\_LHDS\_J105203+572746&  10:52:03.725	&  +57:27:46.97	&6&	0.184$\pm$ 	0.03& --    & 0 &  --   &  B.F.    &  --   \\
244 &	ISO\_LHDS\_J105211+571637&  10:52:11.074	&  +57:16:37.58	&6&	0.179$\pm$ 	0.02& --    &0  &  0.93 &  21.75   &  2.97 \\
245 &	ISO\_LHDS\_J105316+571937&  10:53:16.852	&  +57:19:37.54	&7&	0.177$\pm$ 	0.02& --    & 0 &  0.58 &  24.00   &  1.67 \\
246 &	ISO\_LHDS\_J105044+571806&  10:50:44.456	&  +57:18:06.82	&5&	0.174$\pm$ 	0.03& --    & 0 &  0.86 &  22.82   &  2.42 \\
247 &	ISO\_LHDS\_J105246+571747&  10:52:46.136	&  +57:17:47.34	&5&	0.173$\pm$ 	0.03& --    & 0 &  0.97 &  20.65   &  2.01 \\
248 &	ISO\_LHDS\_J105054+571641&  10:50:54.843	&  +57:16:41.72	&6&	0.172$\pm$ 	0.02& --    & 0 &  0.99 &  21.75   &  1.30 \\
249 &	ISO\_LHDS\_J105231+572458&  10:52:31.639	&  +57:24:58.77	&6&	0.172$\pm$ 	0.02& --    &0  &   --  &  B.F.    &   --  \\ 
250 &	ISO\_LHDS\_J105301+571711&  10:53:01.910	&  +57:17:11.61	&5&	0.171$\pm$ 	0.03& --    & 0 &  0.82 &  22.73   &  2.68 \\
251 &	ISO\_LHDS\_J105103+571419&  10:51:03.867	&  +57:14:19.77	&5&	0.170$\pm$ 	0.03& --    & 0 &  0.93 &  22.46   &  2.50 \\
252 &	ISO\_LHDS\_J105254+571456&  10:52:54.063	&  +57:14:56.91	&6&	0.170$\pm$ 	0.02& --    &0  &  0.99 &  19.87   &  1.90 \\
253 &	ISO\_LHDS\_J105115+572528&  10:51:15.654	&  +57:25:28.98	&5&	0.166$\pm$ 	0.03& --    &0  &  0.66 &  24.15   &  0.71 \\
254 &	ISO\_LHDS\_J105135+572550&  10:51:35.535	&  +57:25:50.52	&6&	0.166$\pm$ 	0.02& --    &0  &  0.90 &  21.34   &  4.32 \\
255 &	ISO\_LHDS\_J105308+571721&  10:53:08.845	&  +57:17:21.61	&5&	0.165$\pm$ 	0.02& --    & 0 &  0.98 &  20.85   &  0.94 \\
256 &	ISO\_LHDS\_J105126+572453&  10:51:26.447	&  +57:24:53.81	&6&	0.165$\pm$ 	0.02& --    &0  &  --   &  B.F.    &   --  \\
257 &	ISO\_LHDS\_J105242+572157&  10:52:42.161	&  +57:21:57.83	&5&	0.163$\pm$ 	0.02& --    & 0 &  --   &  B.F.    &   --  \\ 
258 &	ISO\_LHDS\_J105256+572743&  10:52:56.150	&  +57:27:43.16	&5&	0.163$\pm$ 	0.02& --    & 0 &  0.98 &  21.09   &  2.33 \\
259 &	ISO\_LHDS\_J105206+572301&  10:52:06.176	&  +57:23:01.05	&7&	0.160$\pm$ 	0.02& --    & 0 &  0.99 &  20.71   &  1.09 \\
260 &	ISO\_LHDS\_J105256+572345&  10:52:56.086	&  +57:23:45.69	&5&	0.156$\pm$ 	0.02& --    & 0 &  --   &  B.F.    &   --  \\ 
261 &	ISO\_LHDS\_J105043+571545&  10:50:43.697	&  +57:15:45.37	&5&	0.154$\pm$ 	0.02& --    &0  &  0.50 &  23.35   &  3.92 \\
262 &	ISO\_LHDS\_J105149+571740&  10:51:49.589	&  +57:17:40.46	&5&	0.154$\pm$ 	0.02& --    & 0 &  0.97 &  22.20   &  1.04 \\
263 &	ISO\_LHDS\_J105155+572512&  10:51:55.717	&  +57:25:12.20	&5&	0.154$\pm$ 	0.02& --    & 1 &  0.99 &  16.31   &  3.95 \\
264 &	ISO\_LHDS\_J105219+571826&  10:52:19.155	&  +57:18:26.54	&5&	0.153$\pm$ 	0.02& --    & 0 &  0.57 &  22.71   &  4.87 \\
265 &	ISO\_LHDS\_J105213+571748&  10:52:13.290	&  +57:17:48.79	&6&	0.151$\pm$ 	0.02& --    & 0 &  --   &  B.F.    &   --  \\ 
266 &	ISO\_LHDS\_J105049+571511&  10:50:49.500	&  +57:15:11.71	&5&	0.149$\pm$ 	0.02& --    &0  &  0.96 &  22.08   &  2.09 \\
267 &	ISO\_LHDS\_J105303+572330&  10:53:03.037	&  +57:23:30.51	&5&	0.149$\pm$ 	0.02& --    & 0 &  --   &  B.F.    &   --  \\ 
268 &	ISO\_LHDS\_J105115+571142&  10:51:15.325	&  +57:11:42.84	&6&	0.148$\pm$ 	0.02&  --   &0  &  0.65 &  24.28   &  2.29 \\
269 &	ISO\_LHDS\_J105112+572946&  10:51:12.227	&  +57:29:46.32	&5&	0.148$\pm$ 	0.02& --    & 0 &  0.51 &  22.30   &  4.56 \\
270 &	ISO\_LHDS\_J105256+571132&  10:52:56.350	&  +57:11:32.05	&5&	0.147$\pm$ 	0.02&  --   & 0 &  0.75 &  23.52   &  1.98 \\
271 &	ISO\_LHDS\_J105214+572704&  10:52:14.191	&  +57:27:04.86	&5&	0.145$\pm$ 	0.02& --    & 0 &  --   &  B.F.    &   --  \\ 
272 &	ISO\_LHDS\_J105115+572501&  10:51:15.906	&  +57:25:01.36	&6&	0.139$\pm$ 	0.02& --    & 0 &  0.98 &  21.95   &  0.84 \\
273 &	ISO\_LHDS\_J105236+571844&  10:52:36.396	&  +57:18:44.71	&5&	0.135$\pm$ 	0.02& --    & 0 &  0.98 &  21.64   &  1.28 \\
274 &	ISO\_LHDS\_J105251+571120&  10:52:51.086	&  +57:11:20.60	&5&	0.132$\pm$ 	0.02&  --   &0  &  --   &  B.F.    &   --  \\
275 &	ISO\_LHDS\_J105101+572005&  10:51:01.692	&  +57:20:05.20	&5&	0.132$\pm$ 	0.02& --    & 0 &  --   &  B.F.    &   --  \\ 
276 &	ISO\_LHDS\_J105304+572322&  10:53:04.482	&  +57:23:22.69	&5&	0.121$\pm$ 	0.02& --    & 0 &  0.63 &  24.99   &  0.75 \\
277 &	ISO\_LHDS\_J105242+571251&  10:52:42.361	&  +57:12:51.42	&8&	*0.118$\pm$ 	0.01&  --   & 0 &  --   &  B.F.    &   --  \\
278 &	ISO\_LHDS\_J105150+573240&  10:51:50.650	&  +57:32:40.77	&5&	0.103$\pm$ 	0.01& --    & 0 &  0.60 &  18.69   &  4.99 \\
279 &	ISO\_LHDS\_J105324+572903&  10:53:24.881	&  +57:29:03.59	&5&	0.085$\pm$ 	0.01& --    & 0 &  0.87 &  23.33   &  2.14 \\
280 &	ISO\_LHDS\_J105044+571725&  10:50:44.228	&  +57:17:25.51	&5&	0.059$\pm$ 	0.01& --    & 0 &  --   &  B.F.    &   --  \\ 
281 &	ISO\_LHDS\_J105209+572532&  10:52:09.942	&  +57:25:32.29	&5&	0.058$\pm$ 	0.01& --    & 0 &  0.88 &  22.63   &  2.94 \\
282 &	ISO\_LHDS\_J105043+571722&  10:50:43.731	&  +57:17:22.28	&5&	0.055$\pm$ 	0.01& --    & 0 &  --   &  B.F.    &  3.12 \\
283 &	ISO\_LHDS\_J105241+571917&  10:52:41.206	&  +57:19:17.77	&5&	0.045$\pm$ 	0.01& --    & 0 &  0.80 &  22.56   &  3.15 \\
\hline 														     
\hline 														     
\end{tabular} 													     
\normalsize													     
\label{tabella_faint}												     
\end{table*}

\end{document}